\documentclass[final,3p,times,authoryear]{elsarticle}
\usepackage{amsmath}
\usepackage{amsfonts}
\usepackage{amssymb}
\usepackage[titletoc,title]{appendix}
\usepackage{booktabs}
\usepackage[mmddyyyy]{datetime}
\usepackage{dutchcal}
\usepackage{enumitem}
\usepackage{float}
\usepackage{graphicx}
\usepackage{hyperref}
\usepackage[utf8]{inputenc}
\usepackage{lineno}
\usepackage{natbib}
\usepackage[percent]{overpic}
\usepackage{pgfplots}
\usepackage{relsize}
\usepackage{subfigure}
\usepackage{siunitx}
\usepackage{tikz}

\newcommand{\ra}[1]{\renewcommand{\arraystretch}{#1}}
\modulolinenumbers[5]
\numberwithin{equation}{section}

\usetikzlibrary{calc}
\usetikzlibrary{plotmarks}
\usetikzlibrary{positioning}
\usetikzlibrary{fit}
\usetikzlibrary{decorations.pathmorphing}
\usetikzlibrary{backgrounds}
\makeatother
\modulolinenumbers[5]
\numberwithin{equation}{section}
\pgfplotsset{compat = newest}
\pgfplotsset{ legend style={font=\tiny} }
\definecolor{bgreen}{rgb}{0.0,0.5,0.0}
\definecolor{bblue}{rgb}{0.0,0.0,0.9}
\definecolor{bgold}{rgb}{0.7,0.5,0.0}
\definecolor{bred}{rgb}{0.9,0.0,0.0}
\begin{document}

\begin{frontmatter}



\title{Self-similar orbit-averaged Fokker-Planck equation for isotropic spherical dense clusters  (i) accurate pre-collapse solution}


\author[address1,address2]{Yuta Ito}
\ead{yito@gradcenter.cuny.edu}

\address[address1]{Department of Physics, CUNY Graduate Center, \tnoteref{footnote1}} 
\address[address2]{Department of Engineering and Physics,
	CUNY College of Staten Island\tnoteref{footnote2}}
\tnotetext[footnote1]{ 365 Fifth Avenue, New York, NY 10016, USA}
\tnotetext[footnote2]{2800 Victory Boulevard, Staten Island, NY 10314, USA}

\begin{abstract}
This is the first paper of a series of our works on the self-similar orbit-averaged Fokker-Planck (OAFP) equation for distribution function of stars in dense isotropic star clusters. At the late stage of relaxation evolution of the clusters, standard stellar dynamics predicts that the clusters evolve in a self-similar fashion forming collapsing cores. However, the corresponding mathematical model, the self-similar OAFP equation, has never been solved on the whole energy domain $(-1< E < 0)$. The existing works based on kinds of finite difference methods provide solutions only on the truncated domain $-1< E<-0.2$. To broaden the range of the truncated domain, the present work resorts to a (highly accurate and efficient) Gauss-Chebyshev pseudo-spectral method. We provide a spectral solution, whose number of significant figures is four, on the whole domain. Also, the solution can reduce to a semi-analytical form whose degree of polynomials is only eighteen holding three significant figures. We also provide the new eigenvalues; $c_{1}=9.0925\times10^{-4}$, $c_{2}=1.1118\times10^{-4}$, $c_{3}=7.1975\times10^{-2}$ and $c_{4}=3.303\times10^{-2}$, corresponding to the core collapse rate $\xi=3.64\times10^{-3}$, scaled escape energy $\chi_\text{esc}=13.881$ and power-law exponent $\alpha=2.2305$.  Since the solution on the whole domain is unstable against degree of Chebyshev polynomials, we also provide spectral solutions on truncated domains ( $-1< E<E_\text{max}$, where $-0.35<E_\text{max}<-0.03$) to explain how to handle the instability. By reformulating the OAFP equation in several ways, we improve the accuracy of the spectral solution and reproduce an existing self-similar solution, which infers that existing solutions have only one significant figure at best.
\end{abstract}


\end{frontmatter}

\section{Introduction}\label{sec:intro}

The present paper is the first paper for a series of our works on the self-similar orbit-averaged Fokker-Planck (ss-OAFP) equation and shows an accurate Gauss-Chebyshev spectral solution for pre-collapse stage of relaxation evolution of isotropic star clusters. In the second paper \citep{Ito_2020_2} and third paper \citep{Ito_2020_3}, we will discuss the physical properties of the ss-OAFP model focusing on the negative heat capacity of the core and application to the observed structural profiles of Galactic globular clusters with resolved cores.

The relaxation evolution of core-collapsing dense star clusters (e.g. globular clusters) can not result in a state of thermal equilibrium of stars due to the `negative' heat capacity; as relaxation processes mostly in the core cause stars and kinetic energy to flow from the core to the halo, the core heats up and halo cools down. Once the core density reaches so high that the cluster undergoes the gravothermal instability \citep{Antonov_1985}, it begins to show a self-similar density profile in the core and inner halo \citep{Lynden_Bell_1980}.  Without existence of primordial binary stars or formation and growth of binary stars, the most probable distribution of stars, still in sense of increment of Boltzmann entropy for distribution function (DF) of stars, is a core-collapsed (infinite-density) profile that can be achieved during a finite time duration at the late stage of the self-similar evolution  \citep{Henon_1961, Cohn_1980}.  While the core-collapsing self-similar profile of stars is just a mathematical idealization, it has been one's concern \citep[e.g.][]{Baumgardt_2003,Szell_2005,Pavl_k_2018} since it can provide, in addition to conceptual understandings of the late stage of relaxation evolution, the (asymptotic value of) physical parameters to characterize the evolution; the core collapse rate $\xi$, the power-law index $\alpha$ in spatial density profile and scaled escape energy $\chi_\text{esc}$ in energy distribution function \citep{Heggie_1988}. In the rest of the present section, we explain OAFP equation (Section \ref{subsec:OAFP_eqn}), its self-similar form (Section \ref{subsec:ss_OAFP_eqn}) and numerical difficulty in integration of the ss-OAFP equation (Section \ref{subsec:numeric_OAFP_eqn}).

\subsection{Orbit-averaged Fokker-Planck (OAFP) equation}\label{subsec:OAFP_eqn}

The ideal model of a dense star cluster would be a collection of $N$ equal-mass stars that is isotropic in velocity space and spherical in configuration space; the model can provide a good qualitative understandings of relaxation evolution \citep{Cohn_1980, Takahashi_1995}.\footnote{More realistic star clusters must be modeled as anisotropic systems in velocity space based on statistical and dynamical principles \citep{Polyachenko_1982, Luciani_1987}, numerical results \citep{Cohn_1979, Takahashi_1995, Giersz_1994_2, Baumgardt_2002_2} and observation \citep{Meylan_1987,Meylan_1997}.} Since the total number $N$ of stars in a typical globular cluster is relatively high $\left( N\approx 10^4 \sim10^6 \right)$, one may assume that the orbits of stars are dominated by the self-consistent mean field (m.f.) Newtonian potential $\psi(r)$ on crossing time scales $(t_\text{dyn})$ in zeroth-order of $1/N$ $(N\to\infty)$ or \emph{collisionless} limit \citep{Jeans_1902}. Due to the nature of the long-range interacting stars, the DF of stars may be considered to reach a state of quasi-stationary equilibrium (Virial-equilibrium) through the rapid fluctuation in m.f. potential ('violent relaxation', phase- or chaotic- mixing, ... ). Hence, by assuming that the m.f. potential is regular, the strong Jeans theorem \citep[e.g.][]{Binney_2011} may allow one to simplify the phase-space probability DF at time $t$ as $f(\textbf{r},\textbf{v},t)\approx f(\epsilon)$ for the isotropic system in which the energy of star per unit mass is as follows $\epsilon=\psi(\textbf{r},t)+\frac{\textbf{v}^{2}}{2}$ where $\textbf{r}$ and $\textbf{v}$ are stellar position and velocity. 

This collisionless dynamical-evolution scenario breaks down on relaxation time scales ($t_\text{sec}\sim N t_\text{dyn}/\ln\left[N\right]$) due to the effect of finiteness of total number $N$ of stars; the `smooth' orbits of stars are gradually changed due to (stochastic) irregular forces via many-body Newtonian interaction and the system could reach various quasi-stationary states. In this sense, the explicit time-dependence of DF may be retrieved $\left(f(\textbf{r},\textbf{v},t)\approx f(\epsilon,t)\right)$ and the m.f. potential is to be determined by Poisson equation 
\begin{align}
\frac{\partial^{2} \psi }{\partial r^{2} }+\frac{2}{r}\frac{\partial \psi }{\partial r }=\rho\left[\psi(r,t)\right]\equiv16\pi^{2}Gm\int^{0}_{\psi(r,t)}f(\epsilon,t)\sqrt{2\epsilon'-2\psi(r,t)}\text{d}\epsilon'.\label{Poisson_eqn}
\end{align}
Stellar dynamicists have conventionally modeled the effect of many-body interaction, in first-order approximation of $1/N$, \footnote{See \citep{Gilbert_1968, Ito_2018_2, Ito_2018_3} for more statistically-exact treatment of $1/N$-expansion in $N$-body Liouville equation which includes the effects of the inhomogeneity in encounter, gravitational polarization, statistical acceleration and/or strong encounter.} as a (cumulative) weak two-body encounter with a homogeneous background approximation. The corresponding time-evolution model of DF is a (nonlinear) Fokker-Planck equation averaged over the radial period between the apocenter and pericenter of the orbits, which is known as the orbit-averaged Fokker-Planck (OAFP) equation \cite[e.g.][]{Henon_1961, Spitzer_1988}
\begin{subequations}
	\begin{align}
	&\frac{\partial f(\epsilon,t)}{\partial t}\frac{\partial q(\epsilon,t)}{\partial \epsilon}+\frac{\partial q(\epsilon,t)}{\partial t}\frac{\partial f(\epsilon,t)}{\partial \epsilon}=\Gamma \frac{\partial}{\partial \epsilon}\left\{f(\epsilon,t)\left[f(\epsilon,t)q(\epsilon,t)-j(\epsilon,t)\right]+\frac{\partial f(\epsilon,t)}{\partial \epsilon}\left[i(\epsilon,t)+q(\epsilon,t)g(\epsilon,t)\right]\right\},\label{OAFP_eqn}\\
	&\Gamma\equiv(4\pi G m)^{2}\ln{N}
	\end{align}
\end{subequations}
where $G$ is the gravitational constant, and $m$ the stellar mass. The $q$-integral  (the integral associated with the radial action) reads
\begin{align}
q(\epsilon,t)=\frac{1}{3}\int^{r_\text{max}(\epsilon,t)}_{0}\left[2\epsilon-2\psi(r',t)\right]^{3/2}r^{2}\text{d}r',
\end{align}
where $r_\text{max}(\epsilon,t)=\psi^{-1}(\epsilon)$. The integrals associated with dynamical friction and energy diffusion read
\begin{subequations}
\begin{align}
&i(\epsilon,t)\equiv\int^{\epsilon}_{-1}f\left(\epsilon',t\right)q\left(\epsilon',t\right)\text{d}\epsilon',\\
&j(\epsilon,t)\equiv\int^{\epsilon}_{-1}\frac{\partial f\left(\epsilon',t\right)}{\partial \epsilon'}q\left(\epsilon',t\right)\text{d}\epsilon',\\
&g(\epsilon,t)\equiv\int^{0}_{\epsilon}f\left(\epsilon',t\right)\text{d}\epsilon',
\end{align}
\end{subequations}
where $\psi(0)=-1$ is assumed.

\subsection{Self-similar OAFP equation}\label{subsec:ss_OAFP_eqn}

The OAFP system (i.e. the system of OAFP equation \eqref{OAFP_eqn} and Poisson equation \eqref{Poisson_eqn}) predicts that at the early stage of relaxation evolution the DF of stars may be characterized by a lowered-Maxwellian while at the late stage the cluster may undergo a self-similar evolution \citep{Cohn_1980}. To reflect the self-similar evolution of a core-collapsing isotropic cluster, the following self-similar variables are employed in equations \eqref{OAFP_eqn} and \eqref{Poisson_eqn} for independent variables concerned
\begin{subequations}
\begin{align}
&E=\epsilon/E_{c}(t),\\
&R=r/r_{c}(t),
\end{align}
\end{subequations}
and for dependent variables concerned
\begin{subequations}
\begin{align}
&F(E)=f(\epsilon,t)/f_c(t),\\
&Q(E)=q(\epsilon,t)/q_c(t),\\
&\Psi(R)=\psi(r,t)/\psi_c(t),\\
&I(E)=i(\epsilon,t)/i_c(t),\\
&J(E)=j(\epsilon,t)/j_c(t),\\
&G(E)=g(\epsilon,t)/g_c(t),
\end{align}
\end{subequations}
where suffice $c$ means that the variables depend only on time $t$. Following \citep{Heggie_1988}, one can obtain the ss-OAFP system; a system of four ordinary differential equations (4ODEs)
\begin{subequations}
\begin{align}
\left[I(E)+G(E)Q(E)\right]&\frac{\text{d}F}{\text{d} E}=c_{1}Q(E)F(E)+\frac{2c_{1}-3c_{2}}{4}J(E)-F(E)\left[F(E)Q(E)-J(E)\right],\label{Eq.ss_OAFP_F}\\
&\frac{\text{d}G}{\text{d} E}=-F(E),\label{Eq.ss_OAFP_G}\\
&\frac{\text{d}I}{\text{d} E}=Q(E)F(E),\label{Eq.ss_OAFP_I}\\
&\frac{\text{d}J}{\text{d} E}=Q(E)\frac{\text{d}F}{\text{d} E},\label{Eq.ss_OAFP_J}
\end{align}\label{Eq.ss_OAFP_4ODEs}
\end{subequations}
the $Q$-integral
\begin{align}
Q(E)=\frac{1}{3}\int^{R_\text{max}(E)}_{0}\left[2E-2\Psi\left(R'\right)\right]^{3/2}R'^{2}\text{d}R',\hspace{2cm}\left(R_\text{max}(E)=\Psi^{-1}(E)\right),
\end{align}\label{Eq.ss_OAFP_Q_int}
and Poisson equation
\begin{align}
\frac{\partial^{2} \Psi }{\partial R^{2} }+\frac{2}{R}\frac{\partial \Psi }{\partial R }=D\left[\Psi(R)\right]\equiv\int^{0}_{\Psi(R)}F\left(E'\right)\sqrt{2E'-2\Psi(R)}\text{d}E'.\label{Eq.ss_OAFP_Poisson}
\end{align}
The self-similar parameters read
\begin{subequations}
\begin{align}
&c_{1}=\frac{1}{\Gamma f_{c}(t)}\frac{\text{d}}{\text{d}t}f_{c}(t),\\
&c_{2}=\frac{1}{\Gamma f_{c}(t)}\frac{\text{d}}{\text{d}t}E_{c}(t),
\end{align}
\end{subequations}
and the corresponding physical parameters concerned are
\begin{subequations}        
\begin{align}
&\alpha=\frac{2(3+2\beta)}{2\beta+1},\\
&\xi=\frac{c_{1}+c_{2}}{0.167\sqrt{\pi}},\\
&\chi_\text{esc}=\frac{F_\text{BC}(F_\text{BC}-c_1)}{c_{3}},
\end{align}
\end{subequations}
where $c_{3}(\equiv G(E=-1))$ is the third eigenvalue and the value $F_\text{BC}$ is a boundary value to be assigned.\footnote{Although the boundary condition for the DF in \citep{Heggie_1988, Takahashi_1992} was set to $F(-1)=1$, the present work  specifies the value of $F_\text{BC}$ when it is necessary.} The new eigenvalue $\beta\left(\equiv c_{1}/c_{2}\right)$ characterizes the power-law profile of stars in the halo for each of dependent variables in the following boundary conditions for the 4ODEs \eqref{Eq.ss_OAFP_F}-\eqref{Eq.ss_OAFP_J} and $Q$-integral (equation \eqref{Eq.ss_OAFP_Q_int})
\begin{subequations}
\begin{align}
&F(E\to 0)=c_{4}(\beta+1)(-E)^{\beta},&&F(E=-1)=F_\text{BC},\label{Eq.ss_BCs_4ODE_F}\\
&G(E\to 0)=c_{4}(-E)^{\beta+1},&&G(E=-1)=c_3,\\
&I(E\to 0)=c_{4}\frac{4(\beta+1)}{2\beta-7}(-E)^{\beta+1}Q(E\to 0),&&I(E=-1)=0,\\
&J(E\to 0)=-c_{4}\frac{4\beta(\beta+1)}{2\beta-3}(-E)^{\beta}Q(E\to 0),&&J(E=-1)=0,\label{Eq.ss_BCs_4ODE_J}\\
&Q(E\to 0)\propto(-E)^{\sigma},\hspace{1cm} &&Q(E=-1)=0,
\end{align}\label{Eq.ss_BCs_4ODE}
\end{subequations}
where $\sigma=-3(2\beta-1)/4$ and $c_{4}$ is the fourth eigenvalue. The boundary condition for Poisson equation is
\begin{align}
   \frac{\text{d}\Psi(R=0)}{\text{d}R}=0,\hspace{2cm} \Psi(R=0)=-1.\label{Eq.ss_BCs_Poisson}
\end{align}

\subsection{Numerical problems in integration of ss-OAFP equation and spectral methods}\label{subsec:numeric_OAFP_eqn}

Solving the ss-OAFP system, i.e. solving equations \eqref{Eq.ss_OAFP_4ODEs}-\eqref{Eq.ss_OAFP_Poisson} for the set of dependent variables $\{F, G, J, I, \Phi\}$ and four eigenvalues $\{c_{1}(\text{or}\hspace{0.1cm}\beta), c_{2}, c_{3}, c_{4}\}$ based on the boundary conditions \eqref{Eq.ss_BCs_4ODE}-\eqref{Eq.ss_BCs_Poisson}, is \emph{supposed} to be a simple task compared to more exact models (e.g. time-dependent OAFP model and $N$-body direct simulations). However, it was studied only in a few works \citep{Heggie_1988, Takahashi_1992, Takahashi_1993} in which clear difficulties in numerical integration of the ss-OAFP system were reported. Although \cite{Heggie_1988, Takahashi_1993} found their self-similar solutions, their works are not complete due to the following reasons. First, the domains of their solutions are truncated in energy space, which means the solutions may depend on the extrapolation of power-law profile; they did not discuss the relationship between their solutions and a solution obtained on the whole domain. Second, \cite{Heggie_1988} reported the value of scaled escape energy $\chi_\text{esc}$ is $13.85$ while this value is not compatible with a result of \citep{Cohn_1980} in which, at the same epoch of the energy (=13.85), the central density reaches \emph{only} $10^{20}$ times higher than the initial density; if the value $13.85$ is correct, the \citep{Cohn_1980}'s time-evolution model is supposed to reach an infinite density; one has yet to discuss which of their works is a more accurate result. Third, \cite{Takahashi_1992, Takahashi_1993} tried to reproduce the result of \citep{Heggie_1988} based on a variational principle though it was not a plentiful result; both the works reported that Newton iteration method did not well work unless the initial guess for solution was very close to the `true' solution. 

In the present work, we employ a Gauss-Chebyshev pseudo-spectral method to overcome the numerical difficulties associated with the ss-OAFP model and to obtain a solution on the whole domain. Spectral methods are a very accurate and efficient numerical scheme compared to finite difference (deferred correction) methods, also they can provide a closed form of solution different from finite element methods. Especially, Chebyshev spectral method has the advantages over other spectral methods in the sense that the explicit expression of Chebyshev nodes, numerical differentiation and integrals are known and that its numerical stability and efficiency have been extensively studied \citep[e.g.][]{Boyd_2001}. The ss-OAFP system is associated with infinite-domain problems through Poisson equation; the infinite domain problems have been a matter of concern in applied-mathematics and computational-physics communities as an end-point singularity problem last decades, especially which was discussed for Lane-Emden equations and the variants in astrophysical context \citep[e.g.][]{Parand_2010, Caruntu_2013,Ito_2018}. The present work also aims at extending the numerical scheme developed in \citep{Ito_2018} to the ss-OAFP system.

The present paper is organized as follows. Section \ref{sec:math_formul} explains the transformation of functions and change of variables for the ss-OAFP system that we made to avoid singularities of the functions and to adjust their domains for the spectral method. Section \ref{sec:spectral} explains the Gauss-Chebyshev pseudo-spectral method and also the numerical arrangements that we made to make the Newton iteration method converge. Sections \ref{sec:ss_soln} and \ref{sec:domain_trunc} show the spectral solutions and eigenvalues obtained on whole- and truncated- domains respectively; the former provides the main result of the present work while the latter details the mathematical structure of the ss-OAFP system to validate the spectral solution on the whole domain.  Section \ref{sec:mod_change} reproduces the Heggie-Stevenson's (HS's) solution using the spectral method to see the consistency of our solution. Section \ref{sec:conclusion} makes a conclusion.
\section{Mathematical formulation}\label{sec:math_formul}

The domains of 4ODEs \eqref{Eq.ss_OAFP_F} - \eqref{Eq.ss_OAFP_J} and $Q$-integral (equation \eqref{Eq.ss_OAFP_Q_int}) are finite $(E\in[-1,0))$ while the domain of Poisson equation \eqref{Eq.ss_OAFP_Poisson} is semi-infinite $(R\in[0,\infty))$. To employ the Chebyshev spectral method throughout the present work, in Section \ref{subsec:Inv_Poiss}, we convert the domain of the latter to the same domain as the former employing an inverse function of the m.f. potential $\Psi(R)$ following the inverse-mapping method  \citep{Ito_2018}. Also, since all the dependent variables have power-law profiles forming large-scale gaps between terms in the 4ODEs and Poisson equation, we regularize the variables by the factor $(-E)^{\beta}$, DF $F(E)$ and/or the integral $Q(E)$ in Section \ref{subsec:regulariz}. Lastly, the truncation of the domain is essential following \citep{Heggie_1988,Takahashi_1993}, hence Section \ref{subsec:domain_trunc_integrals} provides the explicit expression of the $Q$- and $D$- integrals on the whole- and truncated- domains.

\subsection{Inverse form of Poisson equation}\label{subsec:Inv_Poiss}

Using the inverse mapping $R$ of $\Psi$ through the local theorem
\begin{align}
\frac{\text{d}\Psi }{\text{d} R }=\frac{1}{\frac{\text{d}R }{\text{d} \Psi }},\hspace{3cm}\frac{\text{d}^{2}\Psi }{\text{d} R^{2} }=-\frac{\text{d}^{2}R }{\text{d} \Psi^{2} }\left(\frac{1}{\frac{\text{d}R }{\text{d} \Psi }}\right)^{3},
\end{align}
we reduced Poisson equation \eqref{Eq.ss_OAFP_Poisson} to
\begin{align}
R(\Psi)\frac{\text{d}^{2}R }{\text{d} \Psi^{2} }-2\left(\frac{\text{d}R }{\text{d} \Psi }\right)^{2}+R(\Psi)\left(\frac{\text{d}R }{\text{d} \Psi }\right)^{3}D(\Psi)=0.\label{Eq.ss_OAFP_Poisson_inv}
\end{align}
The asymptotic approximation of  the inverse form of Poisson equation \eqref{Eq.ss_OAFP_Poisson_inv}  near $\Psi=-1$ (corresponding to the boundary condition \eqref{Eq.ss_BCs_Poisson} at $R=0$) reads
\begin{align}
R(\Psi\to-1)=\left(1+\Psi\right)^{1/2}.\label{Eq.asymp_R}
\end{align} 
Also, the asymptotic approximation of the dependent variable $R$ near $\Psi=0$  is
\begin{align}
R(\Psi\to 0)\propto (-\Psi)^{\nu},\hspace{1cm} \left(\nu=-\frac{2\beta+1}{4}\right).\label{Eq.BC_behave_R}
\end{align}

\subsection{Regularization of ss-OAFP system}\label{subsec:regulariz}

We introduced the following independent variables $x$ and $y$ to employ Chebyshev polynomials (which are defined on $(-1,1)$ to be explained in Section \ref{sec:spectral})
\begin{align}
x\equiv2(-E)^{1/L}-1, \hspace{1cm} y\equiv2(-\Psi)^{1/L}-1,
\end{align}
where $L$ is a numerical parameter introduced to deal with a certain kind of end-point singularities of Chebyshev polynomials \citep{Ito_2018}. Making use of the known asymptotic approximation of dependent variables  (i.e. equations \eqref{Eq.ss_BCs_4ODE}, \eqref{Eq.asymp_R} and \eqref{Eq.BC_behave_R}), we regularized the dependent variables as follows
\begin{subequations}
\begin{align}
&\varv_{R}(y)\equiv\ln\left[R(y)\left(\frac{1-y}{2}\right)^{-1/2}\left(\frac{1+y}{2}\right)^{-L\nu}\right],  && \left(\nu=-\frac{2\beta+1}{4}\right),\\
&\varv_{S}(y)\equiv\ln\left[-S(y)\left(\frac{1-y}{2}\right)^{1/2}\left(\frac{1+y}{2}\right)^{1-L\nu}\right],  &&\\
&\varv_{Q}(x)\equiv\left[Q(x)\left(\frac{1+x}{2}\right)^{-L\sigma}\right]^{1/3},&&\left(\sigma=-\frac{6\beta-3}{4}\right),\label{Eq.reg_Q}\\
&\varv_{F}(x)\equiv\ln\left[F(x)\left(\frac{1+x}{2}\right)^{-\beta L}\right],&&\\
&\varv_{G}(x)\equiv\frac{G(x)}{F(x)},\\
&\varv_{I}(x)\equiv\frac{I(x)}{F(x)Q(x)},\\
&\varv_{J}(x)\equiv\frac{(2\beta-3)J(x)}{4\beta F(x)Q(x)},
\end{align}
\end{subequations}
where the following new dependent variable was introduced for convenience
\begin{align}
S(y)\equiv 2\frac{\text{d}R }{\text{d} y }.
\end{align}

The regularized variables provide more straightforward boundary conditions to understand the relation between the conditions and eigenvalues, compared to the original ones (equations \eqref{Eq.ss_BCs_4ODE_F}-\eqref{Eq.ss_BCs_4ODE_J});
 \begin{subequations}
 \begin{align}
 &\varv_{F}(x\to-1)=\ln{c_{4}^{*}},&& \varv_{F}(x=1)=\ln{F_\text{BC}}\label{BC1},\\
 &\varv_{I}(x\to-1)=0,&&\varv_{I}(x=1)=0,\\
 &\varv_{G}(x\to-1)=0, &&\varv_{G}(x=1)=c_{3},\\
 &\varv_{J}(x\to-1)=-1, &&\varv_{J}(x=1)=0,
 \end{align}
 \end{subequations}
 where $c_{4}^{*}$ is a newly-introduced eigenvalue for convenience and the relation of the eigenvalue $c^{*}_{4}$ with $c_{4}$ in \citep{Heggie_1988}'s work is
 \begin{align}
 c_{4}^{*}\equiv c_{4}(\beta+1).
  \end{align}
 Since all the 4ODEs \eqref{Eq.ss_OAFP_F}-\eqref{Eq.ss_OAFP_J} are first order in differentiation, the eigenvalues (end-point values at $x=-1$) $c_{4}^{*}$ and $c_{3}$ would be directly associated with the boundary conditions at the opposite ends ($\varv_{F}(x=1)$ and $\varv_{G}(x\to-1)=0$) while the eigenvalues $c_1$ and $c_{2}$ would be determined by two of the boundary conditions for $\varv_{I}(x)$ and $\varv_{J}(x)$.

      The inverse form of Poisson equation \eqref{Eq.ss_OAFP_Poisson_inv} reduces to a system of the following two ODEs
\begin{subequations}
\begin{align}
&2(1-y)(1+y)\frac{\text{d}\varv_{R}}{\text{d} y }+2\nu L(1-y)-(1+y)+4e^{\varv_{S}(y)-\varv_{R}(y)}=0,\label{Eq.ss-Poisson_R}\\
&2(1-y)(1+y)\frac{\text{d}\varv_{S}}{\text{d} y }+2L(\nu-1)(1-y)+(1+y)+8e^{\varv_{S}(y)-\varv_{R}(y)}-\frac{4}{L}e^{\varv_{S}(y)}\varv_{D}(y)=0,\label{Eq.ss-Poisson_S}
\end{align}\label{Eq.ss-Poisson}
\end{subequations}
where the regularized density $\varv_{D}(y)$ is
\begin{align}
\varv_{D}(y)\equiv D(y)\left(\frac{1+y}{2}\right)^{-(\beta+3/2)L}=\frac{L}{2}\left(\frac{1+y}{2}\right)^{-L(\beta+3/2)}\int_{-1}^{y}A_{L}(y,x')e^{\varv_{F}(x')}\sqrt{\frac{x-x'}{2}}\left(\frac{1+x'}{2}\right)^{L(\beta+1)-1}\text{d}x',\label{Eq.vD}
\end{align}
where the factor $A_{L}(x,x')$ is
\begin{align}
A_{L}(x,x')\equiv\sqrt{\left(\frac{1+x}{2}\right)^{L}-\left(\frac{1+x'}{2}\right)^{L}}\left(\frac{x-x'}{2}\right)^{-1/2}.
\end{align}
We did not need to employ any boundary conditions for equations \eqref{Eq.ss-Poisson_R}-\eqref{Eq.ss-Poisson_S} since the equations are completely regularized at each end point of the domains of $\varv_{R}(x)$ and $\varv_{S}(x)$; in other words; the equations themselves include their boundary conditions, which appears after the limits of $x\to\pm$ are taken at equation level. 

We regularized the integral $Q(x)$ (equation \eqref{Eq.ss_OAFP_Q_int}) as follows
\begin{align}
\left[\varv_{Q}(x)\right]^{3}=\frac{L}{6\sqrt{2}}\left(\frac{1+x}{2}\right)^{-\sigma L}\int_{x}^{1}\left(\frac{1-y'}{2}\right)^{3/2}A_{L}(y',x)e^{3\varv_{R}(y')}\sqrt{y'-x}\left(\frac{1+y'}{2}\right)^{(3\nu+1)L-1}\text{d}y'.\label{Eq.vQ}
\end{align}
4ODEs \eqref{Eq.ss_OAFP_F} - \eqref{Eq.ss_OAFP_J} reduce to
\begin{subequations}
\begin{align}
&\frac{2}{L}\left(\frac{1+x}{2}\right)^{(\beta-1) L+1}\frac{\text{d}\varv_{F}}{\text{d}x}\left[\varv_{I}(x)+\varv_{G}(x)\right]+\left(\frac{1+x}{2}\right)^{(\beta-1)L}\left\{\beta\left[\varv_{I}(x)+\varv_{G}(x)\right]+\left(\frac{1+x}{2}\right)^{L}\left(\frac{4\beta\varv_{J}(x)}{2\beta-3}-1\right)\right\}+c_{1}e^{-\varv_{F}(x)}\left[1+\varv_{J}(x)\right]=0,\label{Eq.ss-4ODE-vF}\\
&\frac{1+x}{L}\varv_{Q}(x)\left(\frac{\text{d}\varv_{I}}{\text{d}x}+\varv_{I}(x)\frac{\text{d}\varv_{F}}{\text{d}x}\right)+\varv_{I}(x)\left[\frac{-2\beta+3}{4}\varv_{Q}(x)+\frac{3(1+x)}{L}\frac{\text{d}\varv_{Q}}{\text{d}x}\right]+\left(\frac{1+x}{2}\right)^{L}\varv_{Q}(x)=0,\label{Eq.ss-4ODE-vI}\\
&\frac{1+x}{L}\frac{\text{d}\varv_{G}}{\text{d}x}+\varv_{G}(x)\left(\frac{1+x}{L}\frac{\text{d}\varv_{F}}{\text{d}x}+\beta\right)-\left(\frac{x+1}{2}\right)^{L}=0,\label{Eq.ss-4ODE-vG}\\
&\frac{1+x}{L}\varv_{Q}(x)\left\{\frac{\text{d}\varv_{J}}{\text{d}x}+\frac{\text{d}\varv_{F}}{\text{d}x}\left[\varv_{J}(x)-\frac{2\beta-3}{4\beta}\right]\right\}+\varv_{J}(x)\left[\varv_{Q}(x)\frac{-2\beta+3}{4}+3\frac{1+x}{L}\frac{\text{d}\varv_{Q}}{\text{d}x}\right]-\frac{2\beta-3}{4}\varv_{Q}(x)=0.\label{Eq.ss-4ODE-vJ}
\end{align}\label{Eq.ss-4ODE}
\end{subequations}

\subsection{The integral formulations on the whole- and truncated- domains}\label{subsec:domain_trunc_integrals}

When we solved the ss-OAFP system on the whole-domains $x,y\in(-1,1)$, we numerically integrated the integrals $\varv_{D}(y)$ (equation \eqref{Eq.vD}) and  $\varv_{Q}(x)$ (equation \eqref{Eq.vQ}) using Fej$\acute{\mathrm{e}}$r's first rule quadrature
\begin{subequations}
\begin{align}
&\varv_{D}(y)=\frac{L}{2\sqrt{2}}\left(\frac{1+y}{2}\right)^{\frac{1-L}{2}}\int_{-1}^{1}A_{L}\left[y,\frac{(y+1)\left(x'+1\right)}{2}-1\right]e^{\varv_{F}\left[\frac{(y+1)\left(x'+1\right)}{2}-1\right]}\sqrt{1-x'}\left(\frac{1+x'}{2}\right)^{L(\beta+1)-1}\text{d}x',\label{Eq.ss-vD-whole}\\
&\left[\varv_{Q}(x)\right]^{3}=\frac{L}{24}\left(\frac{1-x}{2}\right)^{3}\left(\frac{1+x}{2}\right)^{-\sigma L}\int_{-1}^{1}A_{L}\left[1-\frac{(1-x)\left(1-y'\right)}{2},x\right]e^{3\varv_{R}\left[1-\frac{(1-x)\left(1-y'\right)}{2}\right]}\sqrt{1+y'}\left(1-y'\right)^{3/2}\left[1-\frac{(1-x)(1-y')}{4}\right]^{(3\nu+1)L-1}\text{d}y'.\label{Eq.ss-vQ-whole}
\end{align}
\end{subequations}

On one hand, for numerical integration of the regularized ss-OAFP system on truncated domains $x,y\in(x_\text{min},1)$ where $-1<x_\text{min}<1$, we introduced new independent variables
\begin{align}
z\equiv-2\frac{1-x}{1-x_\text{min}}+1, \hspace{2.5cm}w\equiv-2\frac{1-y}{1-x_\text{min}}+1. \label{Eq.trunc_vari}
\end{align}
Since the original domain of $\varv_{F}(x)$ in the integral $\varv_{D}(w)$ is $(-1,1)$, one must extrapolate $\varv_{F}(x)$ on $(-1,x_\text{min})$. The present work employed the following extrapolated DFs
\begin{align}
\varv_{F}^{\text{(ex)}}(x)\equiv\begin{cases}\ln\left\{c^{*}_{4}+\frac{2c^{*}_{4}}{1-x_\text{min}}\frac{\,\text{d}\varv_{F}(w=-1)}{\,\text{d}w}e^{d(1+x_\text{min })^{-c}\left[1-\left(\frac{1+x_\text{min}}{1+x}\right)^{c}\right]}(x-x_\text{min})\right\},\hspace{2cm} (\text{smooth at }x=x_\text{min})\\
\ln[c^{*}_{4}],\hspace{7cm} (\text{non-smooth at }x=x_\text{min}\quad \text{or}\quad c\to \infty)
\end{cases}\label{Eq.vF_extrap}
\end{align}
where $c$ and $d$ are numerical parameters. Hence, $\varv_{D}(w)$ is composed of the total of integrals $\varv_{D}^{(\text{nonex})}(w)$ and $\varv_{D}^{(\text{ex})}(w)$; the former is contribution to $\varv_{D}(w)$ from the (non-extrapolated) DF $\varv_{F}(x)$ and the latter is from an extrapolated DF $\varv_{F}^\text{(ex)}(x)$ as follows
\begin{align}
\varv_{D}(w)=\varv_{D}^{(\text{nonex})}(w)+\varv_{D}^{(\text{ex})}(w).\label{Eq.ss-vD-domain}
\end{align}
The non-extrapolated $\varv_{F}(x)$ on $x_\text{min}<x<1$ contributes to $\varv_{D}(w)$ as follows
\begin{align}
&\varv_{D}^{(\text{nonex})}(w)=
\frac{1}{2\sqrt{2}}\left(\frac{1-x_\text{min}}{2}\right)^{3/2}\left[\frac{1}{2}+\frac{2+x_\text{min}(1-w)}{2(1+w)}\right]^{-3/2}\nonumber\\
&\hspace{2cm}\times\int_{-1}^{1}e^{\varv_{F}\left[\frac{(w+1)\left(z'+1\right)}{2}-1\right]}\sqrt{1-z'}\left\{1-\frac{(1-z')(1+w)(1-x_\text{min})}{2\left[3+x_\text{min}+z'(1-x_\text{min})\right]}\right\}^{\beta}\,\text{d}z',
\end{align}
where $L=1$ is assumed for simplicity. On one hand, the contribution of $\varv_{F}^{\text{(ex)}}(x)$ on $-1<x<x_\text{min}$ to $\varv_{D}(w)$ reads
\begin{align}
\varv_{D}^{\text{(ex)}}(w)\equiv\frac{1}{2}\int^{1}_{-1}\left(\frac{1+x_\text{min}}{1+y}\right)^{\beta+3/2}e^{\varv_{F}^{\text{(ex)}}\left[\frac{(1+x_\text{min})(1+z'')}{2}-1\right]}\sqrt{\frac{1+y}{1+x_\text{min}}-\frac{1+z''}{2}}\left(\frac{1+z''}{2}\right)^{\beta}\,\text{d}z'',
\end{align}
where $y=\frac{1}{2}\left[1+x_\text{min}+w(1-x_\text{min})\right]$.  
Lastly, the integral $\varv_{Q}(z)$ on the truncated domain is
\begin{align}
&\left[\varv_{Q}(z)\right]^{3}=
\frac{1}{24}\left(\frac{1-x}{2}\right)^{3}\left(\frac{1+x}{2}\right)^{-3/2}\nonumber\\
&\hspace{2cm}\times\int_{-1}^{1}e^{3\varv_{R}\left[1-\frac{(1-z)\left(1-w'\right)}{2}\right]}\sqrt{1+w'}\left(1-w'\right)^{3/2}\left\{1+\frac{(1-z)(1+w')(1-x_\text{min})}{4\left(1+x\right)}\right\}^{3\nu}\,\text{d}w',\label{Eq.ss-vQ-domain}
\end{align}
where $L=1$ is assumed and $x=\frac{1}{2}\left[1+x_\text{min}+z(1-x_\text{min})\right]$. 

\section{Gauss-Chebyshev spectral method and numerical treatments of ss-OAFP system}\label{sec:spectral}

Sections \ref{Chebyshev_spectral} and \ref{subsec:numeric_treat} explain Gauss-Chebyshev pseudo-spectral method and numerical treatment of the ss-OAFP system respectively.

\subsection{The Gauss-Chebyshev pseudo-spectral method}\label{Chebyshev_spectral}

Chebyshev polynomials of the first kind is defined on domain $x\in [-1,1]$ as \citep[e.g.][]{Boyd_2001, Mason_2002}
\begin{align}
T_{n}(x)=\cos\left[n \cos^{-1}(x)\right],  \qquad\quad(n=0,1,2,\cdots,\mathcal{N})
\end{align}
Due to the singularities in 4ODES \eqref{Eq.ss-4ODE-vF}-\eqref{Eq.ss-4ODE-vF} and Poisson equation \eqref{Eq.ss-Poisson} at the endpoints $x=\pm 1$, we had to solve the equations as an open-interval problem $x\in (-1,1)$\footnote{The Poisson equation is regular singular at both the end points of the domain and the $D$- and $Q$-integral also have a singular property as $x\to-1$. In this sense, to handle the singularities, we employed Gauss-Chebyshev nodes by considering the domain to be an open interval \citep[See e.g.][for application of Gauss-Chabyshev spectral methods ]{Bhrawy_2012,Boyd_2013}.}. Hence, the discretized domain of the polynomials at Gauss-Chebyshev points is
\begin{equation}
x_{k}=\cos(t_{k})\equiv\cos\left(\frac{2k-1}{2\mathcal{N}}\right). \qquad (k=1,2,3 \cdots \mathcal{N})
\end{equation}
The discrete Gauss-Chebyshev polynomials $T_{n}\left(x_{j}\right)$ of the first kind  satisfy the orthogonality condition (e.g. \cite{Mason_2002}) 
\begin{equation}
\sum^{\mathcal{N}}_{j=1}T_{n}\left(x_j\right)T_{m}\left(x_j\right)=\left\{\begin{matrix}
&0,&\\
&\mathcal{N},&\\
&\frac{\mathcal{N}}{2}.&
\end{matrix}\right.\hspace{1cm}
\begin{matrix}
&(1<n\neq m<\mathcal{N})&\\
&(n=m=0)&\\
&(0<n=j\leq \mathcal{N})&
\end{matrix}
\end{equation}
Hence, the discrete Gauss-Chebyshev polynomial expansion of any function $h(x)$ and its derivative are
\begin{align}
h(x_j)=\sum_{n=1}^{\mathcal{N}}a_{n}T_{n-1}(x_j),\label{Eq.Cheb_expan}\hspace{3cm}
\frac{\text{d}h\left(x_j\right)}{\text{d}x}=\sum_{n=1}^{\mathcal{N}}a_{n}\frac{n\sin\left({n\cos^{-1}{x_j}}\right)}{\sin\left(\cos^{-1}{x_j}\right)},
\end{align}\label{DCP}
and the Chebyshev-Gauss expansion can be inverted to
\begin{subequations}
	\begin{align}
	&a_{1}=\frac{1}{\mathcal{N}}\sum^{\mathcal{N}}_{j=1}T_{0}(x_{j})h(x_j),\label{DCC1}\\
	&a_{n}=\frac{2}{\mathcal{N}}\sum^{\mathcal{N}}_{j=1}T_{n-1}(x_{j})h(x_j). \label{DCC2} \hspace{1cm}(2\leq n\leq \mathcal{N})
	\end{align}\label{Eq.inv_Cheb_expan}
\end{subequations}

\subsection{Numerical treatments of the ss-OAFP equation}\label{subsec:numeric_treat}

In a similar way to \citep{Heggie_1988}'s work, we had to carry out many numerical arrangements. First, Newton iteration method for the whole-domain formulation did not work at all. Hence we truncated the domain of $\varv_{Q}$ and differentiation $\frac{\,\text{d}}{\,\text{d}x}$ in the 4ODEs employing equation \eqref{Eq.trunc_vari}. Then, this arrangement provided spectral solutions on $x_\text{min}\approx-0.96<x<1$. Also, truncated-domain formulation did not work, hence, we regularized $\varv_{I}(x)$ and $\varv_{J}(x)$ by the factor $(1+x)/2$ so that $\lim_{x\to x_\text{min}}2\varv_{I}(x)/(1+x)=4/(2a-7)$ and $\lim_{x\to x_\text{min}}2\varv_{G}(x)/(1+x)=1/(a+1)$. This arrangement provided solutions on $x_\text{min}\approx-0.2<x<1$. To broaden the range of the effective interval $(x_\text{min}<x<1)$, following \citep{Heggie_1988}, we shortened the Newton steps in the iteration process though, it did not work. 

To overcome the difficulty in convergence of Newton method, we fixed the eigenvalue $\beta$ to a certain value during iteration process. For the fixed $\beta$-value, once we found a solution at a specific $x_\text{min}$, we chose a new $\beta$ that is close to the old $\beta$. Then, we found a new solution for the new $\beta$ using Newton iteration method. We repeated this process until $\varv_{I}(x=1)$ reached its minimum. Then, at a new $x_\text{min}$ that is \emph{very} close to the old $x_\text{min}$ with new $\beta$ that is \emph{very} close to old $\beta$,\footnote{For example, to find the whole-domain solution, the change $\delta\beta$ in $\beta$ was $0.03$ from $x_\text{min}=-0.94$ to $-0.96$, $\delta\beta\approx0.001$ from  $x_\text{min}=-0.9994$ to $-0.9996$, and $\delta\beta\approx0.000001$ from  $x_\text{min}=-0.999994$ to $-0.999995$.} we repeated the whole process above. As a result, $x_\text{min}$ reached $-1$ for the whole-domain formulation while $x_\text{min}$ reached $-0.96$ for the domain-truncated formulation. 

Also, since the eigenvalue $\beta$ was fixed during the iteration process, we speeded up the numerical integration of the integrals $\varv_{Q}(x)$ and $\varv_{D}(x)$ by applying the Fej$\acute{\mathrm{e}}$r's first rule quadrature to the integrals \emph{before} the iteration process starts. For example, we discretized $\varv_{D}(y)$ as follows
\begin{align}
\varv_{D}(y_{j})=\sum_{n=1}^{\mathcal{N}}F_{n}^{\text{linear}}\mathcal{D}_{n}(y_{j})
\end{align}
where $\{F_{n}^{\text{linear}}\}$ is the Chebyshev coefficients of function $\varv_{F}^{\text{linear}}(x)(\equiv\exp{\left[\varv_{F}(x)\right]}$). One can obtain $\{F_{n}^{\text{linear}}\}$ from $\varv_{F}(x)$ by using equations \eqref{Eq.Cheb_expan} and \eqref{Eq.inv_Cheb_expan}. The matrix $\mathcal{D}_{n}(y_{j})$ is a preset matrix to be integrated before the Newton-iteration (loop) process begins and explicitly reads
\begin{align}
&\mathcal{D}_{n}(y_{j})=\frac{L}{2\sqrt{2}}\left(\frac{1+y_{j}}{2}\right)^{\frac{1-L}{2}}\nonumber\\
&\hspace{2cm}\times\int_{-1}^{1}A_{L}\left[y_{j},\frac{(y_{j}+1)(x'+1)}{2}-1\right]T_{n}\left[\frac{(y+1)(x'+1)}{2}-1\right]\sqrt{1-x'}\left(\frac{1+x'}{2}\right)^{L(\beta+1)-1}\,\text{d}x'.
\end{align}
We also prepared a similar preset matrix for $\varv_{Q}(x)$. As a result, the two preset matrices made the iteration process $10\sim100$ times more efficient\footnote{Using a 2.4 GHz CPU processor, the resulting CPU time for $10^{4}$ iterations was $\sim$ a few min for $\mathcal{N}=70$, which was needed to find solutions near $x_\text{min}=-1$.} than the original iteration process in which we implemented the Fej$\acute{\mathrm{e}}$r's first rule quadrature for each iteration.

\section{Self-similar solution on the whole domain}\label{sec:ss_soln}

As the main result, we provide the whole-domain solution, its semi-analytical form and eigenvalues (Section \ref{sec:result_whole}). Section \ref{sec:ss_soln_asymp} details the asymptotic approximation of the solution and the characteristics of the Chebyshev coefficients. Section \ref{sec:inst_whole} discusses the numerical stability of the solution and reports that the solution is unstable against degree $\mathcal{N}$.

\subsection{Numerical results (main results of the present paper)}\label{sec:result_whole}

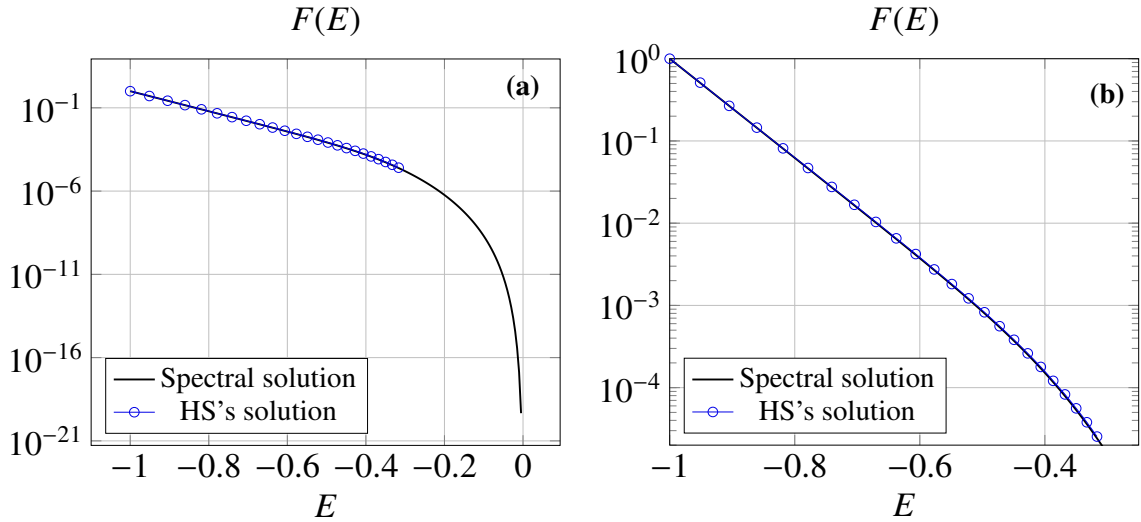
\begin{figure}[H]
	\centering
	\tikzstyle{every node}=[font=\Large]
	\begin{tikzpicture}[scale=0.9]
	\begin{semilogyaxis}[ grid=major,xlabel=\Large{$E$},title=\Large{$F(E)$}, legend pos=south west]
	\addplot [color = black ,mark=no,thick,solid] table[x index=0, y index=1]{SolnF_N70.txt}; 
	\addlegendentry{\large{Spectral solution}}
	\addplot [color = bblue ,mark=o,mark options={solid},smooth] table[x index=0, y index=1]{HS_soln.txt}; 
	\addlegendentry{\large{HS's solution}}
      \node[above,black] at (0.0,1e-1) {\large{$\textbf{(a)}$}};
	\end{semilogyaxis}
	\end{tikzpicture}\hspace{0.3cm}
	\begin{tikzpicture}[scale=0.9]
	\begin{semilogyaxis}[ grid=major,xlabel=\Large{$E$},title=\Large{$F(E)$},xmin=-1e0,xmax=-0.25,ymin=2e-5,ymax=1e0, legend pos=south west]
	\addplot [color = black ,mark=no,thick,solid] table[x index=0, y index=1]{SolnF_N70.txt}; 
	\addlegendentry{\large{Spectral solution}}
	\addplot [color = bblue ,mark=o,mark options={solid},smooth] table[x index=0, y index=1]{HS_soln.txt}; 
	\addlegendentry{\large{HS's solution}}
      \node[above,black] at (-0.30,2e-1) {\large{$\textbf{(b)}$}};
	\end{semilogyaxis}
     \hspace{0.5cm}
	\end{tikzpicture}
	\caption{$\textbf{(a)}$ Distribution function $F(E)$ of stars on the whole domain and $\textbf{(b)}$ its magnified graph on $-1\leq E<-0.25$. ($\mathcal{N}=70$, $F_\text{BC}=1$, $L=1$.)}
	\label{fig:solnF_N70}
\end{figure}

\begin{figure}[H]
	\centering
	\tikzstyle{every node}=[font=\Large]
	\begin{tikzpicture}[scale=0.9]
	\begin{loglogaxis}[ grid=major,xlabel=\Large{$R$},title=\Large{$\mid\Phi(R)\mid$}, legend pos=south west]
	\addplot [color = black ,mark=no,thick,solid] table[x index=0, y index=1]{Phi_N70.txt}; 
	\addlegendentry{\large{Spectral solution}}
	\addplot [color = bblue ,mark=o,mark options={solid},smooth] table[x index=0, y index=1]{HS_Phi.txt}; 
	\addlegendentry{\large{HS's solution}}
\node[above,black] at (1e12,0.8) {\large{$\textbf{(a)}$}};
	\end{loglogaxis}
	\end{tikzpicture}
\hspace{0.5cm}
	\begin{tikzpicture}[scale=0.9]
	\begin{semilogxaxis}[ grid=major,xlabel=\Large{$R$},title=\Large{$\mid\Phi(R)\mid$},xmin=4e0,xmax=7e2,ymin=0.3,ymax=0.95, legend pos=south west]
	\addplot [color = black ,mark=no,thick,solid] table[x index=0, y index=1]{Phi_N70.txt}; 
	\addlegendentry{\large{Spectral solution}}
	\addplot [color = bblue ,mark=o,mark options={solid},smooth] table[x index=0, y index=1]{HS_Phi.txt}; 
	\addlegendentry{\large{HS's solution}}
\node[above,black] at (5e2,0.85) {\large{$\textbf{(b)}$}};
	\end{semilogxaxis}
	\end{tikzpicture}
	\caption{ \textbf{(a)} Self-consistent m.f. potential $\Phi(R)$ of stars on the whole potential range and \textbf{(b)} its magnified graph on $-1<\Phi(R)<-0.25$. ($\mathcal{N}=70$, $F_\text{BC}=1$, $L=1$.)}
	\label{fig:solnPhi_N70}
\end{figure}
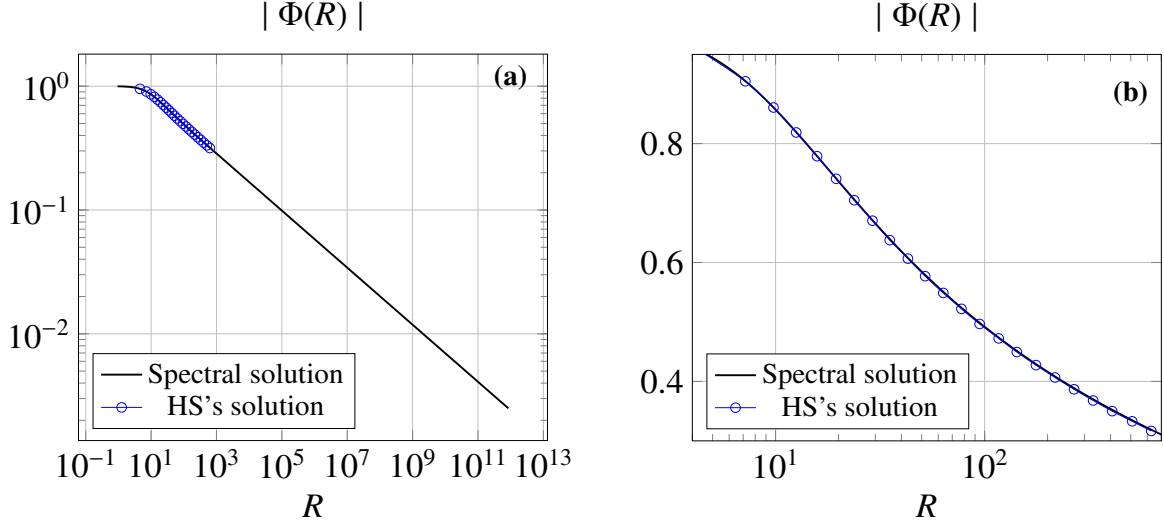

We found the whole-domain solution compatible to the HS's solution. Figures \ref{fig:solnF_N70}\textbf{(a)} and \ref{fig:solnPhi_N70} \textbf{(a)} depict DF $F(E)$ and m.f. potential $\Phi(R)$ obtained from the whole-domain spectral solution. In the figures, the HS's solution is also depicted. The spectral- and HS's solutions are visually almost identical on the figures. For the whole-domain solution, the optimal values of numerical parameters are  $\mathcal{N}=70$, $F_\text{BC}=1$, $L=1$. The optimal eigenvalue of $\beta$ is
\begin{align} 
\beta_\text{o}\equiv8.1783711596581.
\end{align}
We chose the value of $\beta_\text{o}$ so that $\varv_{I}(x=1)$ reached its minimum value $(\sim10^{-12})$. In order to make Newton iteration method work, we needed to correctly specify at least eight significant figures of $\beta_\text{o}$ (Appendix  \ref{sec:stability_eigen}). Also, degree $\mathcal{N}=70$ is the minimum value among $70\leq \mathcal{N}\leq400$ for which Newton iteration method worked (Section \ref{sec:inst_whole}).  Figures \ref{fig:solnF_N70}\textbf{(b)} and \ref{fig:solnPhi_N70} \textbf{(b)} show the magnified figures for the solutions. The spectral solution slightly deviates from the HS's solution around $E=-0.3$  

The eigenvalues we found are the same as one to two significant figures of the HS's eigenvalues. Table \ref{table:Eigenvalues} lists the eigenvalues obtained from the spectral solution. Our eigenvalues $c_1$, $c_2$ and $c_3$ are the same as two significant figures of the HS's values while $c_4$ is only one significant figure with relative error of $6.7\%$. On one hand, the physical parameters $\alpha$,  $\chi_\text{esc}$ and $\xi$ are the same as three significant figures of the HS's values.  The present value of $\chi_\text{esc}$ is greater than the HS's value $13.85$. This is consistent with the result of \citep{Cohn_1980} that predicted $\chi_\text{esc}\approx13.9$ at which a complete core-collapse (an infinite central density) occurs.  

In the rest of sections, we call the following eigenvalues and $\beta_\text{o}$ the \emph{reference} eigenvalues for comparison with other solutions 
\begin{subequations}
\begin{align}
&c_{1\text{o}}=9.09254120455\times10^{-4},\\
&c^{*}_{4\text{o}}=3.03155222\times10^{-1}.
\end{align}
\end{subequations}
The reference eigenvalues were obtained from the whole-domain solution when $\beta=\beta_\text{o}$, $\mathcal{N}=70$, $L=1$ and $F_\text{BC}=1$.

\begin{table*}\large\centering
	\ra{1.3}
	\begin{tabular}{@{}ccccc@{}}\toprule
		\large{Eigenvalues} &Spectral method & HS & T & \% relative error $[\%]$  \\ 
		\midrule
		$c_1$ & $9.0925\times10^{-4}$ &$9.10\times10^{-4}$& $9.1\times10^{-4}$  & 0.1\\
		$c_2$ & $1.1118\times10^{-4}$&$1.12\times10^{-4}$ &$-$ & 0.9\\
		$c_3$ & $7.1975\times10^{-2}$ &$7.21\times10^{-2}$  &$-$ &0.1\\
		$c_4$ & $3.303\times10^{-2}$&$3.52\times10^{-2}$ &$-$ &6.7\\
		$\alpha$ & $2.2305$  & 2.23  & 2.23 &0\\
		$\chi_\text{esc}$ & $13.881$  & $13.85$  & $-$ &0.3\\
		$\xi$ & $3.64\times10^{-3}$  &  $3.64\times10^{-3}$  & $-$ &0\\
		\bottomrule
	\end{tabular}\\
	\caption{Comparison of the present eigenvalues and physical parameters with the results of 'HS' \citep{Heggie_1988}  and 'T' \citep{Takahashi_1993}. The relative error between Heggie-Stevenson(HS)'s eigenvalues and the present ones are also shown. The present eigenvalues are based on the results for various combinations of numerical parameters ($13<\mathcal{N}<560$, $10^{-4}<F_\text{BC}<10^{4}$ and $L=1/2,3/4,1$), different formulations (Sections \ref{sec:domain_trunc} and \ref{sec:mod_change} and Appendix\ref{sec:Stability_L}) and stability analyses (Appendix \ref{sec:stability}).}
	\label{table:Eigenvalues}
\end{table*}

Lastly, we report the semi-analytical solution of the ss-OAFP system. Since spectral-method studies generally provide a solution of equation concerned with a low degree of polynomials,\footnote{Spectral methods can provide `semi-analytical' solutions in the sense that the solutions can be expanded in terms of polynomials with degree of \emph{a few to tens} ; typical base functions are such as Legendre polynomials, Geggenbauer polynomials and Hermite functions. Sections \ref{sec:domain_trunc} and \ref{sec:mod_change} and Appendix \ref{sec:Stability_L} show spectral solutions based on different formulations of the ss-OAFP system, hence we could construct variants of the semi-analytic solutions in the present work. However, they do not have an outstanding property. For example, the corresponding semi-analytical solutions on the truncated domain (Section \ref{sec:domain_trunc})  and contracted domain (Appendix \ref{sec:Stability_L} ) need only $12\sim 13$ degrees to achieve a relative error of $10^{-4}$, but they are not be practical since they depend on parameters $L$ and $x_\text{min}$. Also, the degrees of the exponential of the regularized solution $\exp[\varv_{F}(x)]$ still needs $16$.} Table \ref{table:semi-explicit} lists `semi-analytical' forms of  $F(E)$, $Q(E)$ and $R(\Phi)$. The degrees of polynomials are at most eighteen and $\%$ error is $0.1\%$ compared to the whole-domain solution with degree $\mathcal{N}=70$. 

\begin{table*}\large
	\ra{1.3}
	\scalebox{0.7}{
		\begin{tabular}{@{}crrr@{}}
			\toprule
			index &  &  Coefficients & \\
			\midrule
			\large{$n$} &  $F_n$ &  $Q_n$ &  $R_n$   \\ 
			\midrule
			$1$ &  $-0.9793$ &   $0.7405$& $2.0588$  \\
			$2$ &   $0.4515$ &  $-0.2455$& $0.7337$  \\
			$3$ &   $0.3949$ &  $-0.2598$& $0.1589$  \\
			$4$ &   $0.1751$ &  $-0.1778$&$-0.0066$  \\
			$5$ &  $-0.0171$ &  $-0.0597$&$-0.0182$  \\
			$6$ &  $-0.0381$ &   $0.0003$& $0.0013$  \\			
			$7$ &   $0.0076$ &   $0.0037$& $0.0038$  \\
			$8$ &   $0.0103$ &  $-0.0017$&$-0.0007$  \\
			$9$ &  $-0.0046$ &  $-0.0004$&$-0.0009$  \\
			$10$ & $-0.0023$ &   $0.0006$& $0.0003$  \\
			$11$ &  $0.0023$ &  $-0.0001$& $0.0002$  \\
			$12$ &  $0.0001$ &  $-0.0001$&$-0.0001$  \\
			$13$ & $-0.0009$ &   $0.0001$&  \\
			$14$ &  $0.0002$ &           &   \\
			$15$ &  $0.0002$ &           &   \\
			$16$ & $-0.0002$ &           &   \\
			$17$ &  $0.0000$ &           &   \\
			$18$ &  $0.0001$ &           &   \\
			\bottomrule	
	\end{tabular}}\hspace{1cm}
	\scalebox{0.7}{
		\begin{tabular}{@{}cl@{}}
			\toprule
			Function & semi-analytical expression   \\ 
			\midrule
			$F(E)$&  $\exp\left[\sum_{n=1}^{18}F_{n}T_{n-1}\left(-2E-1\right)\right](-E)^{\beta}$   \\
			$Q(E)$ &  $\left[\sum_{n=1}^{13}Q_{n}T_{n-1}\left(-2E-1\right)\right]^{3}(-E)^{\sigma}$    \\
			$R(E)$ &   $\exp\left[\sum_{n=1}^{12}R_{n}T_{n-1}\left(-2E-1\right)\right](-E)^{\nu}\sqrt{1+E}$   \\
			\midrule
			$E\in(-1,0)$    & $T_{n-1}(-2E-1)=\cos\left[(n-1)\cos^{-1}(-2E-1)\right]$\\
			& $\beta=8.178$\\
			& $\sigma=-\frac3{2\beta-1}{4}=-11.51$ \\
			& $\nu=-\frac{2\beta+1}{4}=-4.339$ \\
			\bottomrule	
	\end{tabular}}
	\caption{Semi-analytical forms of the Chebyshev spectral solution $F(E)$, $R(E)$ and $Q(E)$. The relative error of the semi-analytic form from the whole-domain solution is order of $10^{-4}.$}
	\label{table:semi-explicit}
\end{table*}

\subsection{The detail analyses regarding the whole-domain solution and its asymptotic feature}\label{sec:ss_soln_asymp}

The present section details the mathematical characteristics of the whole-domain solution. We discuss the Chebyshev coefficients of the regularized functions (Section \ref{sec:cheb}),  regularized functions (Section \ref{sec_reg_soln}) and detail structure of $\varv_\text{J}$ (Section \ref{sec_detail_vJ}). 

\subsubsection{Chebyshev coefficients} \label{sec:cheb}

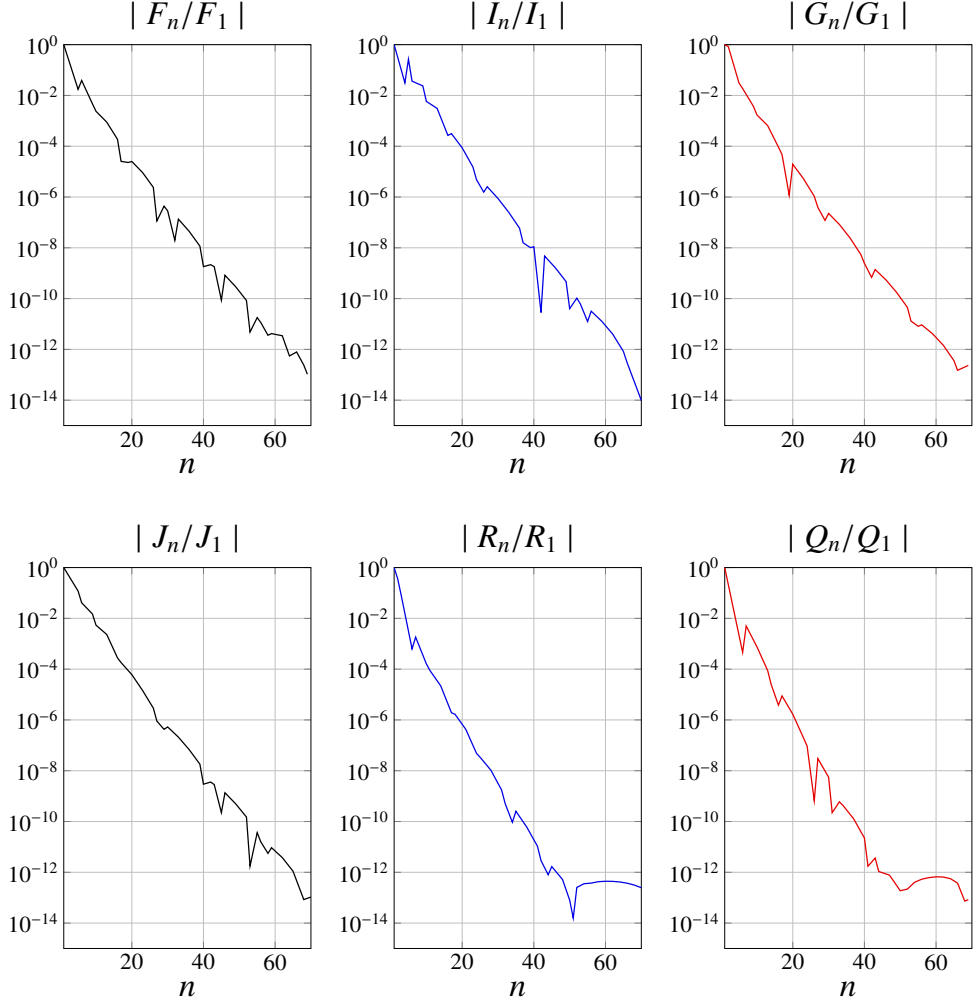
\begin{figure}[H]
	\centering
	\tikzstyle{every node}=[font=\Large]
	\begin{tikzpicture}[scale=0.6]
	\begin{semilogyaxis}[width=7cm, height=10cm, grid=major,xlabel=\huge{$n$},title=\huge{$\mid F_n/F_1\mid$}, xmin=1e0,xmax=70,ymin=1e-15,ymax=1e0, legend pos=south west]
	\addplot [color = black ,mark=no,thick,solid] table[x index=0, y index=1]{cFKGLRQ_N70.txt}; 
	\end{semilogyaxis}
	\end{tikzpicture}\hspace{0.2cm}
	\begin{tikzpicture}[scale=0.6]
	\begin{semilogyaxis}[width=7cm, height=10cm,   grid=major,xlabel=\huge{$n$},title=\huge{$\mid I_n/I_1\mid$},xmin=1e0,xmax=70,ymin=1e-15,ymax=1e0, legend pos=south west]
	\addplot [color = bblue ,mark=no,thick,solid ] table[x index=0, y index=2]{cFKGLRQ_N70.txt}; 
	\end{semilogyaxis}
	\end{tikzpicture}\hspace{0.2cm}
	\begin{tikzpicture}[scale=0.6]
	\begin{semilogyaxis}[width=7cm, height=10cm,   grid=major,xlabel=\huge{$n$},title=\huge{$\mid G_n/G_1\mid$},xmin=1e0,xmax=70,ymin=1e-15,ymax=1e0, legend pos=south west]
	\addplot [color = bred ,mark=no,thick,solid] table[x index=0, y index=3]{cFKGLRQ_N70.txt}; 
	\end{semilogyaxis}
	\end{tikzpicture}

     \vspace{0.5cm}

	\centering
	\begin{tikzpicture}[scale=0.6]
	\begin{semilogyaxis}[width=7cm, height=10cm,   grid=major,xlabel=\huge{$n$},title=\huge{$\mid J_n/J_1\mid$},xmin=1e0,xmax=70,ymin=1e-15,ymax=1e0, legend pos=south west]
	\addplot [color = black ,mark=no,thick,solid] table[x index=0, y index=4]{cFKGLRQ_N70.txt}; 
	\end{semilogyaxis}
	\end{tikzpicture}\hspace{0.2cm}
	\begin{tikzpicture}[scale=0.6]
	\begin{semilogyaxis}[width=7cm, height=10cm,   grid=major,xlabel=\huge{$n$},title=\huge{$\mid R_n/R_1\mid$},xmin=1e0,xmax=70,ymin=1e-15,ymax=1e0, legend pos=south west]
	\addplot [color = bblue ,mark=no,thick,solid ] table[x index=0, y index=5]{cFKGLRQ_N70.txt}; 
	\end{semilogyaxis}
	\end{tikzpicture}\hspace{0.2cm}
	\begin{tikzpicture}[scale=0.6]
	\begin{semilogyaxis}[width=7cm, height=10cm,   grid=major,xlabel=\huge{$n$},title=\huge{$\mid Q_n/Q_1\mid$},xmin=1e0,xmax=70,ymin=1e-15,ymax=1e0, legend pos=south west]
	\addplot [color = bred ,mark=no,thick,solid] table[x index=0, y index=6]{cFKGLRQ_N70.txt}; 
	\end{semilogyaxis}
	\end{tikzpicture}
	\caption{Absolute values of the normalized Chebyshev coefficients for the regularized functions. ($\mathcal{N}=70$, $F_\text{BC}=1$ and $L=1$). The coefficients are divided by their own first coefficients.}
	\label{fig:cFKGLRQ_N70}
\end{figure}

The Chebyshev coefficients of the regularized functions are depicted in Figure \ref{fig:cFKGLRQ_N70} in which the coefficients are divided by their own first $(n=1)$ coefficients. The minimum absolute values of all the coefficients reach $\sim10^{-12}$ around at $n=70$. This implies that possible relative error of the spectral solution is $\sim10^{-10}\%$ at best. The coefficients show geometrical convergences; $\mid F_n/F_1\mid \sim \mid G_n/G_1\mid\sim \mid I_n/I_1\mid \sim \mid J_n/J_1\mid \sim\exp(-0.3n)$  and  $\mid R_n/R_1\mid\sim \mid Q_n/Q_1\mid \sim\exp(-0.4n)$.

\subsubsection{Regularized solution and its asymptotic approximation}\label{sec_reg_soln}

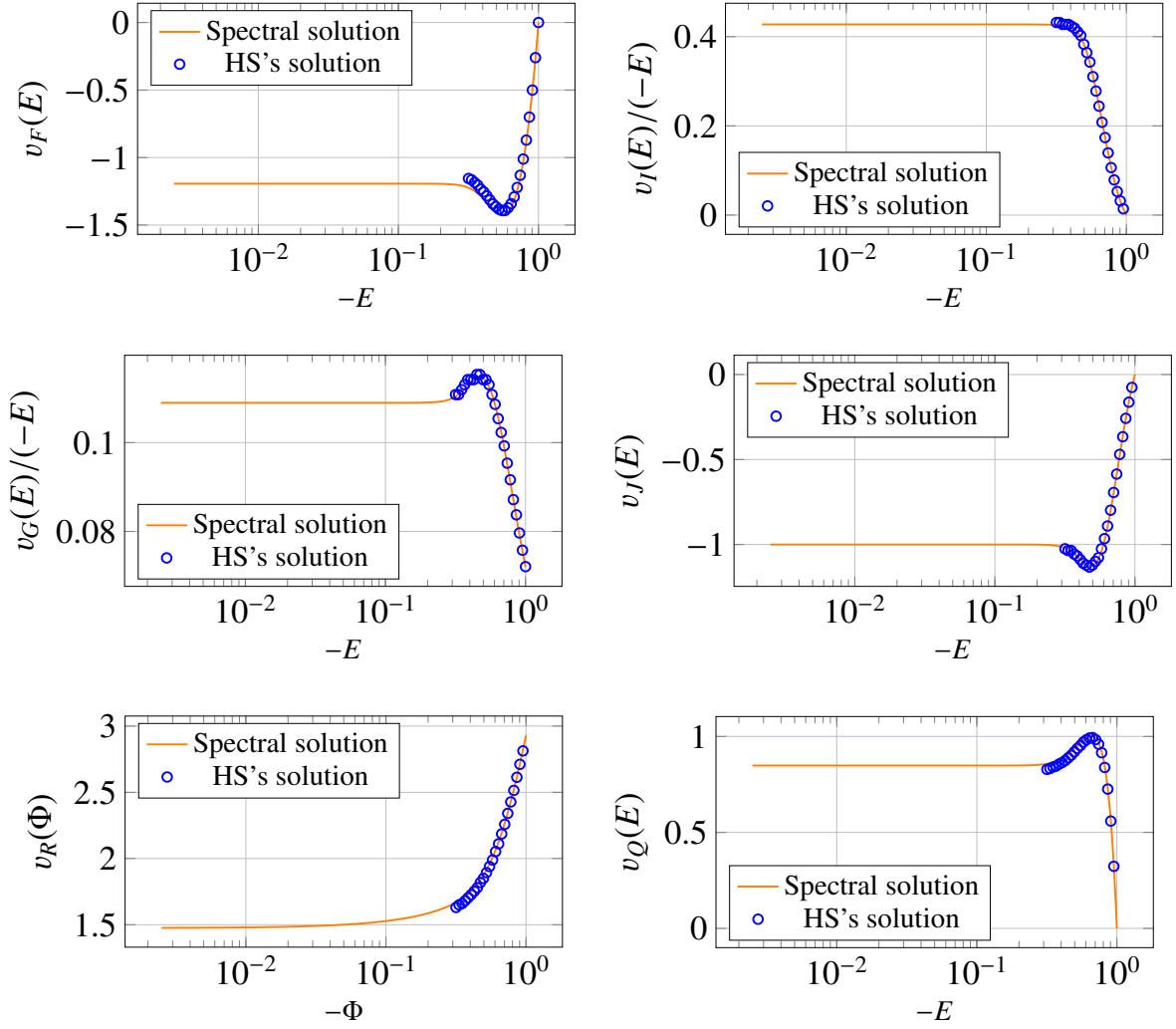
\begin{figure}[H]
	\centering
	\tikzstyle{every node}=[font=\Large]
	\begin{tikzpicture}[scale=0.9]
	\begin{semilogxaxis}[width=8cm,height=5cm, grid=major, xlabel=\large{$-E$},ylabel=\Large{$\varv_F(E)$}, legend pos=north west]
	\addplot [color = orange ,mark=no,thick,solid] table[x index=0, y index=1]{vFKGLRQ2_N70.txt}; 
	\addlegendentry{\large{Spectral solution}}
	\addplot [only marks, color = blue ,mark=o, thick] table[x index=0, y index=1]{HS_vFG_70.txt}; 
	\addlegendentry{\large{HS's solution}}
	\end{semilogxaxis}
	\end{tikzpicture}\hspace{0.5cm}
	\begin{tikzpicture}[scale=0.9]
	\begin{semilogxaxis}[width=8cm,height=5cm, grid=major, xlabel=\large{$-E$},ylabel=\Large{$\varv_I(E)/(-E)$}, legend pos=south west]
     \addplot [color=orange ,mark=no,thick,solid ] table[x index=0, y index=2]{vFKGLRQ2_N70.txt}; 
	\addlegendentry{\large{Spectral solution}}
	\addplot [only marks, color = blue ,mark=o, thick] table[x index=0, y index=1]{HS_vKLRQ_70.txt}; 
	\addlegendentry{\large{HS's solution}}
	\end{semilogxaxis}
	\end{tikzpicture}

     \vspace{0.5cm}

	\begin{tikzpicture}[scale=0.9]
	\begin{semilogxaxis}[width=8cm,height=5cm, grid=major,xlabel=\large{$-E$},ylabel=\Large{$\varv_G(E)/(-E)$}, legend pos=south west,y tick label style={/pgf/number format/fixed,/pgf/number format/precision=5}, scaled ticks=false]
	\addplot [color = orange ,mark=no,thick,solid] table[x index=0, y index=3]{vFKGLRQ2_N70.txt}; 
	\addlegendentry{\large{Spectral solution}}
	\addplot [only marks, color = blue ,mark=o, thick] table[x index=0, y index=2]{HS_vFG_70.txt}; 
	\addlegendentry{\large{HS's solution}}
	\end{semilogxaxis}
	\end{tikzpicture}\hspace{0.5cm}
	\begin{tikzpicture}[scale=0.9]
	\begin{semilogxaxis}[width=8cm,height=5cm, grid=major,xlabel=\large{$-E$},ylabel=\Large{$\varv_J(E)$}, legend pos=north west]
	\addplot [color = orange ,mark=no,thick,solid] table[x index=0, y index=4]{vFKGLRQ2_N70.txt}; 
	\addlegendentry{\large{Spectral solution}}
	\addplot [only marks, color = blue ,mark=o, thick] table[x index=0, y index=2]{HS_vKLRQ_70.txt}; 
	\addlegendentry{\large{HS's solution}}
	\end{semilogxaxis}
	\end{tikzpicture}
     
      \vspace{0.5cm}

	\begin{tikzpicture}[scale=0.9]
	\begin{semilogxaxis}[width=8cm,height=5cm, grid=major,xlabel=\large{$-\Phi$},ylabel=\Large{$\varv_R(\Phi)$}, legend pos=north west]
	\addplot [color =orange ,mark=no,thick,solid ] table[x index=0, y index=5]{vFKGLRQ2_N70.txt}; 
	\addlegendentry{\large{Spectral solution}}
	\addplot [only marks, color = blue ,mark=o, thick] table[x index=0, y index=3]{HS_vKLRQ_70.txt}; 
	\addlegendentry{\large{HS's solution}}
	\end{semilogxaxis}
	\end{tikzpicture}\hspace{0.5cm}
	\begin{tikzpicture}[scale=0.9]
	\begin{semilogxaxis}[width=8cm,height=5cm, grid=major,xlabel=\large{$-E$},ylabel=\Large{$\varv_Q(E)$}, legend pos=south west]
	\addplot [color = orange ,mark=no,thick,solid] table[x index=0, y index=6]{vFKGLRQ2_N70.txt}; 
	\addlegendentry{\large{Spectral solution}}
	\addplot [only marks, color = blue ,mark=o, thick] table[x index=0, y index=4]{HS_vKLRQ_70.txt}; 
	\addlegendentry{\large{HS's solution}}
	\end{semilogxaxis}
	\end{tikzpicture}
	\caption{Regularized spectral functions  ($\mathcal{N}=70$, $F_\text{BC}=1$ and $L=1$.)}
	\label{fig:vFKGLRQ_N70}
\end{figure}

To discuss the fine difference between the spectral and HS's solutions, Figure \ref{fig:vFKGLRQ_N70} compares the regularized functions obtained from the spectral solution and from the HS's work. One can find a discrepancy between the two works as $E\to 0$ for $\varv_{F}(E)$, $\varv_{R}(\Phi)$ and $\varv_{Q}(E)$. The figure indicates that the HS's functions were obtained \emph{outside} the domain on which our functions asymptotically behave as constant functions. This implies that the actual number of significant figures of the HS's solution may not be more than one. This matter is discussed in detail in Section \ref{sec:repro_HS_soln} (and also Appendix \ref{Appendix_Ref_HS}). 

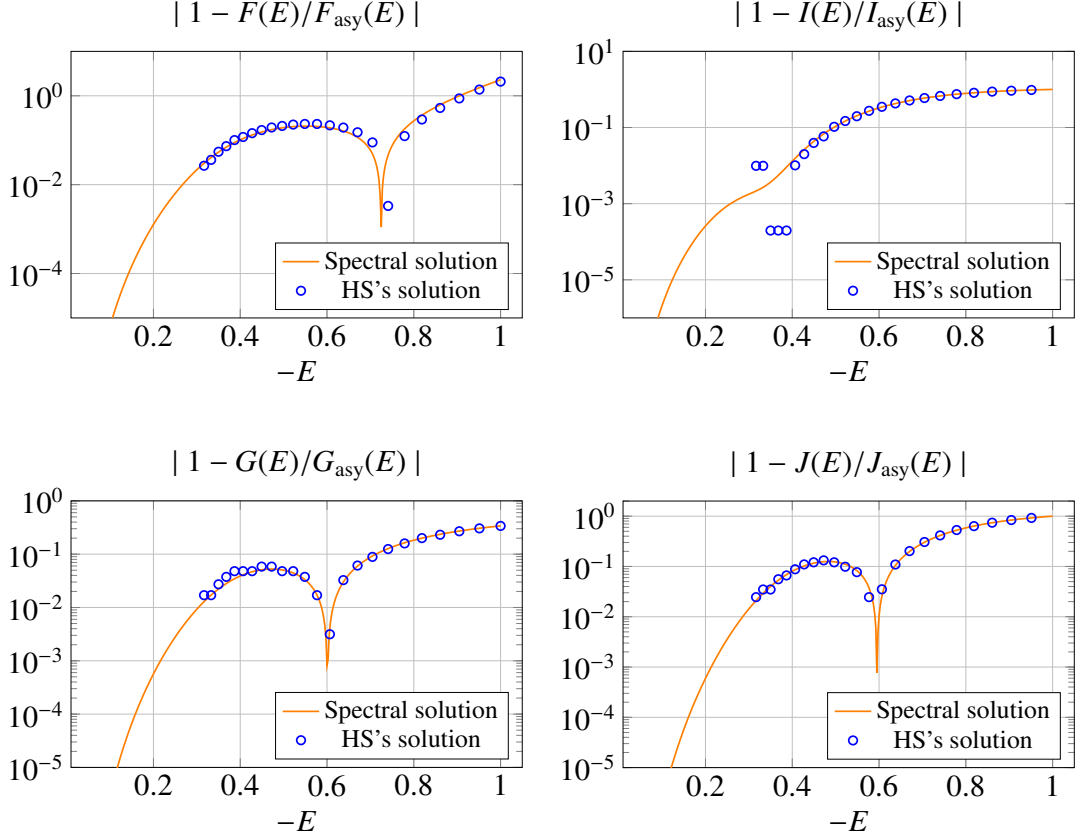
\begin{figure}[H]
	\centering
	\tikzstyle{every node}=[font=\Large]
	\begin{tikzpicture}[scale=0.8]
	\begin{semilogyaxis}[width=9cm, height=6cm, grid=major,xlabel=\Large{$-E$},title=\Large{$\mid1-F(E)/F_\text{asy}(E)\mid$},xmin=1e-2,xmax=1.05,ymin=1e-5,ymax=1e1, legend pos=south east]
	\addplot [color = orange,mark=no,thick,solid] table[x index=0, y index=1]{Del_asymFKGL_70.txt}; 
	\addlegendentry{\large{Spectral solution}}
	\addplot [only marks, color = blue ,mark=o, thick] table[x index=0, y index=1]{Del_HSasym_FG_70.txt}; 
	\addlegendentry{\large{HS's solution}}
	\end{semilogyaxis}
	\end{tikzpicture}\hspace{0.3cm}
	\begin{tikzpicture}[scale=0.8]
	\begin{semilogyaxis}[width=9cm, height=6cm, grid=major,xlabel=\Large{$-E$},title=\Large{$\mid1-I(E)/I_\text{asy}(E)\mid$},xmin=1e-2,xmax=1.05,ymin=1e-6,ymax=1e1, legend pos=south east]
	\addplot [color = orange ,mark=no,thick,solid ] table[x index=0, y index=2]{Del_asymFKGL_70.txt}; 
	\addlegendentry{\large{Spectral solution}}
	\addplot [only marks, color = blue ,mark=o, thick] table[x index=0, y index=2]{Del_HSasym_JI_70.txt}; 
	\addlegendentry{\large{HS's solution}}
	\end{semilogyaxis}
	\end{tikzpicture}\vspace{0.2cm}

     \vspace{0.5cm}

	\begin{tikzpicture}[scale=0.8]
	\begin{semilogyaxis}[width=9cm, height=6cm, grid=major,xlabel=\Large{$-E$},title=\Large{$\mid1-G(E)/G_\text{asy}(E)\mid$},xmin=1e-2,xmax=1.05,ymin=1e-5,ymax=1e0, legend pos=south east]
	\addplot [color =orange ,mark=no,thick,solid] table[x index=0, y index=3]{Del_asymFKGL_70.txt}; 
	\addlegendentry{\large{Spectral solution}}
	\addplot [only marks, color = blue ,mark=o, thick] table[x index=0, y index=2]{Del_HSasym_FG_70.txt}; 
	\addlegendentry{\large{HS's solution}}
	\end{semilogyaxis}
	\end{tikzpicture}\hspace{0.3cm}
	\centering
	\begin{tikzpicture}[scale=0.8]
	\begin{semilogyaxis}[width=9cm, height=6cm, grid=major,xlabel=\Large{$-E$},title=\Large{$\mid1-J(E)/J_\text{asy}(E)\mid$},xmin=1e-2,xmax=1.05,ymin=1e-5,ymax=2e0, legend pos=south east]
	\addplot [color = orange ,mark=no,thick,solid] table[x index=0, y index=4]{Del_asymFKGL_70.txt}; 
	\addlegendentry{\large{Spectral solution}}
	\addplot [only marks, color = blue ,mark=o, thick] table[x index=0, y index=1]{Del_HSasym_JI_70.txt}; 
	\addlegendentry{\large{HS's solution}}
	\end{semilogyaxis}
	\end{tikzpicture}
	\caption{Relative error of the whole-domain spectral solution from the asymptotic approximation ($\mathcal{N}=70$, $F_\text{BC}=1$, $L=1$). The circles are the corresponding relative errors in the HS's solution. \citep{Heggie_1988} listed the numerical values of their solution rounded to the second \emph{decimal places}, meaning the present solution can compare to their solutions at order of $5\times10^{-3}$ at best.}
	\label{fig:Del_asymFKGL_N70}
\end{figure}

Since the asymptotic approximations of $F(E)$, $G(E)$, $I(E)$ and $J(E)$ as $E\to0$ read
\begin{subequations}
\begin{align}
&F_\text{asy}(E)\equiv\ln[c_{4}^{*}](-E)^{\beta},\\
&G_\text{asy}(E)\equiv\frac{1}{a+1}(-E)^{\beta+1},\\
&I_\text{asy}(E)\equiv\frac{4}{2\beta-7}(-E)^{\beta+1}Q(E),\\
&J_\text{asy}(E)\equiv-\frac{4\beta}{2\beta-3}(-E)^{\beta}Q(E),
\end{align} 
\end{subequations}
we computed the relative errors between $\{F,G,I,J\}$ obtained from the spectral solution and $\{F_\text{asy},G_\text{asy}, I_\text{asy},J_\text{asy}\}$ (Figure \ref{fig:Del_asymFKGL_N70}).  The figure also depicts the corresponding errors for the HS's functions. Both our and HS's functions show that $F_\text{asy}(E)$, $G_\text{asy}(E)$ and $J_\text{asy}(E)$ can well approximate $F(E)$, $G(E)$ and $J(E)$ near $E=-0.6$ or $E=-0.7$ and the relative errors between them are order of $10^{-3}$. Since our value of $c_{4}$ is relatively different from the HS's value, one finds a discrepancy between the works for $\mid1-F(E)/F_\text{asy}(E)\mid$. 

\subsubsection{Detail structure of $\varv_{J}$}\label{sec_detail_vJ}

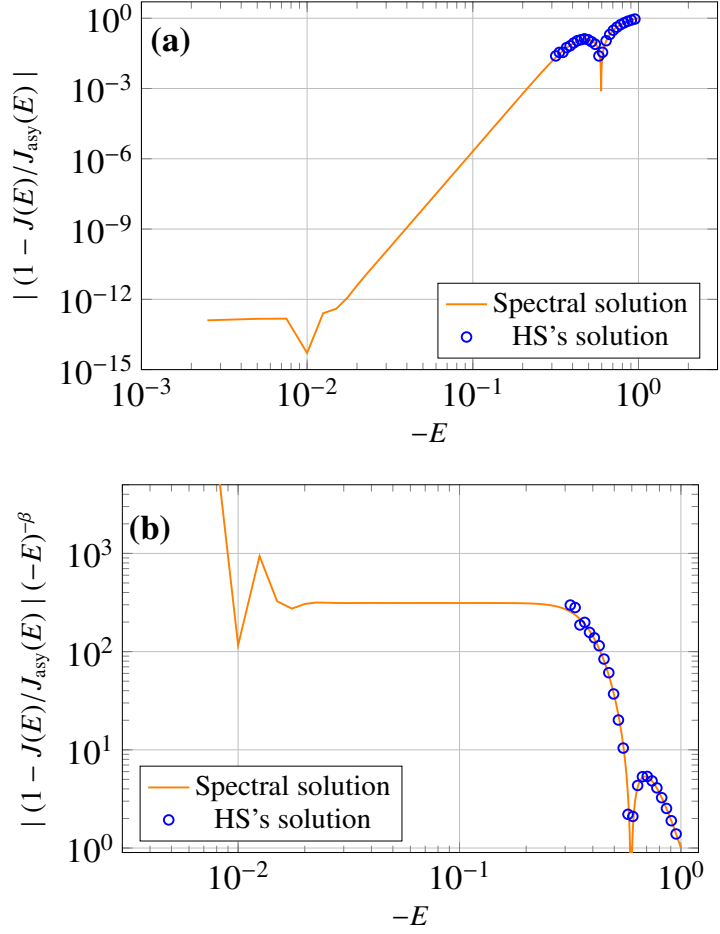
\begin{figure}[H]
	\centering
	\tikzstyle{every node}=[font=\Large]
	\begin{tikzpicture}[scale=0.9]
	\begin{loglogaxis}[width=10cm, height=7cm, grid=major,xlabel=\large{$-E$},ylabel=\large{$\mid(1-J(E)/J_\text{asy}(E)\mid$},xmin=1e-3,xmax=3e0,ymin=1e-15,ymax=5e0, legend pos=south east]
	\addplot [color = orange ,mark=no,thick,solid] table[x index=0, y index=4]{Del_asymFKGL_70.txt}; 
	\addlegendentry{\large{Spectral solution}}
	\addplot [only marks, color = blue ,mark=o, thick] table[x index=0, y index=1]{Del_HSasym_JI_70.txt}; 
	\addlegendentry{\large{HS's solution}}
	\node[above,black] at (1.5e-3,1e-2) {\Large{$\textbf{(a)}$}};
	\end{loglogaxis}
	\end{tikzpicture}
	
	\hspace{0.4cm}
	
	\begin{tikzpicture}[scale=0.9]
	\begin{loglogaxis}[width=10cm, height=7cm,  grid=major,xlabel=\large{$-E$},ylabel=\large{$\mid(1-J(E)/J_\text{asy}(E)\mid(-E)^{-\beta}$},xmin=3e-3,xmax=1.2e0,ymin=9e-1,ymax=5e3, legend pos=south west]
	\addplot [color = orange,mark=no,thick,solid] table[x index=0, y index=1]{Const_power_L_70.txt}; 
	\addlegendentry{\large{Spectral solution}}
	\addplot [only marks, color = blue ,mark=o, thick] table[x index=0, y index=1]{HSConst_power_L_70.txt}; 
	\addlegendentry{\large{HS's solution}}
      \node[above,black] at (4e-3,1e3) {\Large{$\textbf{(b)}$}};
	\end{loglogaxis}
	\end{tikzpicture}
	\caption{\textbf{(a)} Characteristics of $\mid1-J(E)/J_\text{asy}(E)\mid$ and \textbf{(b)} characteristics of $\mid1-J(E)/J_\text{asy}(E)\mid (-E)^{-\beta}$ ($\mathcal{N}=70$, $F_\text{BC}=1$, $L=1$). The circles are the corresponding characteristics of \citep{Heggie_1988}'s work.}
	\label{fig:Const_power_L_N70}
\end{figure}

Since the higher order of asymptotic approximation for $\varv_{J}(x)$ is analytically tractable, we discuss the feature qualitatively and quantitatively. First, we can qualitatively find a consistency of the spectral whole-domain solution by examining the asymptotic approximation of $J(E)$ that can be explicitly found from one of the 4ODEs (Appendix \ref{sec:eqn_vJp1}); 
\begin{align}
\mid1-J(E\to0)/J_\text{asy}(E)\mid=-\frac{c^{*}_{4}}{c_{1}}\frac{(2\beta+7)(6\beta-3)}{(2\beta-7)(2\beta-3)(\beta+1)} (-E)^{\beta}\equiv  \mathcal{C}_{\beta}\left(c_{1},c_{4}^{*}\right)(-E)^{\beta}. \label{Eq.1pJ_asym}
\end{align}
This $ (-E)^{\beta}$-dependence is numerically reproduced in Figure \ref{fig:Const_power_L_N70} \textbf{(a)}. Figure \ref{fig:Const_power_L_N70} \textbf{(b)} depicts the characteristics of $\mid1-J(E)/J_\text{asy}(E)\mid (-E)^{-\beta}$ that is still approximately constant on the interval $-4\times10^{-2}\lesssim E\lesssim -1\times 10^{-1}$. 

\begin{figure}[H]
	\centering
	\tikzstyle{every node}=[font=\Large]
	\begin{tikzpicture}[scale=0.8]
	\begin{loglogaxis}[ grid=major,xlabel=\large{$-E$},ylabel=\large{$\mid1-\,\text{d}\ln\left[1-J(E)/J_\text{asy}(E)\right]\,\text{d}\ln [E]/\beta_\text{o}\mid$},xmin=3e-3,xmax=1.2e0,ymin=1e-8,ymax=1e2, legend pos=south east]
	\addplot [color = bred ,mark=*,thick,solid] table[x index=0, y index=1]{vJ_slope_index.txt}; 
     \node[above,black] at (5e-3,2e-8) {\Large{$\textbf{(a)}$}};
	\end{loglogaxis}
	\end{tikzpicture}\hspace{0.4cm}
	\begin{tikzpicture}[scale=0.8]
	\begin{loglogaxis}[ grid=major,xlabel=\large{$-E$},ylabel=\large{$\mid1-J(E)/J_\text{asy}(E)\mid(-E)^{-\beta}/\mathcal{C}_{\beta}\left(c_{1},c_{4}^{*}\right)$},xmin=3e-3,xmax=1.2e0,ymin=1e-8,ymax=1e2, legend pos=south west]
	\addplot [color = bred,mark=*,thick,solid] table[x index=0, y index=2]{vJ_slope_index.txt}; 
      \node[above,black] at (5e-3,2e-8) {\Large{$\textbf{(b)}$}};
	\end{loglogaxis}
	\end{tikzpicture}
	\caption{\textbf{(a)} Logarithmic derivative of $\mid1-J(E)/J_\text{asy}(E)\mid$ with respect to $E$ and \textbf{(b)} characteristics of $\mid1-J(E)/J_\text{asy}(E)\mid (-E)^{-\beta}$ ($\mathcal{N}=70$, $F_\text{BC}=1$, $L=1$). }
	\label{fig:vJ_slope_index}
\end{figure}
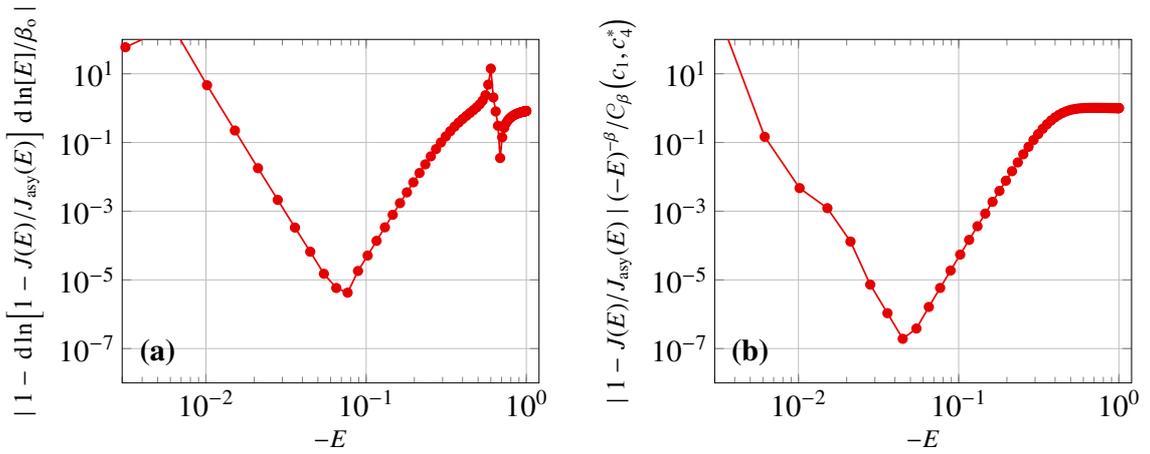

To quantitatively see the consistency of the spectral solution, we numerically calculated the values of $\beta$ and $\mathcal{C}_{\beta}\left(c_{1},c_{4}^{*}\right)$ of equation \eqref{Eq.1pJ_asym}. Figure \ref{fig:vJ_slope_index} \textbf{(a)} shows the relative error between $\beta_\text{o}$ and the logarithmic derivative of  $\mid1-J(E)/J_\text{asy}(E)\mid$ and Figure \ref{fig:vJ_slope_index} \textbf{(b)} depicts the error between $\mid1-J(E)/J_\text{asy}(E)\mid (-E)^{-\beta}$ and $\mathcal{C}_{\beta}\left(c_{1},c_{4}^{*}\right)$. One can find the former is correct at order of $4.3\times10^{-6}$ at best  and the latter $2.0\times10^{-7}$. The logarithmic derivative and asymptotic approximation lose their accuracies at energies greater than $E=-0.07\sim-0.05$. This is since the expression in equation \ref{Eq.1pJ_asym} is correct under the limited condition that the factor $(0.5+0.5x)^{\beta}\frac{\,\text{d}\varv_{F}}{\,\text{d}x}$ in equation \eqref{Eq.ss-4ODE-vF} does not reach order of machine precision (See Appendix \ref{sec:Newton_method} for detail). 

\subsection{Numerical instability of the whole-domain solution}\label{sec:inst_whole}

The present section explains the numerical instability of the whole-domain solution. As explained in detail in Appendix \ref{sec:stability}, the whole-domain solution is stable against various numerical parameters. For a broad range of $\beta$, $F_\text{BC}$, $L$ and the total number of nodes for the Fej$\acute{\mathrm{e}}$r's rule quadrature, the eigenvalues $c_{1}$ and $c^{*}_{4}$ can preserve seven- and five- significant figures compared to  the reference values $c_{1\text{o}}$ and $c^{*}_{4\text{o}}$. On one hand, the whole-domain solution is unstable against degree $\mathcal{N}$. The Newton iteration method well worked only for $70\leq \mathcal{N}\lesssim 400$. It did not work at all for $\mathcal{N}$ less than 70 while it still worked for $\mathcal{N}>400$ but high $\mathcal{N}$ increased $\mid \varv_{I}(x=1)\mid$ and the condition number of the Jacobian Matrix for the 4ODE and $Q$-integral in Newton method, costing an unfeasible CPU time.

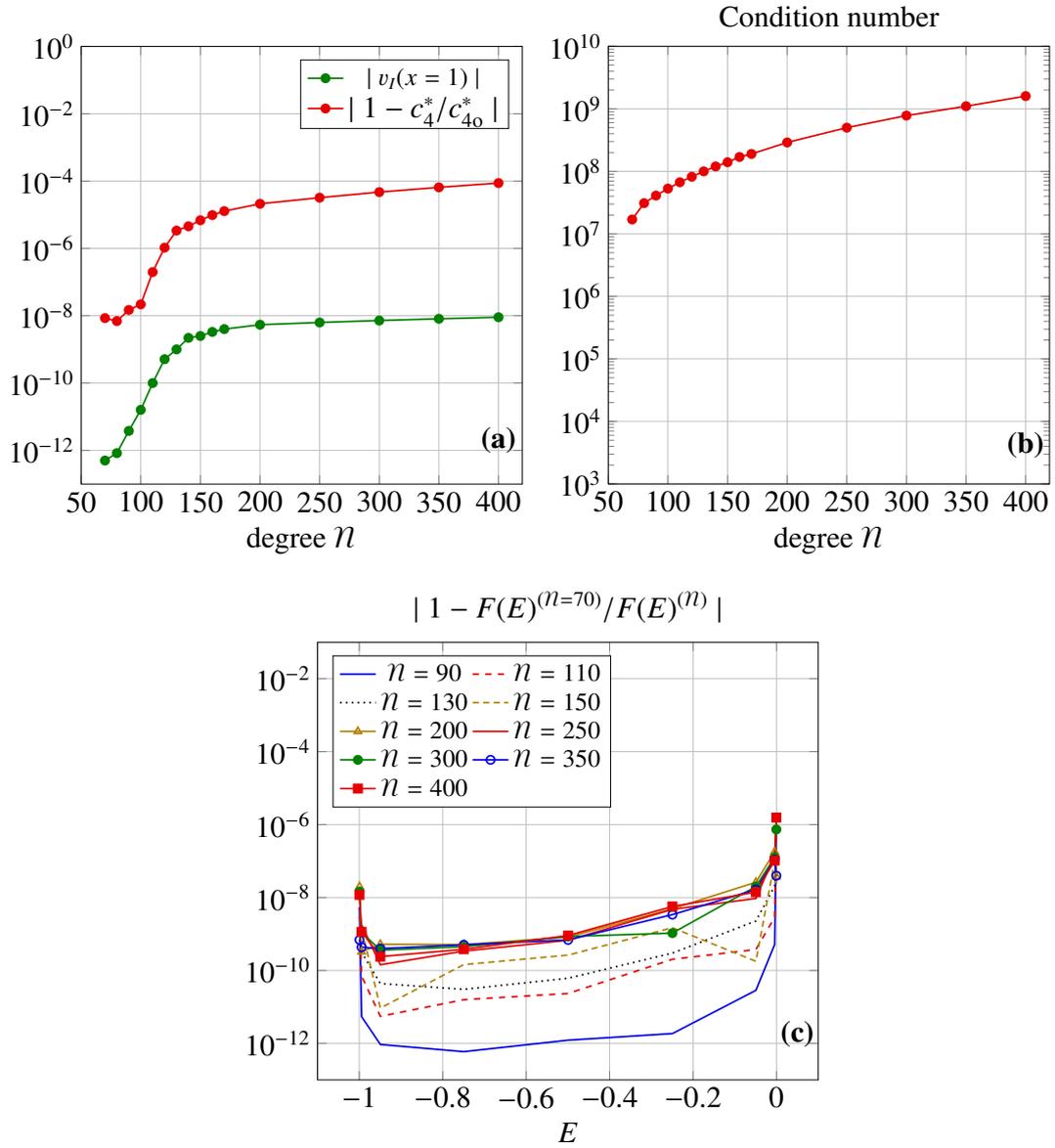
\begin{figure}[H]
	\centering
	\tikzstyle{every node}=[font=\Large]
	\begin{tikzpicture}[scale=0.8]
	\begin{semilogyaxis}[width=9cm,height=9cm, grid=major,xlabel=\Large{degree $\mathcal{N}$},xmin=50,xmax=4.2e2,ymin=1e-13,ymax=1e0, legend pos=north east]
	\addplot [color = bgreen ,mark=*, thick] table[x index=0, y index=1]{Stab_Deg_vKcond.txt}; 
     \addlegendentry{\large{$\mid \varv_{I}(x=1)\mid$}}
     \addplot [color = bred ,mark=*, thick] table[x index=0, y index=3]{Stab_Deg_vKcond.txt}; 
     \addlegendentry{$\mid1-c^{*}_{4}/c^{*}_{4\text{o}}\mid$}
 \node[above,black] at (400,5e-13) {\Large{$\textbf{(a)}$}};
      \end{semilogyaxis}
	\end{tikzpicture}\hspace{0.1cm}
		\begin{tikzpicture}[scale=0.8]
	\begin{semilogyaxis}[width=9cm,height=9cm, grid=major,xlabel=\Large{degree $\mathcal{N}$},title=\Large{Condition number}, xmin=50,xmax=4.2e2,ymin=1e3,ymax=1e10, legend pos=south east]
	\addplot [color = bred ,mark=*, thick] table[x index=0, y index=2]{Stab_Deg_vKcond.txt};
     \node[above,black] at (400,2e3) {\Large{$\textbf{(b)}$}}; 
	\end{semilogyaxis}
	\end{tikzpicture}

    \vspace{0.3cm}

	\begin{tikzpicture}[scale=0.8]
	\begin{semilogyaxis}[width=10cm,height=9cm,legend columns=2, grid=major,xlabel=\Large{$E$},title=\Large{$\mid1-F(E)^{(\mathcal{N}=70)}/F(E)^{(\mathcal{N})}\mid$},xmin=-1.1,xmax=0.1,ymin=1e-13,ymax=1e-1, legend pos=north west]
	\addplot [color = bblue ,mark=no,thick,solid ]  table[x index=0, y index=1]{Stab_Deg_F.txt}; 
       \addlegendentry{\large{$\mathcal{N}=90$}}
      \addplot [color = bred ,mark=no, thick, dashed] table[x index=0, y index=2]{Stab_Deg_F.txt}; 
       \addlegendentry{\large{$\mathcal{N}=110$}}
      \addplot [color = black ,mark=no, thick, dotted] table[x index=0, y index=3]{Stab_Deg_F.txt}; 
       \addlegendentry{\large{$\mathcal{N}=130$}}
      \addplot [color = bgold ,mark=no, thick, densely dashed] table[x index=0, y index=4]{Stab_Deg_F.txt}; 
       \addlegendentry{\large{$\mathcal{N}=150$}}
\addplot [color = bgold ,mark=triangle, thick] table[x index=0, y index=5]{Stab_Deg_F.txt}; 
       \addlegendentry{\large{$\mathcal{N}=200$}}
\addplot [color = bred ,mark=sqaure, thick] table[x index=0, y index=6]{Stab_Deg_F.txt}; 
       \addlegendentry{\large{$\mathcal{N}=250$}}
\addplot [color = bgreen ,mark=*, thick] table[x index=0, y index=7]{Stab_Deg_F.txt}; 
       \addlegendentry{\large{$\mathcal{N}=300$}}
\addplot [color = bblue ,mark=o, thick] table[x index=0, y index=8]{Stab_Deg_F.txt}; 
       \addlegendentry{\large{$\mathcal{N}=350$}}
\addplot [color = bred ,mark=square*, thick] table[x index=0, y index=9]{Stab_Deg_F.txt}; 
       \addlegendentry{\large{$\mathcal{N}=400$}}
       \node[above,black] at (0.05,5e-13) {\Large{$\textbf{(c)}$}};
      \end{semilogyaxis}
	\end{tikzpicture}
	\caption{\textbf{(a)} Relative error between $c_4$ and $c^{*}_{4\text{o}}$ and the value of $\varv_{I}(x=1)$, \textbf{(b)} Condition number of the Jacobian matrix for the 4ODEs and $Q$-integral. The condition number was computed when $\{a_q\}^\text{old}-\{a_q\}^\text{new}$ reached order of $10^{-13}$ in Newton iteration process. \textbf{(c)} Relative error between DFs with different $\mathcal{N}$ and the DF with $\mathcal{N}=70$.($L=1$ and $F_\text{BC}=1$)}
	\label{fig:Stab_Deg}
\end{figure}

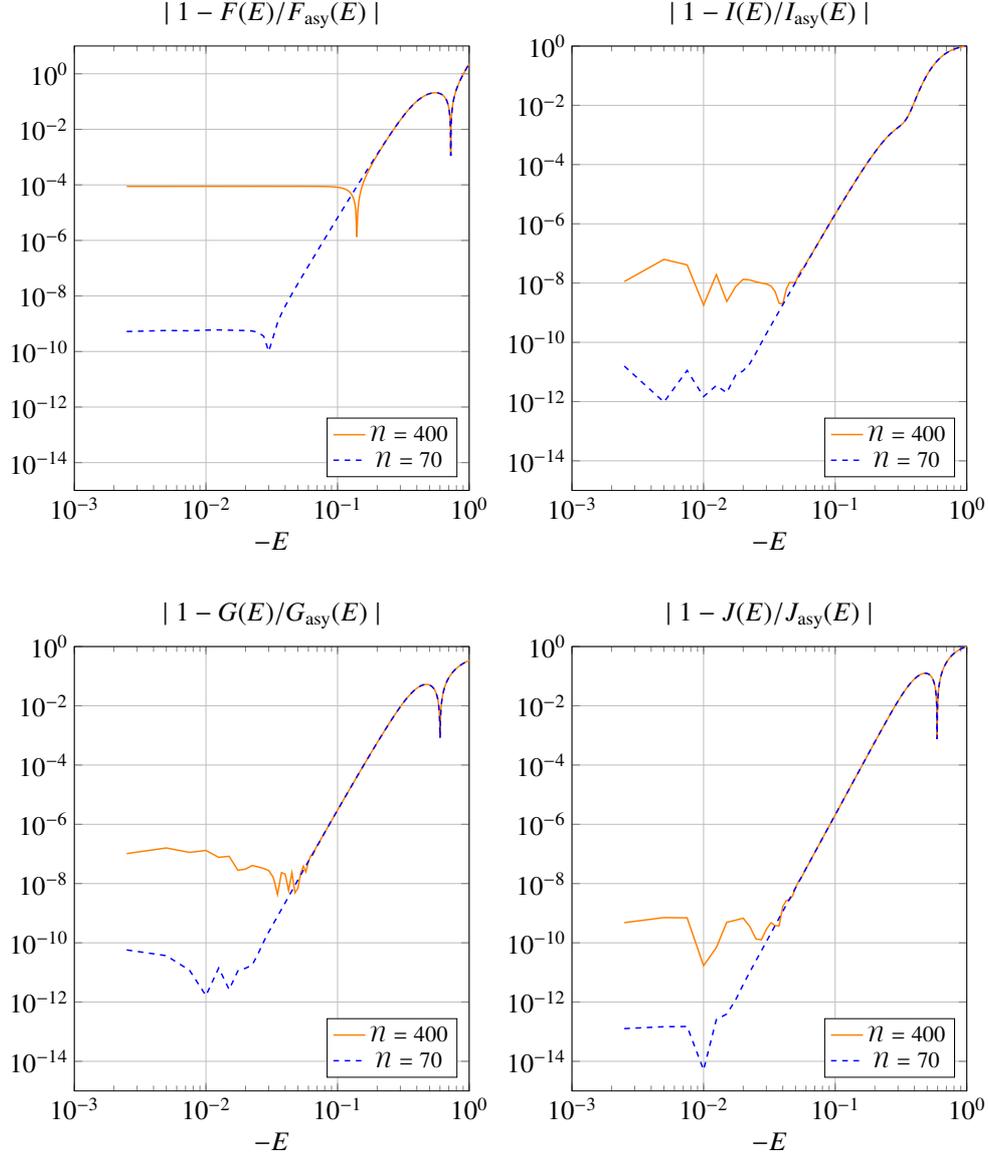
\begin{figure}[H]
	\centering
\tikzstyle{every node}=[font=\Large]
	\begin{tikzpicture}[scale=0.7]
	\begin{loglogaxis}[width=9cm, height=10cm, grid=major,xlabel=\Large{$-E$},title=\Large{$\mid1-F(E)/F_\text{asy}(E)\mid$},xmin=1e-3,xmax=1,ymin=1e-15,ymax=1e1, legend pos=south east]
	\addplot [color = orange,mark=no,thick,solid] table[x index=0, y index=1]{Del_asymFKGL_400.txt}; 
      \addlegendentry{\large{$\mathcal{N}=400$}}
\addplot [color = blue,mark=no,thick,dashed] table[x index=0, y index=1]{Del_asymFKGL_70.txt}; 
      \addlegendentry{\large{$\mathcal{N}=70$}}
	\end{loglogaxis}
	\end{tikzpicture}
\begin{tikzpicture}[scale=0.7]
	\begin{loglogaxis}[width=9cm, height=10cm, grid=major,xlabel=\Large{$-E$},title=\Large{$\mid1-I(E)/I_\text{asy}(E)\mid$},xmin=1e-3,xmax=1,ymin=1e-15,ymax=1e0, legend pos=south east]
	\addplot [color =orange,mark=no,thick,solid ] table[x index=0, y index=2]{Del_asymFKGL_400.txt}; 
      \addlegendentry{\large{$\mathcal{N}=400$}}
\addplot [color = blue,mark=no,thick,dashed] table[x index=0, y index=2]{Del_asymFKGL_70.txt}; 
      \addlegendentry{\large{$\mathcal{N}=70$}}
	\end{loglogaxis}
	\end{tikzpicture}
	
	\vspace{0.5cm}
	
\begin{tikzpicture}[scale=0.7]
	\begin{loglogaxis}[width=9cm, height=10cm, grid=major,xlabel=\Large{$-E$},title=\Large{$\mid1-G(E)/G_\text{asy}(E)\mid$},xmin=1e-3,xmax=1,ymin=1e-15,ymax=1e0, legend pos=south east]
	\addplot [color =orange ,mark=no,thick,solid] table[x index=0, y index=3]{Del_asymFKGL_400.txt}; 
      \addlegendentry{\large{$\mathcal{N}=400$}}
\addplot [color = blue,mark=no,thick,dashed] table[x index=0, y index=3]{Del_asymFKGL_70.txt}; 
      \addlegendentry{\large{$\mathcal{N}=70$}}
	\end{loglogaxis}
	\end{tikzpicture}
	\centering
	\begin{tikzpicture}[scale=0.7]
	\begin{loglogaxis}[width=9cm, height=10cm, grid=major,xlabel=\Large{$-E$},title=\Large{$\mid1-J(E)/J_\text{asy}(E)\mid$},xmin=1e-3,xmax=1,ymin=1e-15,ymax=1e0, legend pos=south east]
	\addplot [color =orange ,mark=no,thick,solid] table[x index=0, y index=4]{Del_asymFKGL_400.txt}; 
      \addlegendentry{\large{$\mathcal{N}=400$}}
\addplot [color = blue,mark=no,thick,dashed] table[x index=0, y index=4]{Del_asymFKGL_70.txt}; 
      \addlegendentry{\large{$\mathcal{N}=70$}}
	\end{loglogaxis}
	\end{tikzpicture}
\caption{ Relative error of the whole-domain spectral solution from the asymptotic approximation for $\mathcal{N}=70$ and $\mathcal{N}=400$.($F_\text{BC}=1$, $L=1$.) }
\label{fig:Del_asymFKGL}
\end{figure}

Figure \ref{fig:Stab_Deg} \textbf{(a)} shows that $\mid \varv_{I}(x=1)\mid$  increases with $\mathcal{N}$ but the rate of change becomes calm for higher $\mathcal{N}$ and Figure \ref{fig:Stab_Deg} \textbf{(b)} depicts the condition number of the Jacobian matrix and the number monotonically increases with $\mathcal{N}$. Both  $\mid \varv_{I}(x=1)\mid$ and the condition number reach their lowest values when  $\mathcal{N}=70$. Hence, we compared the DF $F(E)$ for $\mathcal{N}=70$ to the DFs with different degrees $\mathcal{N}$ (Figure \ref{fig:Stab_Deg} \textbf{(c)}). The figure shows that the accuracy of DFs lowers with increasing $\mathcal{N}$ and it reaches order of $10^{-4}$ at $\mathcal{N}=400$. Also, Figure \ref{fig:Del_asymFKGL} shows the power-law profiles under the asymptotic approximations (that appear if the displayed domain in Figure \ref{fig:Del_asymFKGL_N70} is extended to $E\to0$) lose their characteristics as $\mathcal{N}$ increases. In the figure, the power-law profiles for  $\mathcal{N}=400$ and $\mathcal{N}=70$ are shown.
      
\section{Self-similar solutions on truncated domains}\label{sec:domain_trunc}

The present section provides spectral solutions on several truncated domains and show that the relative error of an optimal truncated-domain solution from the whole-domain solution with $\mathcal{N}=70$ can achieve order of $10^{-9}$ on certain truncated domains. Since the spectral solution on the whole domain is unstable against degree $\mathcal{N}$ and also since the present work relies on a collocation method, it is imperative for us to construct a spectral solution whose accuracy improves with increasing $\mathcal{N}$. To find such a solution, we truncated the domain of the ss-OAFP system following the approach of \citep{Heggie_1988}. According to Section \ref{sec:ss_soln}, we extrapolated the domain of $\varv_{F}(x)$ so that it turns into $\varv_{F}^{\text{(ex)}}(x)$ (equation \eqref{Eq.vF_extrap}) on $-0.2\lesssim E \lesssim 0$ on which the regularized functions obtained from the whole-domain solution behave like constant functions of $E$. This means one may expect to obtain several kinds of solutions for different maximum energy $E_\text{max}$. The following classification lists kinds of solutions based on the absolute value of each term in equation \eqref{Eq.ss-4ODE-vF} (Refer to Appendix \ref{sec:Newton_method} for the details of the classification.)
\begin{align}
&\text{(i) } E_\text{max}\lesssim-0.25\hspace{0.1cm} \text{(Incorrect solution) }\nonumber\\
                  &\hspace{2cm}\text{Solutions and eigenvalues significantly differ from existing results.}\nonumber\\
&\text{(ii) }-0.25\lesssim E_\text{max}\lesssim-0.05\hspace{0.1cm} \text{(Stable solution)}\nonumber\\
                     &\hspace{2cm}\text{Chebyshev coefficients are relatively stable against degree $\mathcal{N}$.}\nonumber\\
&\text{(iii) }-0.05\lesssim E_\text{max}\lesssim-0.005\hspace{0.1cm}\text{(Semi-stable solution)}\nonumber\\
                     &\hspace{2cm}\text{Chebyshev coefficients are stable against up to a certain degree $\mathcal{N}_{c}$.}\nonumber\\
&\text{(iv) } -0.005\lesssim E_\text{max} \hspace{0.1cm}\text{(unstable solution)}\nonumber\\
                      &\hspace{2cm} \text{Chebyshev coefficients are unstable against degree $\mathcal{N}$.}\nonumber
\end{align}
The goal of the present section is, based on four cases (i) - (iv), to show some optimal truncated-domain solution compatible to the whole-domain solution and to explain the cause of numerical instability of the whole-domain solution. First, Section \ref{sec:stable_trunc_solns} explains the condition to obtain a truncated solution by examining cases (i) and (ii).  Sections \ref{sec:semi_stable_trunc_solns} and \ref{sec:trunc_solns_BETAo} discuss cases (ii) and (iii) to find an optimal truncated-domain solution. Especially, Section \ref{sec:semi_stable_trunc_solns} shows solutions on truncated domains with optimal values of $\beta$. Section \ref{sec:trunc_solns_BETAo} discusses the difference between the solutions obtained on whole- and truncated- domains for fixed $\beta=\beta_{\text{o}}$. 

For comparison in the rest of sections, we call the whole-domain solution with $\mathcal{N}=70$, $L=1$, $F_\text{BC}$ and $\beta=\beta_\text{o}$ (shown in Section \ref{sec:result_whole}) the \emph{reference} solution. The solution is labeled with subscript symbol $''\text{o}''$; hence the corresponding functions obtained from the solution are described as $F_{\text{o}}(E)$, $\Phi_\text{o}(R_\text{o})$, $\varv_{F\text{o}}(x)$, $\varv_{G\text{o}}(x)$ ... and so on.

\subsection{Stable solutions on truncated domains with $-0.35<E_\text{max}<-0.05$ }\label{sec:stable_trunc_solns}

\begin{figure}[H]
	\centering
	\tikzstyle{every node}=[font=\large]
	\begin{tikzpicture}[scale=0.8]
	\begin{semilogyaxis}[width=9cm,height=10cm,grid=major,xlabel=\Large{$-E_\text{max}$},xmin=0, xmax=0.4,ymin=1e-5,ymax=1e4, legend pos=north west,x tick label style={/pgf/number format/fixed,/pgf/number format/precision=5}, scaled ticks=false]
	\addplot [color = orange,mark=o,thick,solid] table[x index=0, y index=1]{Non_Soln_vK_Del_c1c4_cond_N40_trunc.txt}; 
      \addlegendentry{$ \large{\mid\varv_{I}(x=1)\mid}$}
\addplot [color = red,mark=square,thick,solid] table[x index=0, y index=2]{Non_Soln_vK_Del_c1c4_cond_N40_trunc.txt}; 
      \addlegendentry{ \large{$\mid1-c_{1}/c_{1\text{o}}\mid$}}
\addplot [color = blue,mark=triangle,thick,solid] table[x index=0, y index=3]{Non_Soln_vK_Del_c1c4_cond_N40_trunc.txt}; 
      \addlegendentry{ \large{$\mid1-c^{*}_{4}/c^{*}_{4\text{o}}\mid$}}
      \node[above,black] at (0.025,2e-5) {\Large{$\textbf{(a)}$}};
	\end{semilogyaxis}
	\end{tikzpicture}
	\begin{tikzpicture}[scale=0.8]
	\begin{semilogyaxis}[width=9cm,height=10cm,grid=major,xlabel=\Large{$-E_\text{max}$},xmin=0, xmax=0.4,ymin=1e6,ymax=1e12, legend pos=north west,x tick label style={/pgf/number format/fixed,/pgf/number format/precision=5}, scaled ticks=false]
\addplot [color = red,mark=square,thick,solid] table[x index=0, y index=4]{Non_Soln_vK_Del_c1c4_cond_N40_trunc.txt}; 
      \addlegendentry{ \large{Condition number}}
      \node[above,black] at (0.025,2e6) {\Large{$\textbf{(b)}$}};
	\end{semilogyaxis}
	\end{tikzpicture}
	\caption{ \textbf{(a)} Characteristics of $\mid\varv_{I}(x=1)\mid$  and relative error of $c_{1}$ and $c^{*}_{4}$ from their reference eigenvalues for different $E_\text{max}$. ($\mathcal{N}=40$, $\beta=8.1783$, $F_\text{BC}=1$ and $L=1$.) \textbf{(b)} Condition number of the Jacobian Matrix calculated when $\{a_q \}^\text{new}-\{a_q \}^\text{old}$  reached order of $10^{-13}$ in Newton iteration. }
	\label{fig:Non_Soln_vK_Del_c1c4_cond_N40_trunc}
\end{figure}
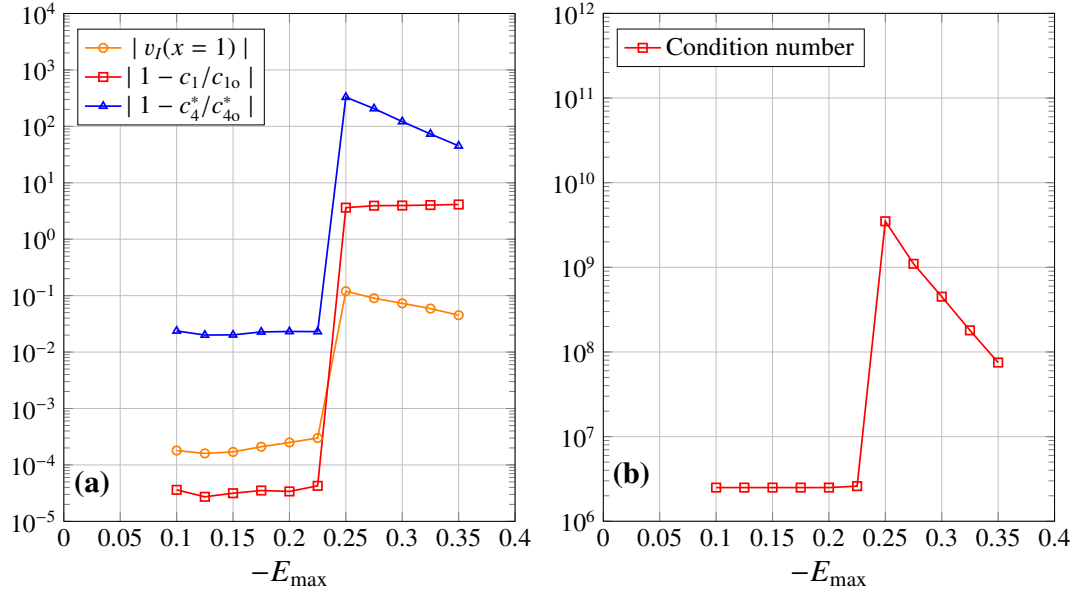

While Newton iteration method itself worked for $E_\text{max}<-0.1$, spectral solutions obtained on the truncated domains have a transition point around at $E_\text{soln}(=-0.225)$ that separates the solutions into incorrect and stable solutions. To see this, the present section shows truncated-domain solutions obtained near $E_\text{soln}$ for $L=1$, $F_\text{BC}=1$ and $\beta=8.1783$.\footnote{The vale of $\beta$ is set to five digits, meaning if one applies the same accuracy relation discussed in Appendix \ref{sec:stability} to this case, the relative error of solutions would be $\mid1-c^{*}_{4}/c^{*}_{4\text{o}}\mid\sim10^{-1}$.} Figure \ref{fig:Non_Soln_vK_Del_c1c4_cond_N40_trunc} \textbf{(a)} shows the values of $\mid1-c_{1}/c_{1\text{o}}\mid$, $\mid1-c^{*}_{4}/c^{*}_{4\text{o}}\mid$ and  $\mid \large{\varv_{I}(x=1)\mid}$ for $-0.35<E_\text{max}<-0.1$ and Figure \ref{fig:Non_Soln_vK_Del_c1c4_cond_N40_trunc} \textbf{(b)} depicts the condition number of the Jacobian matrix of the 4ODEs and $Q$-integral. All the values show significant changes around at $E_\text{soln}$. \cite{Heggie_1988} reported this transition as a difficulty in convergence of Newton method. Since the eigenvalues for $E_\text{soln}>E_\text{max}$  significantly deviate from both the previous and reference eigenvalues and the value of $\mid\varv_{I}(x=1)\mid$ is large $(\gtrapprox 10^{-2})$, solutions on $-1<E<E_\text{soln}$ may be considered as incorrect solutions. 

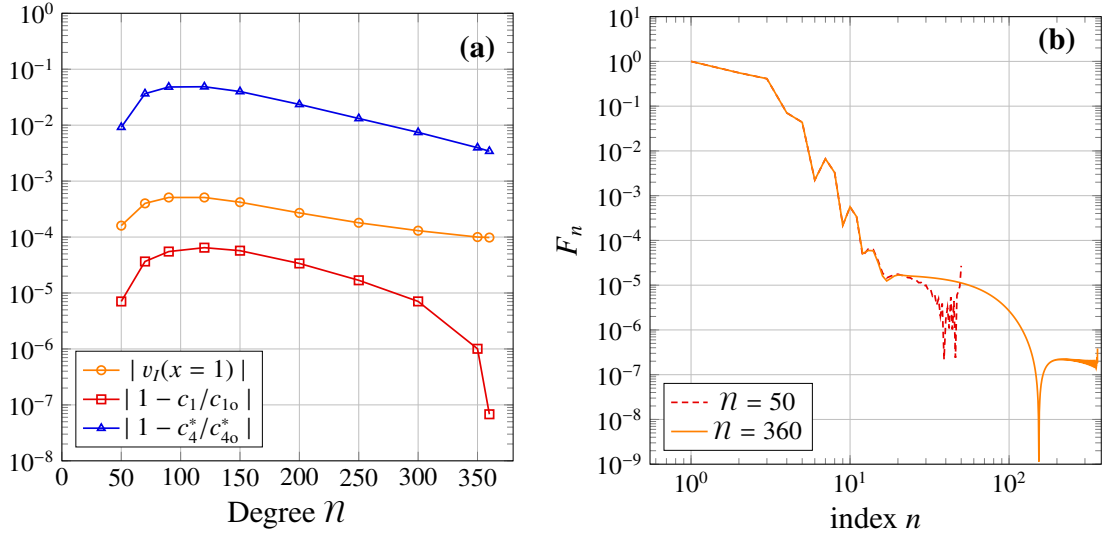
\begin{figure}[H]
	\centering
	\tikzstyle{every node}=[font=\large]
	\begin{tikzpicture}[scale=0.8]
	\begin{semilogyaxis}[width=9cm,height=9cm,grid=major,xlabel=\Large{Degree $\mathcal{N}$},xmin=0, xmax=380,ymin=1e-8,ymax=1e0, legend pos=south west, x tick label style={/pgf/number format/fixed,/pgf/number format/precision=5}, scaled ticks=false]
	\addplot [color = orange,mark=o,thick,solid] table[x index=0, y index=1]{N_vK_Del_c1c4_xmin_055_trunc.txt}; 
      \addlegendentry{$ \large{\mid\varv_{I}(x=1)\mid}$}
\addplot [color = bred,mark=square,thick,solid] table[x index=0, y index=2]{N_vK_Del_c1c4_xmin_055_trunc.txt}; 
      \addlegendentry{ \large{$\mid1-c_{1}/c_{1\text{o}}\mid$}}
\addplot [color = bblue,mark=triangle,thick,solid] table[x index=0, y index=3]{N_vK_Del_c1c4_xmin_055_trunc.txt}; 
      \addlegendentry{ \large{$\mid1-c^{*}_{4}/c^{*}_{4\text{o}}\mid$}}
      \node[above,black] at (350,1e-1) {\Large{$\textbf{(a)}$}};
	\end{semilogyaxis}
	\end{tikzpicture}
     \hspace{0.3cm}
	\begin{tikzpicture}[scale=0.8]
	\begin{loglogaxis}[width=9cm,height=9cm,grid=major,xlabel=\Large{index $n$},ylabel=\Large{$F_{n}$},xmin=0, xmax=380,ymin=1e-9,ymax=1e1, legend pos=south west]
\addplot [color = bred,mark=no,thick,densely dashed] table[x index=0, y index=1]{n_cF_N50_xmin055.txt}; 
      \addlegendentry{ \large{$\mathcal{N}=50$}}
\addplot [color = orange ,mark=no,thick,solid] table[x index=0, y index=1]{n_cF_N360_xmin055.txt}; 
      \addlegendentry{ \large{$\mathcal{N}=360$}}
     \node[above,black] at (2e2,1e0) {\Large{$\textbf{(b)}$}};
	\end{loglogaxis}
	\end{tikzpicture}
	\caption{\textbf{(a)} Relative error of $c_{1}$ and $c^{*}_{4}$ from the reference eigenvalues and the absolute value of $\varv_{I}(x=1)$ for different $\mathcal{N}$. \textbf{(b)} Chebyshev coefficients for $\varv_{F}$ with $\mathcal{N}=50$ and $\mathcal{N}=360$.  ($E_\text{max}=-0.225$ and $\beta=8.17837$. )  }
	\label{fig:Soln_trunc_xmin055}
\end{figure}

Spectral solutions for $-0.225\leq E_\text{max}<-0.05$ are relatively stable against degree $\mathcal{N}$. Especially at $E_\text{max}=-0.225$ the solution is the most stable. For $E_\text{max}=-0.225$ and $\beta=8.17837$,  Figure \ref{fig:Soln_trunc_xmin055}\textbf{(a)} shows the characteristics of $\mid1-c_{1}/c_{1\text{o}}\mid$, $\mid1-c^{*}_{4}/c^{*}_{4\text{o}}\mid$ and  $\mid \large{\varv_{I}(x=1)\mid}$ against degree $\mathcal{N}$. The Newton iteration worked well even for $\mathcal{N}=360$ and the accuracy improves with increasing $\mathcal{N}$ in the sense that the eigenvalues approach the reference eigenvalues.  Also, higher $\mathcal{N}$ provides smaller absolute values of Chebyshev coefficients. Figure \ref{fig:Soln_trunc_xmin055} \textbf{(b)} shows the coefficients for $\mathcal{N}=50$ and $\mathcal{N}=360$. The reason why the coefficients do not decay rapidly with high index $n$ would be that the rapid decay was hindered by the discontinuous behavior of the $Q$-integral (Appendix \ref{sec:Fixed_cR}).  Possible causes of the discontinuity are that one can not correctly specify the value of $\beta$ with high accuracy or even an accurate solution does not exist when $E_\text{max}$ is not close to zero. In fact, the discontinuous behavior disappears for semi-stable solutions with $E_\text{max}\approx 0.05$ (Section \ref{sec:semi_stable_trunc_solns}).

\subsection{Optimal eigenvalues of semi-stable solutions on truncated domains with $-0.08<E_\text{max}<-0.03$}\label{sec:semi_stable_trunc_solns}

On truncated domains with $-0.08<E_\text{max}<-0.03$, Newton iteration method well worked for different $\mathcal{N}$ and we found truncated-domain solutions and eigenvalues that are close to the reference- solution and eigenvalues. Especially, the present section provides an optimal value of $\beta$ on the truncated domains. Table \ref{table:trunc_soln} shows optimal eigenvalues for the semi-stable solutions with $-0.08<E_\text{max}<-0.03$ and the maximum significant figures of $\beta$ is limited to ten.\footnote{The condition number of the Jacobian is order of $10^{7}$ in Newton iteration process for the whole-domain solution. This  means one can obtain approximately five significant-figure solution since the minimum of the `practical' machine precision is order of $10^{-12}$ (Appendix \ref{sec:Newton_method}). Considering that the gap in accuracy is order of $10^{-5}$ between $c_{1}$ (or $\beta$) and $c_{4}$, the ten maximum significant figures are a reasonable choice for $\beta$.} Also, the table  presents the values of $\mid1-c_{1}/c_{1\text{o}}\mid$, $\mid1-c^{*}_{4}/c^{*}_{4\text{o}}\mid$ and $\mid\varv_{I}(x=1)\mid$ to show the accuracy of the solutions. On $-0.07<E_\text{max}<-0.03$, the optimal value of $\beta$ was the same as up to eight significant figures $(=8.1783712)$ of $\beta_{\text{o}}$. On one hand, at $E_\text{max}=-0.08$ the optimal value of $\beta$ is relatively large.  This would be since $\mathcal{N}=25$ is not large enough to provide an accurate solution (while Newton iteration did not work over $\mathcal{N}=25$). Also, we could not find solutions for small $E_\text{max}<-0.07$ whose eigenvalues are as close to $\beta_\text{o}$ as the solutions for $-0.07<E_\text{max}<-0.03$.

The truncated-domain solutions have an advantage over the whole-domain solution. The former needs low degrees ($\mathcal{N}=25\sim55$) of polynomials to make Newton method work. Also, even the lowest degrees provide reasonable results in accuracy; $\varv_{I}(x=1)=10^{-8}\sim 10^{-9}$ and $\mid1-c^{*}_{4}/c^{*}_{4\text{o}}\mid 10^{-4}\sim 10^{-6}$. The result of the present section confirms that the eigenvalues of the truncated-domain solutions for $-0.08<E_\text{max}<-0.03$ are the same as those of the whole-domain solution with the prescribed accuracies.  

\begin{table*}\centering\large
	\ra{1.3}
	\scalebox{0.7}{
	\begin{tabular}{@{}clccc@{}}
            \multicolumn{2}{l}{ \large{$E_\text{max}=-0.03$}} & \multicolumn{2}{l}{\large{or $\quad (-E_\text{max})^{\beta}\approx 3.5\times10^{-13}$}}    &  \\
              \toprule
		 \large{$\mathcal{N}$} & Eigenvalue $\beta$ & $\mid1-c_{1}/c_{1\text{o}} \mid$ &  $\mid1-c^{*}_{4}/c^{*}_{4\text{o}} \mid$  & $\varv_{I}(x=1)$  \\ 
		\midrule
             $55$ & $8.178371160$ & $6.5\times10^{-10}$& $2.6\times10^{-6}$  &$ 1.6\times10^{-8}$\\
		$50$ & $8.178371160$ & $7.1\times10^{-10}$& $2.9\times10^{-6}$  &$ 1.7\times10^{-8}$\\
		$40$ & $8.178371165$ & $2.4\times10^{-9}$& \underline{$6.9\times10^{-7}$}  &\underline{$ 4.3\times10^{-9}$}\\
		$30$ & $8.178370376$ & $1.3\times10^{-7}$& $9.8\times10^{-6}$  & $8.1\times10^{-10}$\\
		$25$ & $8.1783436$ & $8.1\times10^{-7}$& $1.3\times10^{-4}$  & $1.1\times10^{-9}$\\
		\bottomrule
	\end{tabular}}
\hspace{0.5cm}
	\scalebox{0.7}{
	\begin{tabular}{@{}clccc@{}}
            \multicolumn{2}{l}{ \large{$E_\text{max}=-0.04$}} & \multicolumn{2}{l}{\large{or $\quad (-E_\text{max})^{\beta}\approx 3.7\times10^{-12}$}}    &  \\
              \toprule
		 \large{$\mathcal{N}$} & Eigenvalue $\beta$ & $\mid1-c_{1}/c_{1\text{o}} \mid$ &  $\mid1-c^{*}_{4}/c^{*}_{4\text{o}} \mid$  & $\varv_{I}(x=1)$  \\ 
		\midrule
             $55$ & $8.178371160$ & $5.1\times10^{-10}$& $2.1\times10^{-6}$  &$ 1.3\times10^{-8}$\\
		$50$ & $8.178371160$ & $1.9\times10^{-10}$& $8.3\times10^{-7}$  &$ 5.1\times10^{-9}$\\
		$40$ & $8.178371170$ & $1.4\times10^{-10}$& \underline{$7.1\times10^{-7}$}  &\underline{$ 4.0\times10^{-9}$}\\
		$30$ & $8.17837159$ & $1.8\times10^{-7}$& $4.7\times10^{-6}$  & $6.7\times10^{-9}$\\
		$25$ & $8.1783585$ & $1.7\times10^{-6}$& $6.8\times10^{-5}$  & $5.2\times10^{-9}$\\
		\bottomrule
	\end{tabular}}

\vspace{0.5cm}

\scalebox{0.7}{
\begin{tabular}{@{}clccc@{}}
            \multicolumn{2}{l}{ \large{$E_\text{max}=-0.05$}} & \multicolumn{2}{l}{\large{or $\quad (-E_\text{max})^{\beta}\approx 2.2\times10^{-11}$}}   &  \\
              \toprule
		 \large{$\mathcal{N}$} & Eigenvalue $\beta$ & $\mid1-c_{1}/c_{1\text{o}} \mid$ &  $\mid1-c^{*}_{4}/c^{*}_{4\text{o}} \mid$  & $\varv_{I}(x=1)$  \\ 
		\midrule
		$50$ & $8.178371158$ & $2.8\times10^{-9}$& $1.0\times10^{-5}$  &$ 6.8\times10^{-8}$\\
		$40$ & $8.178371165$ & $5.5\times10^{-9}$& \underline{$9.7\times10^{-7}$}  & \underline{$ 3.0\times10^{-9}$}\\
		$30$ & $8.1783723$ & $2.1\times10^{-8}$& $3.7\times10^{-6}$  & $3.1\times10^{-8}$\\
		$25$ & $8.178373$ & $1.5\times10^{-6}$& $1.9\times10^{-5}$  & $4.0\times10^{-9}$\\
		\bottomrule
	\end{tabular}}
\hspace{0.5cm}
\scalebox{0.7}{
\begin{tabular}{@{}clccc@{}}
            \multicolumn{2}{l}{ \large{$E_\text{max}=-0.06$} }& \multicolumn{2}{l}{\large{or $\quad (-E_\text{max})^{\beta}\approx 1.0\times10^{-10}$}}                                    &  \\
              \toprule
		 \large{$\mathcal{N}$} & Eigenvalue $\beta$ & $\mid1-c_{1}/c_{1\text{o}} \mid$ &  $\mid1-c^{*}_{4}/c^{*}_{4\text{o}} \mid$  & $\varv_{I}(x=1)$  \\ 
		\midrule
		$40$ & $8.178371160$ & $8.3\times10^{-10}$& \underline{$8.7\times10^{-7}$}  & \underline{$ 2.2\times10^{-9}$}\\
		$30$ & $8.17837225$ & $1.5\times10^{-7}$& $5.0\times10^{-6}$  & $8.1\times10^{-9}$\\
		$25$ & $8.1783813$ & $2.4\times10^{-7}$& $1.6\times10^{-5}$  & $2.3\times10^{-9}$\\
		\bottomrule
	\end{tabular}}

\vspace{0.5cm}

\scalebox{0.7}{
\begin{tabular}{@{}clccc@{}}
           \multicolumn{2}{l}{  \large{$E_\text{max}=-0.07$}} & \multicolumn{2}{l}{\large{or $\quad (-E_\text{max})^{\beta}\approx 3.6\times10^{-10}$}}                    &  \\
              \toprule
		 \large{$\mathcal{N}$} & Eigenvalue $\beta$ & $\mid1-c_{1}/c_{1\text{o}} \mid$ &  $\mid1-c^{*}_{4}/c^{*}_{4\text{o}} \mid$  & $\varv_{I}(x=1)$  \\ 
		\midrule
		$35$ & $8.178371159$ & $8.6\times10^{-9}$& \underline{$2.3\times10^{-6}$}  & \underline{$ 1.4\times10^{-9}$}\\
		$30$ & $8.1783717$ & $2.2\times10^{-7}$& $1.9\times10^{-5}$  & $9.7\times10^{-8}$\\
		$25$ & $8.178382$ & $1.1\times10^{-6}$& $5.0\times10^{-5}$  & $9.8\times10^{-8}$\\
		\bottomrule
	\end{tabular}}
\hspace{0.5cm}
\scalebox{0.7}{
\begin{tabular}{@{}clccc@{}}
            \multicolumn{2}{l}{ \large{$E_\text{max}=-0.08$}} & \multicolumn{2}{l}{\large{or $\quad (-E_\text{max})^{\beta}\approx 1.1\times10^{-9}$}}      &  \\
              \toprule
		 \large{$\mathcal{N}$} & Eigenvalue $\beta$ & $\mid1-c_{1}/c_{1\text{o}} \mid$ &  $\mid1-c^{*}_{4}/c^{*}_{4\text{o}} \mid$  & $\varv_{I}(x=1)$  \\ 
		\midrule
		$25$ & $8.1783768$ & $2.8\times10^{-6}$& $\underline{4.7\times10^{-5}}$  & \underline{$ 7.9\times10^{-9}$}\\
		\bottomrule
	\end{tabular}}
	\caption{ Numerical results for the truncated-domain formulation at different $E_\text{max}$ ($L=1$ and $F_\text{BC}=1$). The minimum values of $\mid1-c^{*}_{4}/c^{*}_{4\text{o}}\mid$ and $\varv_{I}(x=1)$ are underlined to highlight the accuracy for each $\mathcal{N}$. The results do not include some data in which the value of $\varv_{I}(x=1)$ are greater than of the order of $\sim10^{-7}$ for convenience. The Newton iteration method was very hard to work for $\mathcal{N}<25$ and the degrees beyond the maximum values of $\mathcal{N}$ for each $E_\text{max}$; those conditions provided the change $\mid\{a_n\}^\text{(new)}-\{a_n\}^\text{(old)}\mid\gtrapprox10^{-9}$ while the data in the table are the results when it reached order of $\sim10^{-12}$.}
	\label{table:trunc_soln}
\end{table*}

\subsection{Optimal semi-stable solutions on truncated domains with $-0.1<E_\text{max}<-0.03$ for fixed $\beta=\beta_{\text{o}}$}\label{sec:trunc_solns_BETAo}

To see the direct relationship between the reference and truncated-domain solutions, we show the truncated-domain solutions with fixed $\beta=\beta_{\text{o}}$ for $-0.1<E_\text{max}<-0.03$. The result of Section \ref{sec:semi_stable_trunc_solns} shows that the optimal eigenvalues for semi-stable solutions are close to the reference eigenvalues, hence we fix $\beta$ to the reference value $\beta_\text{o}$. We report semi-stable solutions with $\beta=\beta_\text{o}$ (Section \ref{sec:semi_sta_fixed_beta}), show the relation of the solutions with numerical instability against change in degree $\mathcal{N}$ (Section \ref{sec:trunc_insta}) and propose an optimal semi-stable solution that is compatible to the reference solution in accuracy (Section \ref{sec:opt_trunc}).

\subsubsection{Semi-stable solutions with $\beta=\beta_\text{o}$}\label{sec:semi_sta_fixed_beta}

\begin{figure}[H]
	\centering
\tikzstyle{every node}=[font=\Large]
	\begin{tikzpicture}[scale=0.7]
\begin{semilogyaxis}[width=10cm,height=7.5cm,grid=major,xlabel=\Large{$\mathcal{N}$},xmin=0, xmax=71,ymin=1e-17,ymax=1e-0, legend pos=south west]
\addplot [color = bred,mark=square,thick,solid] table[x index=0, y index=1]{N_vK_Del_c1c4_cF_xmin_080_trunc.txt}; 
\addlegendentry{\large{$\mid\varv_{I}(x=1)\mid$}}
\addplot [color = blue,mark=*,thick] table[x index=0, y index=2]{N_vK_Del_c1c4_cF_xmin_080_trunc.txt}; 
\addlegendentry{\large{$\mid1-c_{1}/ c_{1\text{o}}\mid$}}
\addplot [color = orange,mark=triangle,thick] table[x index=0, y index=3]{N_vK_Del_c1c4_cF_xmin_080_trunc.txt}; 
\addlegendentry{\large{$\mid1-c^{*}_{4}/ c^{*}_{4\text{o}}\mid$}}
%\addplot [color = blue,mark=o,thick] table[x index=0, y index=4]{N_vK_Del_c1c4_cF_xmin_082_trunc.txt}; 
%\addlegendentry{\large{min. of $Fn$}}
\node[above,black] at (55,1e-17) {$E_\text{max}=-0.10$};
\end{semilogyaxis}
\end{tikzpicture}
	\begin{tikzpicture}[scale=0.7]
\begin{semilogyaxis}[width=10cm,height=7.5cm,grid=major,xlabel=\Large{$\mathcal{N}$},xmin=0, xmax=71,ymin=1e-17,ymax=1e-0, legend pos=south west]
\addplot [color = bred,mark=square,thick,solid] table[x index=0, y index=1]{N_vK_Del_c1c4_cF_xmin_082_trunc.txt}; 
\addlegendentry{\large{$\mid\varv_{I}(x=1)\mid$}}
\addplot [color = blue,mark=*,thick] table[x index=0, y index=2]{N_vK_Del_c1c4_cF_xmin_082_trunc.txt}; 
\addlegendentry{\large{$\mid1-c_{1}/ c_{1\text{o}}\mid$}}
\addplot [color = orange,mark=triangle,thick]  table[x index=0, y index=3]{N_vK_Del_c1c4_cF_xmin_082_trunc.txt}; 
\addlegendentry{\large{$\mid1-c^{*}_{4}/ c^{*}_{4\text{o}}\mid$}}
%\addplot [color = blue,mark=o,thick] table[x index=0, y index=4]{N_vK_Del_c1c4_cF_xmin_082_trunc.txt}; 
%\addlegendentry{\large{min. of $Fn$}}
\node[above, black] at (55,1e-17) {$E_\text{max}=-0.09$};
\end{semilogyaxis}
\end{tikzpicture}

\vspace{0.4cm}

	\begin{tikzpicture}[scale=0.7]
	\begin{semilogyaxis}[width=10cm,height=7.5cm,grid=major,xlabel=\Large{$\mathcal{N}$},xmin=0, xmax=71,ymin=1e-17,ymax=1e-0, legend pos=south west]
\addplot [color = bred,mark=square,thick,solid] table[x index=0, y index=1]{N_vK_Del_c1c4_cF_xmin_084_trunc.txt}; 
\addlegendentry{\large{$\mid\varv_{I}(x=1)\mid$}}
\addplot [color = blue,mark=*,thick] table[x index=0, y index=2]{N_vK_Del_c1c4_cF_xmin_084_trunc.txt}; 
\addlegendentry{\large{$\mid1-c_{1}/ c_{1\text{o}}\mid$}}
\addplot [color = orange,mark=triangle,thick] table[x index=0, y index=3]{N_vK_Del_c1c4_cF_xmin_084_trunc.txt}; 
\addlegendentry{\large{$\mid1-c^{*}_{4}/ c^{*}_{4\text{o}}\mid$}}
%\addplot [color = blue,mark=o,thick] table[x index=0, y index=4]{N_vK_Del_c1c4_cF_xmin_084_trunc.txt}; 
%\addlegendentry{\large{min. of $Fn$}}
\node[above, black] at (55,1e-17) {$E_\text{max}=-0.08$};
	\end{semilogyaxis}
	\end{tikzpicture}
	\begin{tikzpicture}[scale=0.7]
	\begin{semilogyaxis}[width=10cm,height=7.5cm,grid=major,xlabel=\Large{$\mathcal{N}$},xmin=0, xmax=71,ymin=1e-17,ymax=1e-0, legend pos=south west]
\addplot [color = bred,mark=square,thick,solid] table[x index=0, y index=1]{N_vK_Del_c1c4_cF_xmin_086_trunc.txt}; 
\addlegendentry{\large{$\mid\varv_{I}(x=1)\mid$}}
\addplot [color = blue,mark=*,thick] table[x index=0, y index=2]{N_vK_Del_c1c4_cF_xmin_086_trunc.txt}; 
\addlegendentry{\large{$\mid1-c_{1}/ c_{1\text{o}}\mid$}}
\addplot [color = orange,mark=triangle,thick] table[x index=0, y index=3]{N_vK_Del_c1c4_cF_xmin_086_trunc.txt}; 
\addlegendentry{\large{$\mid1-c^{*}_{4}/ c^{*}_{4\text{o}}\mid$}}
%\addplot [color = blue,mark=o,thick] table[x index=0, y index=4]{N_vK_Del_c1c4_cF_xmin_086_trunc.txt}; 
%\addlegendentry{\large{min. of $Fn$}}
\node[above, black] at (55,1e-17) {$E_\text{max}=-0.07$};
	\end{semilogyaxis}
	\end{tikzpicture}
	
	\vspace{0.4cm}
	
	\begin{tikzpicture}[scale=0.7]
	\begin{semilogyaxis}[width=10cm,height=7.5cm,grid=major,xlabel=\Large{$\mathcal{N}$},xmin=0, xmax=71,ymin=1e-17,ymax=1e-0, legend pos=south west]
\addplot [color = bred,mark=square,thick,solid] table[x index=0, y index=1]{N_vK_Del_c1c4_cF_xmin_088_trunc.txt}; 
\addlegendentry{\large{$\mid\varv_{I}(x=1)\mid$}}
\addplot [color = blue,mark=*,thick] table[x index=0, y index=2]{N_vK_Del_c1c4_cF_xmin_088_trunc.txt}; 
\addlegendentry{\large{$\mid1-c_{1}/ c_{1\text{o}}\mid$}}
\addplot [color = orange,mark=triangle,thick] table[x index=0, y index=3]{N_vK_Del_c1c4_cF_xmin_088_trunc.txt}; 
\addlegendentry{\large{$\mid1-c^{*}_{4}/ c^{*}_{4\text{o}}\mid$}}
%\addplot [color = blue,mark=o,thick] table[x index=0, y index=4]{N_vK_Del_c1c4_cF_xmin_088_trunc.txt}; 
%\addlegendentry{\large{min. of $Fn$}}
\node[above, black] at (55,1e-17) {$E_\text{max}=-0.06$};
	\end{semilogyaxis}
	\end{tikzpicture}
	\begin{tikzpicture}[scale=0.7]
	\begin{semilogyaxis}[width=10cm,height=7.5cm,grid=major,xlabel=\Large{$\mathcal{N}$},xmin=0, xmax=71,ymin=1e-17,ymax=1e-0, legend pos=south west]
\addplot [color = bred,mark=square,thick,solid] table[x index=0, y index=1]{N_vK_Del_c1c4_cF_xmin_090_trunc.txt}; 
\addlegendentry{\large{$\mid\varv_{I}(x=1)\mid$}}
\addplot [color = blue,mark=*,thick] table[x index=0, y index=2]{N_vK_Del_c1c4_cF_xmin_090_trunc.txt}; 
\addlegendentry{\large{$\mid1-c_{1}/ c_{1\text{o}}\mid$}}
\addplot [color = orange,mark=triangle,thick] table[x index=0, y index=3]{N_vK_Del_c1c4_cF_xmin_090_trunc.txt}; 
\addlegendentry{\large{$\mid1-c^{*}_{4}/ c^{*}_{4\text{o}}\mid$}}
%\addplot [color = blue,mark=o,thick] table[x index=0, y index=4]{N_vK_Del_c1c4_cF_xmin_090_trunc.txt}; 
%\addlegendentry{\large{min. of $Fn$}}
\node[above,black] at (55,1e-17) {$E_\text{max}=-0.05$};
	\end{semilogyaxis}
	\end{tikzpicture}

\vspace{0.4cm}

	\begin{tikzpicture}[scale=0.7]
	\begin{semilogyaxis}[width=10cm,height=7.5cm,grid=major,xlabel=\Large{$\mathcal{N}$},xmin=0, xmax=71,ymin=1e-17,ymax=1e-0, legend pos=south west]
\addplot [color = bred,mark=square,thick,solid] table[x index=0, y index=1]{N_vK_Del_c1c4_cF_xmin_092_trunc.txt}; 
\addlegendentry{\large{$\mid\varv_{I}(x=1)\mid$}}
\addplot [color = blue,mark=*,thick] table[x index=0, y index=2]{N_vK_Del_c1c4_cF_xmin_092_trunc.txt}; 
\addlegendentry{\large{$\mid1-c_{1}/ c_{1\text{o}}\mid$}}
\addplot [color = orange,mark=triangle,thick]table[x index=0, y index=3]{N_vK_Del_c1c4_cF_xmin_092_trunc.txt}; 
\addlegendentry{\large{$\mid1-c^{*}_{4}/ c^{*}_{4\text{o}}\mid$}}
%\addplot [color = blue,mark=o,thick] table[x index=0, y index=4]{N_vK_Del_c1c4_cF_xmin_092_trunc.txt}; 
%\addlegendentry{\large{min. of $Fn$}}
\node[above,black] at (55,1e-17) {$E_\text{max}=-0.04$};
	\end{semilogyaxis}
	\end{tikzpicture}
	\begin{tikzpicture}[scale=0.7]
	\begin{semilogyaxis}[width=10cm,height=7.5cm,grid=major,xlabel=\Large{$\mathcal{N}$},xmin=0, xmax=71,ymin=1e-17,ymax=1e-0, legend pos=south west]
\addplot [color = bred,mark=square,thick,solid] table[x index=0, y index=1]{N_vK_Del_c1c4_cF_xmin_094_trunc.txt}; 
\addlegendentry{\large{$\mid\varv_{I}(x=1)\mid$}}
\addplot [color = blue,mark=*,thick] table[x index=0, y index=2]{N_vK_Del_c1c4_cF_xmin_094_trunc.txt}; 
\addlegendentry{\large{$\mid1-c_{1}/ c_{1\text{o}}\mid$}}
\addplot [color = orange,mark=triangle,thick] table[x index=0, y index=3]{N_vK_Del_c1c4_cF_xmin_094_trunc.txt}; 
\addlegendentry{\large{$\mid1-c^{*}_{4}/ c^{*}_{4\text{o}}\mid$}}
%\addplot [color = blue,mark=o,thick] table[x index=0, y index=4]{N_vK_Del_c1c4_cF_xmin_094_trunc.txt}; 
%\addlegendentry{\large{min. of $Fn$}}
\node[above,black] at (55,1e-17) {$E_\text{max}=-0.03$};
	\end{semilogyaxis}
	\end{tikzpicture}
\caption{ Relative errors of  $c_{1}$ and $c^{*}_{4}$ from the reference eigenvalues and characteristics of $\varv_{I}(x=1)$ against $\mathcal{N}$ for the truncated-domain solutions with $\beta=\beta_{\text{o}}$. ($-0.1\leq E_\text{max}\leq-0.03$,  $L=1$ and $F_\text{BC}=1$.) }
\label{N_vK_Del_c1c4_cF_trunc}
\end{figure}
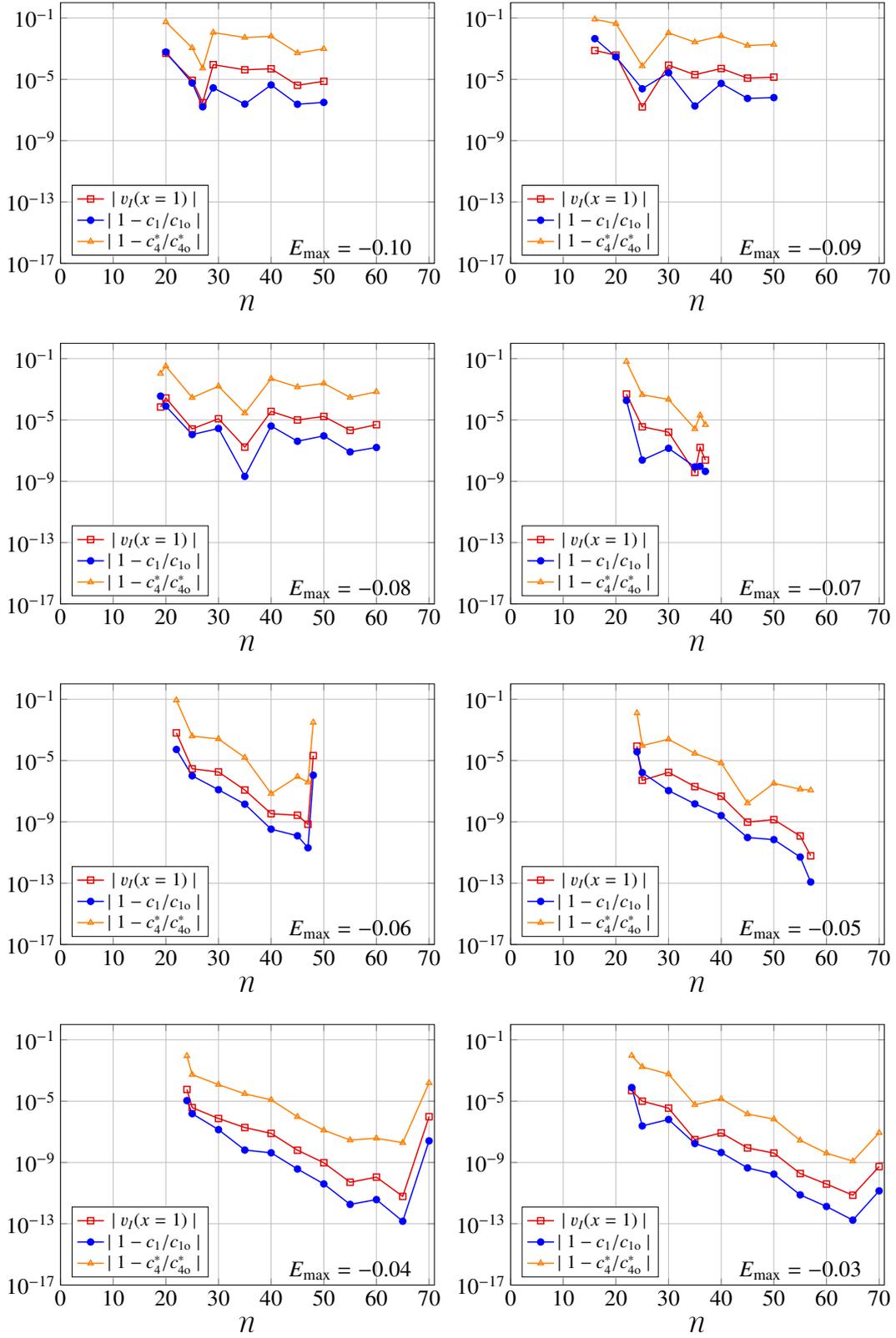

Semi-stable solutions with $\beta=\beta_\text{o}$ approach the reference solution as the degree of polynomials increases but they lose accuracy beyond certain degrees. We found truncated-domain solutions with $\beta=\beta_{\text{o}}$ for $-0.1\leq E_\text{max}<-0.03$. In order to see the accuracy of the solutions, Figure \ref{N_vK_Del_c1c4_cF_trunc} shows the characteristics of $\varv_{I}(x=1)$ against $\mathcal{N}$ and the relative errors of $c_{1}$ and $c^{*}_{4}$ from the reference values. For $E_\text{max}>-0.07$, $\mid\varv_{I}(x=1)\mid$, $\mid1-c_{1}/ c_{1\text{o}}\mid$ and $\mid1-c^{*}_{4}/ c^{*}_{4\text{o}}\mid$ show an ideal characteristics under change in $\mathcal{N}$. They decrease with increasing $\mathcal{N}$ and can reach very small values ($\approx10^{-9}\sim10^{-13}$) at certain degrees. Beyond the degrees, the Newton iteration method, however, did not work or the solutions significantly lose their accuracies. On one hand, for $E_\text{max}<-0.07$ the characteristics of the $\mathcal{N}$-dependence are less ideal. The values of $\mid\varv_{I}(x=1)\mid$, $\mid1-c_{1}/ c_{1\text{o}}\mid$ and $\mid1-c^{*}_{4}/ c^{*}_{4\text{o}}\mid$ stall with increasing $\mathcal{N}$ while the minimum values still can be found at relatively-low degrees  $(\mathcal{N}=27\sim35)$ . This would be since the optimal value of $\beta$ is not close to $\beta_\text{o}$ as found in Table \ref{table:trunc_soln}; near $E_\text{max}=-0.08$ the optimal value may be larger than $\beta_\text{o}$. 

\subsubsection{Truncated domains and numerical instability against change in degree $\mathcal{N}$}\label{sec:trunc_insta}

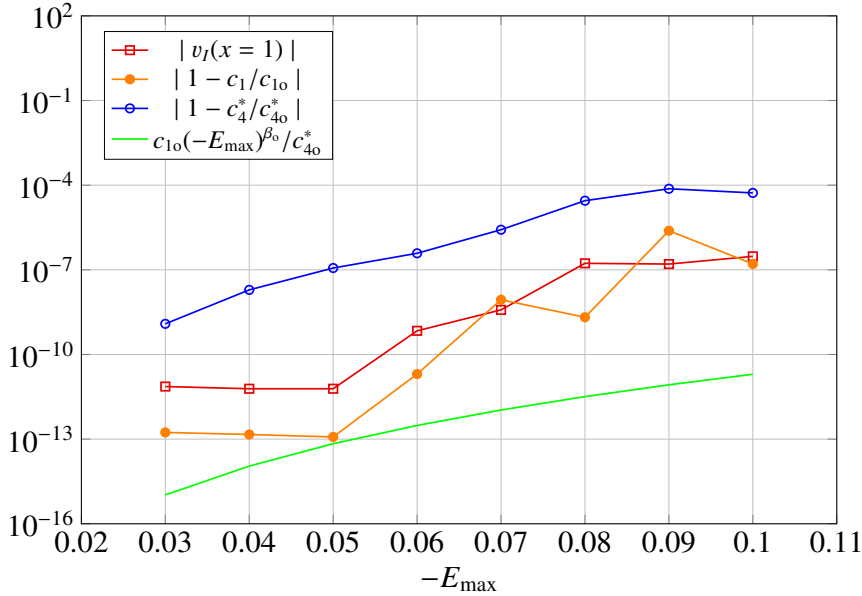
\begin{figure}[H]
	\centering
	\tikzstyle{every node}=[font=\Large]
	\begin{tikzpicture}[scale=0.8]
	\begin{semilogyaxis}[width=14cm,height=10cm,grid=major,xlabel=\Large{$-E_\text{max}$},xmin=0.02, xmax=0.11,ymin=1e-16,ymax=1e2, legend pos=north west,x tick label style={/pgf/number format/fixed,/pgf/number format/precision=5}, scaled ticks=false]
	\addplot [color = bred,mark=square,thick,solid] table[x index=0, y index=1]{Emin_vK_Del_c1c4_BETAo_trunc.txt}; 
	\addlegendentry{\large{$\mid\varv_{I}(x=1)\mid$}}
	\addplot [color =orange,mark=*,thick] table[x index=0, y index=2]{Emin_vK_Del_c1c4_BETAo_trunc.txt}; 
	\addlegendentry{\large{$\mid1-c_{1}/ c_{1\text{o}}\mid$}}
	\addplot [color = blue,mark=o,thick] table[x index=0, y index=3]{Emin_vK_Del_c1c4_BETAo_trunc.txt}; 
	\addlegendentry{\large{$\mid1-c^{*}_{4}/ c^{*}_{4\text{o}}\mid$}}
	\addplot [color = green,mark=no,thick,solid] table[x index=0, y index=4]{Emin_vK_Del_c1c4_BETAo_trunc.txt}; 
	\addlegendentry{\large{$c_{1\text{o}}(-E_\text{max})^{\beta_\text{o}}/c^{*}_{4\text{o}}$}}
	%\addplot [color = blue,mark=no,thick,dashed] table[x index=0, y index=5]{xmin_vK_Del_c1c4_N200_trunc.txt}; 
	%\addlegendentry{\large{eps}}
	\end{semilogyaxis}
	\end{tikzpicture}
	\caption{Relative error of $c_{1}$ and $c^{*}_{4}$ from the reference values and characteristics of $\mid \varv_{I}(x=1)\mid$ that are obtained at each $\mathcal{N}_\text{best}$. The guideline $c_{1\text{o}}(-E_\text{max})^{\beta_{\text{o}}}/c_{4\text{o}}^{*}$ is shown for comparison. ($L=1$, $F_\text{BC}=1$ and $\beta=\beta_{\text{o}}$.)}
	\label{fig:Best_xmin_094_N65}
\end{figure}

We consider the numerical stability that occurred to the semi-stable solutions and reference solution originates from the property that ss-OAFP system may not have a solution when the terms of the system reach order of machine precision at equation level and beyond the accuracy. In Figure \ref{N_vK_Del_c1c4_cF_trunc}, the minimum of $\mid\varv_{I}(x=1)\mid$ occurs at degrees $\mathcal{N}_\text{best}=\{$ 27, 25, 35, 35, 47, 57, 65, 65$\}$ for $E_\text{max}=\{$ -0.10, -0.09, -0.08, -0.07, -0.06, -0.05, -0.04, -0.03 $\}$. To consider why the truncated-domain solutions lose their accuracy beyond $\mathcal{N}_\text{best}$,  Figure \ref{fig:Best_xmin_094_N65} depicts the $E_\text{max}$-dependence of  $\mid\varv_{I}(x=1)\mid$, $\mid1-c_{1}/ c_{1\text{o}}\mid$ and $\mid1-c^{*}_{4}/ c^{*}_{4\text{o}}\mid$ obtained at each $\mathcal{N}_\text{best}$. The value of $\mid1-c^{*}_{4}/ c^{*}_{4\text{o}}\mid$ decreases in a power-law-like fashion with increasing $E_\text{max}$. One may understand this characteristics by introducing a power-law profile $c_{1\text{o}}(-E_\text{max})^{\beta_{\text{o}}}/c_{4\text{o}}^{*}$. This profile originates from the power-law dependence of the last term in equation \eqref{Eq.ss-4ODE-vF} (See Appendix \ref{sec:Newton_method}). In Figure \ref{fig:Best_xmin_094_N65}, the decrease of $c_{1\text{o}}(-E_\text{max})^{\beta_{\text{o}}}/c_{4\text{o}}^{*}$ is similar to that of $\mid1-c^{*}_{4}/ c^{*}_{4\text{o}}\mid$. On one hand,  $\mid1-c_{1}/ c_{1\text{o}}\mid$ and $\mid\varv_{I}(x=1)\mid$ stops decreasing at $E_\text{max}$ larger than $-0.05$. This may be understood as the limit of double precision.  In addition to $c_{1\text{o}}(-E_\text{max})^{\beta_{\text{o}}}/c_{4\text{o}}^{*}$ that characterizes the accuracy of $c_{4\text{o}}^{*}$ (correspondingly the solution), the infinity norm of $\mid\{F_{n}\}^{\text{(old)}}-\{F_{n}\}^{\text{(new)}}\mid$ for Newton method reaches order of $10^{-13}$ at best (Appendix \ref{sec:Newton_method}). Under these circumstances, $c_{1\text{o}}(-E_\text{max})^{\beta_{\text{o}}}/c_{4\text{o}}^{*}$reaches order of $10^{-13}$ at $E_\text{max}\approx-0.0523$ that is the maximum value of $E_\text{max}$ to preserve numerical accuracy. This result implies that, for the truncated-domain solutions for $E_\text{max}\lesssim-0.05$, machine precision is not enough precise to obtain more accurate solution. On one hand, in case of the reference solution, the solution is not truncated on large $E$, meaning the power-law boundary conditions in the ss-OAFP system can be satisfied only when they reach order of machine precision. Hence, the reference solution does not improve accuracy with increasing low degrees of polynomials unlike the semi-stable solution and it only loses its accuracy with increasing large degree. One can find more detail discussion for machine precision and the convergence of Newton iteration method in Appendix \ref{sec:Newton_method}.

\subsubsection{An optimal semi-stable solution}\label{sec:opt_trunc}

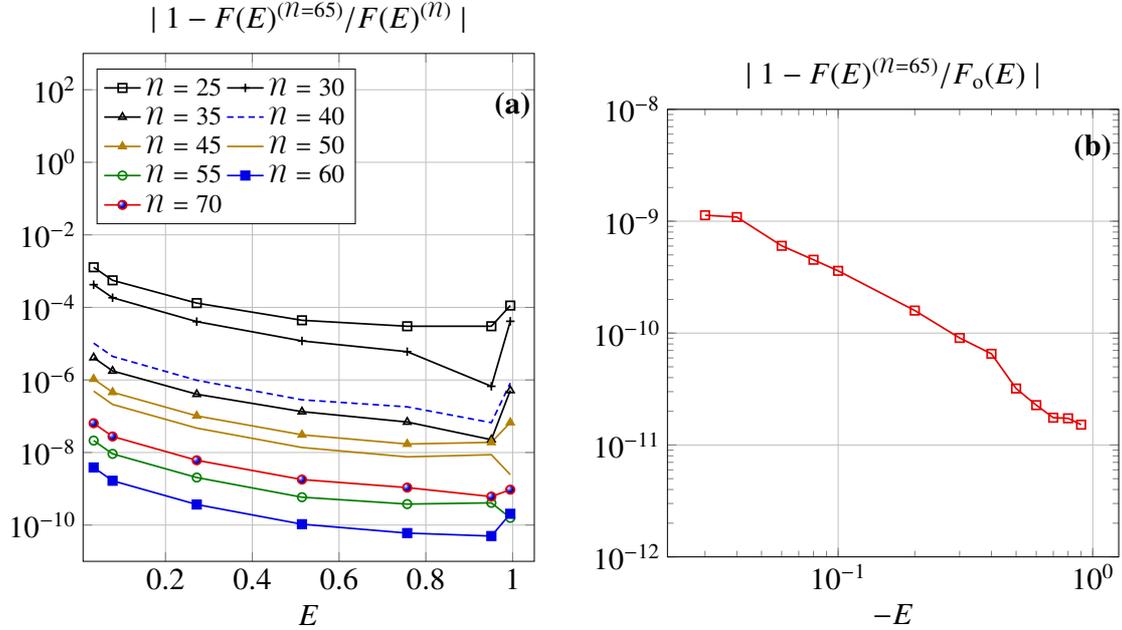
\begin{figure}[H]
	\centering
	\tikzstyle{every node}=[font=\Large]
	\begin{tikzpicture}[scale=0.8]
	\begin{semilogyaxis}[width=9cm,height=10cm,legend columns=2, grid=major,xlabel=\Large{$E$},title=\Large{$\mid1-F(E)^{(\mathcal{N}=65)}/F(E)^{(\mathcal{N})}\mid$},xmin=0.01,xmax=1.05,ymin=1e-11,ymax=1e3, legend pos=north west]
	\addplot [color = black ,mark=square,thick,solid ]  table[x index=0, y index=1]{Stab_Deg_F_xmin_094_N65.txt}; 
       \addlegendentry{\large{$\mathcal{N}=25$}}
      \addplot [color = black ,mark=+, thick, solid] table[x index=0, y index=2]{Stab_Deg_F_xmin_094_N65.txt}; 
       \addlegendentry{\large{$\mathcal{N}=30$}}
      \addplot [color = black ,mark=triangle, thick, solid] table[x index=0, y index=3]{Stab_Deg_F_xmin_094_N65.txt}; 
       \addlegendentry{\large{$\mathcal{N}=35$}}
      \addplot [color = bblue ,mark=no, thick,densely dashed] table[x index=0, y index=4]{Stab_Deg_F_xmin_094_N65.txt}; 
       \addlegendentry{\large{$\mathcal{N}=40$}}
\addplot [color = bgold ,mark=triangle*, thick] table[x index=0, y index=5]{Stab_Deg_F_xmin_094_N65.txt}; 
       \addlegendentry{\large{$\mathcal{N}=45$}}
\addplot [color = bgold ,mark=no, thick, solid] table[x index=0, y index=6]{Stab_Deg_F_xmin_094_N65.txt}; 
       \addlegendentry{\large{$\mathcal{N}=50$}}
\addplot [color = bgreen ,mark=o, thick] table[x index=0, y index=7]{Stab_Deg_F_xmin_094_N65.txt}; 
       \addlegendentry{\large{$\mathcal{N}=55$}}
\addplot [color = bblue ,mark=square*, thick] table[x index=0, y index=8]{Stab_Deg_F_xmin_094_N65.txt}; 
       \addlegendentry{\large{$\mathcal{N}=60$}}
\addplot [ color = bred ,mark=ball, thick] table[x index=0, y index=9]{Stab_Deg_F_xmin_094_N65.txt}; 
       \addlegendentry{\large{$\mathcal{N}=70$}}
\node[above,black] at (1,1e1) {$\textbf{(a)}$};
      \end{semilogyaxis}
	\end{tikzpicture}
	\hspace{0.5cm}
	\begin{tikzpicture}[scale=0.8]
	\begin{loglogaxis}[width=9cm,height=9cm,grid=major,xlabel=\Large{$-E$},title=\Large{$\mid1-F(E)^{(\mathcal{N}=65)}/F_\text{o}(E)\mid$},xmin=-0.94, xmax=-0.001,ymin=1e-12,ymax=1e-8, legend pos=north west]
	\addplot [color = bred,mark=square,thick,solid] table[x index=0, y index=1]{compari_N70_N65_xm094.txt}; 
\node[above,black] at (1e0,3e-9) {$\textbf{(b)}$};
	\end{loglogaxis}
	\end{tikzpicture}
	\caption{\textbf{(a)} Relative errors between DFs with different $\mathcal{N}$ and DF with $\mathcal{N}=65$ for $E_\text{max}=-0.03$. ($L=1$ and $F_\text{BC}=1$) \textbf{(b)} Relative error between the reference solution $F_\text{o}(E)$ and the optimal truncated-domain solution. ($L=1$, $F_\text{BC}=1$ and $\beta=\beta_{\text{o}}$.)}
	\label{fig:Stab_Deg_F_xmin_094_N65}
\end{figure}

Lastly, we propose an optimal truncated-domain solution that is compatible to the reference solution. For $E_\text{max}=-0.03$ in Figure \ref{N_vK_Del_c1c4_cF_trunc}, the order of $\mid1-c^{*}_{4}/ c^{*}_{4\text{o}}\mid$  reaches $10^{-9}$. This is the same order as the minimum value of $\mid1-c^{*}_{4}/ c^{*}_{4\text{o}}\mid$ computed against different $\beta$ in Section \ref{sec:stability_eigen}. Also, $\mid \varv_{I}(x=1)\mid\approx6.1\times10^{-12}$ for $\mathcal{N}=65$ is one of the least values among the values of $\mid\varv_{I}(x=1)\mid$ calculated for semi-stable truncated solutions. Hence we calculated the relative errors between the DF with $\mathcal{N}=65$ and DFs with different $\mathcal{N}$ for $E_\text{max}=-0.03$ (Figure \ref{fig:Stab_Deg_F_xmin_094_N65} \textbf{(a)}). We obtained the ideal tendency that as $\mathcal{N}$ increases the DFs gradually converge to the DF with $\mathcal{N}=65$. Hence the truncated-domain solution with $E_\text{max}=-0.03$, $\mathcal{N}=65$ and $\beta=\beta_\text{o}$ is the optimal truncated-domain solution in the present work. Figure \ref{fig:Stab_Deg_F_xmin_094_N65} \textbf{(b)} shows the relative error between DFs between the optimal solution and the reference solution at points that are less associated with the Gauss-Chebyshev nodes. The optimal truncated-domain solution validates the reference solution and the largest relative error between them is order of $10^{-9}$ at the prescribed points.

\section{Discussion: Modifying the mathematical formulation of the ss-OAFP system to reproduce the HS's solution}\label{sec:mod_change}

The present section discusses how to improve the whole-domain solution and reproduce the solution of \citep{Heggie_1988} using the spectral method to discuss the accuracy of the reference solution. The result of Section \ref{sec:stable_trunc_solns} shows that the stable truncated-domain solutions on $-0.35<E<-0.1$ are closer to the reference solution rather than the HS's solution that was obtained on almost the same domain. Our goal of the present section is to show that the discrepancy between our and the HS's solutions originates from the difference in mathematical formulation of the ss-OAFP system between the two works. In order to explain the discrepancy and also see the consistency of our result compared to the HS's solution, we discuss several classes of ss-OAFP solutions by modifying the regularized independent variables. We found that only modification of $\varv_{J}(x)$,  $\varv_{R}(x)$ and $\varv_{F}(x)$ provides significant change in ss-OAFP solution while that of the rest of the regularized function does not change the solution.  Sections \ref{sec:mod_vJ} and \ref{sec:mod_vR} detail the effect of modifying the regularization of $\varv_{J}(x)$ and $\varv_{R}(x)$ to improve the asymptotic behavior of $\varv_{J}(x)$ and to discuss the effect of discontinuity in $\varv_{R}(x)$. Based on the modification of $\varv_{R}(x)$, Section \ref{sec:repro_HS_soln} reproduces the HS's solution with limited degrees of Chebyshev polynomials and shows that the formulation can provide both the HS's and reference- solutions only by controlling $x_\text{min}$ (or $E_\text{max}$). For brevity, further detail discussion on reproducing the HS's solution is included into Appendix \ref{Appendix_Ref_HS} in which we discuss how to take off the limitation on the degrees of polynomial; this can be done by modifying the regularization of $\varv_{R}(x)$ and $\varv_{F}(x)$.

\subsection{Modification of function $\varv_{J}(x)$ and its asymptotic behavior}\label{sec:mod_vJ}

The present section shows that one can improve the reference solution by modifying the regularization of the regularized function $\varv_{J}(x)$. Even after all the independent variables of the ss-OAFP system are completely regularized (so that the variables reach certain constant values at the end points of the domain), the terms of the regularized ss-OAFP system significantly change at equation level. All the terms in the 4ODEs (equations \eqref{Eq.ss-4ODE-vF} and \eqref{Eq.ss-4ODE-vJ}) change like at least $\sim(0.5+0.5x)^{\beta}$ as $x\to-1$. The result of Section \ref{sec:ss_soln} shows the consequence of the large-scale gap in Figure \ref{fig:vJ_slope_index} in which the accuracy in the logarithmic derivative and higher order of $\sim(0.5+0.5x)^{\beta}$ in $\varv_{J}$ are divergent as $x\to-1$. One can weaken the divergence by modifying the regularization of $\varv_{J}(x)$ as follows
\begin{align}
\varv_{J}^{\text{(m)}}(x)\equiv(\varv_{J}(x)+1)\left(\frac{1+x}{2}\right)^{-b L}\label{Eq_Modi_vJ},
\end{align}
where $b$ is a real number. In equation \eqref{Eq.ss-4ODE-vF} the highest order of $(0.5+0.5x)^{\beta}$ is the term $\frac{\,\text{d}\varv_{F}(x)}{\,\text{d}x}\left(\frac{1+x}{2}\right)^{\beta}$, hence the function $\varv_{J}^{\text{(m)}}(x)$ can reduce it to $\frac{\,\text{d}\varv_{F}(x)}{\,\text{d}x}\left(\frac{1+x}{2}\right)^{\beta-b}$. We solved the ss-OAFP system again following the procedure of Section \ref{subsec:numeric_treat}, but this time for $\varv_{J}^{\text{(m)}}(x)$ (in place of $\varv_{J}$) and the rest of unchanged regularized independent variables on both truncated and whole domains. 

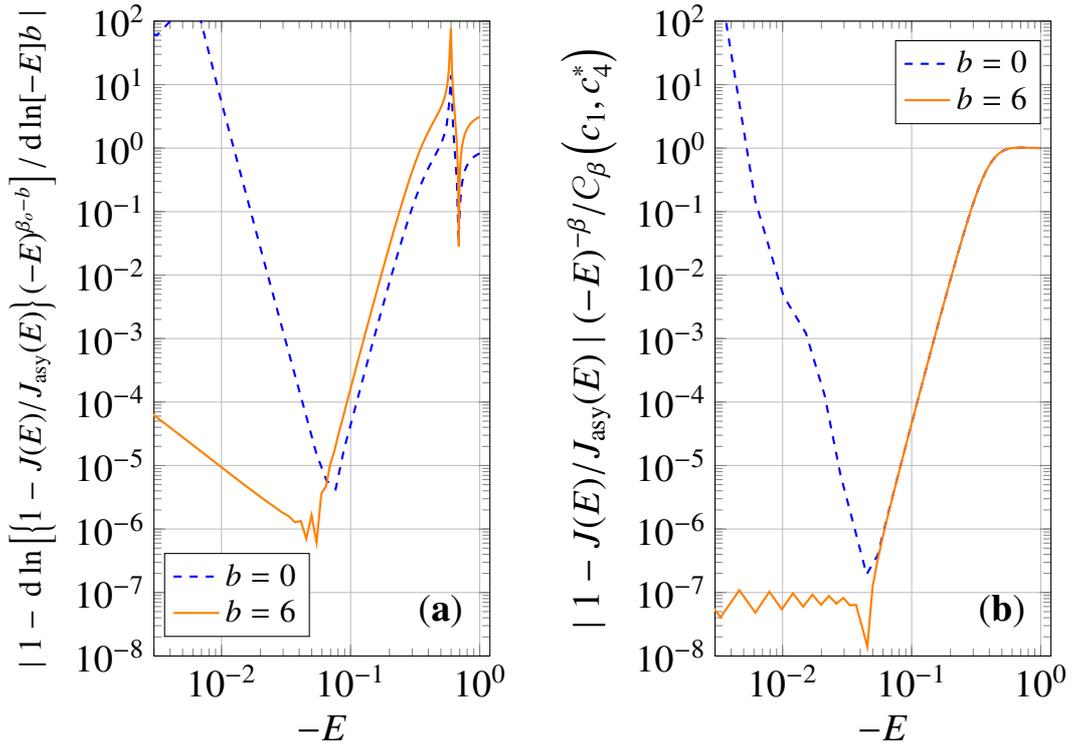
\begin{figure}[H]
	\centering
\tikzstyle{every node}=[font=\Large]
\begin{tikzpicture}[scale=1]
	\begin{loglogaxis}[width=6cm, height=10cm, grid=major,xlabel=\Large{$-E$},ylabel=\large{$\mid1-\,\text{d}\ln\left[\left\{1-J(E)/J_\text{asy}(E)\right\}(-E)^{\beta_{o}-b}\right]/\,\text{d}\ln [-E]b\mid$},xmin=3e-3,xmax=1.2e0,ymin=1e-8,ymax=1e2, legend pos=south west]
	\addplot [color = blue,mark=no,thick,dashed] table[x index=0, y index=1]{vJ_slope_index.txt}; 
    \addlegendentry{\large{$b=0$}}
\addplot [color = orange,mark=no,thick,solid] table[x index=0, y index=1]{vJ_slope_index_interm_a6_N150.txt}; 
 \addlegendentry{\large{$b=6$}}
\node[above,black] at (5e-1,2e-8) {\Large{$(\textbf{a})$}};
\end{loglogaxis}
	\end{tikzpicture}\hspace{0.4cm}
	\begin{tikzpicture}[scale=1]
	\begin{loglogaxis}[width=6cm, height=10cm, grid=major,xlabel=\Large{$-E$},ylabel=\Large{$\mid1-J(E)/J_\text{asy}(E)\mid(-E)^{-\beta}/\mathcal{C}_{\beta}\left(c_{1},c_{4}^{*}\right)$},xmin=3e-3,xmax=1.2e0,ymin=1e-8,ymax=1e2, legend pos=north east]
	\addplot [color = blue,mark=no,thick,dashed] table[x index=0, y index=2]{vJ_slope_index.txt}; 
    \addlegendentry{\large{$b=0$}}
\addplot [color = orange,mark=no,thick,solid] table[x index=0, y index=2]{vJ_slope_index_interm_a6_N150.txt}; 
 \addlegendentry{\large{$b=6$}}
\node[above,black] at (5e-1,2e-8) {\Large{$(\textbf{b})$}};
	\end{loglogaxis}
	\end{tikzpicture}
\caption{$(\textbf{a})$ Logarithmic derivative of $\mid1-J(E)/J_\text{asy}(E)\mid$ with respect to $E$ and $(\textbf{b})$Characteristics of $\mid1-J(E)/J_\text{asy}(E)\mid (-E)^{-\beta}$ on the whole domain for $b=6$.  ($\mathcal{N}=150$, $F_\text{BC}=1$, $L=1$). The solution with $b=0$ corresponds to the reference solution.}
\label{fig:vJ_slope_index_interm_a6_N150}
\end{figure}

\begin{figure}[H]
	\centering
	\tikzstyle{every node}=[font=\Large]
	\begin{tikzpicture}[scale=1]
	\begin{semilogyaxis}[width=9cm, height=6cm, grid=major,xlabel=\Large{$-E$},title=\Large{$\mid1-F(E)^{(b=6)}_\text{trunc}/F_\text{o}(E)\mid$},xmin=0.01, xmax=0.95,ymin=1e-10,ymax=1e-7]
	\addplot [color = bred,mark=square,thick,solid] table[x index=0, y index=1]{compari_N70_N55_xm095_interm.txt}; 
	\end{semilogyaxis}
	\end{tikzpicture}
	\caption{ Relative error between DFs obtained from the reference solution and from the truncated-domain solution with $x_\text{min}=-0.95$, $b=6$ and $\mathcal{N}=55$.  ($F_\text{BC}=1$ and $L=1$ ).}
	\label{fig:diff_interm_a6_N150}
\end{figure}
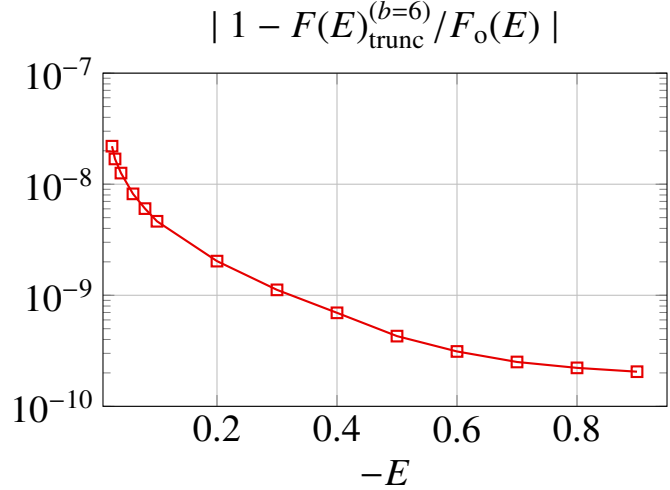

We found solutions for $1\leq b\leq 6$ on both whole and truncated domains and $b=6$ provided the best result in accuracy. The results are quite well; the modification of $\varv_{J}$ improved the asymptotic behaviors of the logarithmic derivative of $\varv_{J}(x)$ and the approximation $\mathcal{C}_{\beta}\left(c_{1},c_{4}^{*}\right)$ as $x\to-1$ on the whole domain (Figure \ref{fig:vJ_slope_index_interm_a6_N150}). Also, the eigenvalues that were obtained for $b=6$ on the truncated and whole domains are almost identical to the reference eigenvalues (Table \ref{table:interm_results}). The relative error between stellar DFs obtained from the reference solution and the truncated-domain solution is at most order of $\sim 3\times10^{-8}$ for $b=6$ at $E_\text{max}=-0.025$  (Figure \ref{fig:diff_interm_a6_N150}). This result infers that one can obtain a \emph{suitable} solution that is less divergent in higher order of $\sim(0.5+0.5x)^{\beta}$ as $x\to-1$ by correctly regularizing function $J(E)$.

\begin{table*}\centering\large
	\ra{1.3}
	\scalebox{0.8}{
		\begin{tabular}{@{}lclcccc@{}}
			\toprule
			domain&$\mathcal{N}$&optimal \large{$\beta$}  & $\mid1-c_{1}/c_{1\text{o}} \mid$ &  $\mid1-c^{*}_{4}/c^{*}_{4\text{o}} \mid$  & $\varv_{I}(x=1)$ &condition number  \\ 
			\midrule
                whole & $150$ & $\beta_\text{o}$ & $2.4\times10^{-11}$ & $7.7\times10^{-8}$& $1.6\times10^{-9}$ & $8.8\times10^{7}$ \\
    truncated  & $55$ & $\beta_\text{o}+3.0\times10^{-12}$ & $1.6\times10^{-12}$ & $2.6\times10^{-8}$& $3.0\times10^{-12}$  & $1.2\times10^{8}$ \\
                ($x_\text{min}=-0.95$) &&&&&&\\
			\bottomrule
	\end{tabular}}
	\caption{ Numerical results for the integration of the ss-OAFP system for $b=6$. ($L=1$ and $F_\text{BC}=1$). The eigenvalues are compared to the reference eigenvalues.}
	\label{table:interm_results}
\end{table*}

\subsection{Modification of function $\varv_{R}(x)$ and its discontinuous behavior}\label{sec:mod_vR}

The asymptotic behavior of $\varv_{R}(x)$ as $x\to-1$ is important to see the effect of discontinuity in the ss-OAFP solutions and the discontinuity clearly appears in the solutions that are obtained without the assumption that $\varv_{R}(x)$ is regular at $x=-1$ (this assumption is made implicitly in Section \ref{sec:math_formul} by regularizing $R$ with $(1-E)$.). We show this by modifying the regularization of $\varv_{R}(x)$ as follows
\begin{align}
    \varv_{R}^\text{(m)}(x)=[\varv_{R}(x)]^{2}\left(\frac{1-x}{2}\right).
\end{align}
The square of $\varv_{R}(x)$ can avoid the endpoint singularity at the branch point $x=1$. Again we solved the ss-OAFP system but this time for $\varv_{R}^\text{(m)}(x)$ and for the rest of independent variables (without including  $\varv_{J}^\text{(m)}(x)$), following the procedure of Section \ref{subsec:numeric_treat}. 

\begin{figure}[H]
	\centering
	\tikzstyle{every node}=[font=\large]
		\begin{tikzpicture}
	\begin{loglogaxis}[width=7cm, height=8cm, grid=major,xlabel=\Large{$-E$},xmin=1e-2,xmax=1.2e0,ymin=1e-18,ymax=6e2, legend pos=south east]
	\addplot [color = blue,mark=no,thick,densely dashed] table[x index=0, y index=1]{E_Diff_vRasymp_N540.txt};
	\addlegendentry{\large{$\mid1-\varv_{R\text{o}}/\varv_{R\text{o}}(-1)\mid$}} 
	\addplot [color = orange,mark=no,thick,solid] table[x index=0, y index=1]{Ec_Diff_vRasymp_N540_trunc.txt};
	\addlegendentry{\large{$\mid1-\varv_{R}^\text{(m)}/\varv_{R}^\text{(m)}(-0.9)\mid$}} 
       \node[above,black] at (8e-1,8e-1) {\Large{$(\textbf{a})$}};
	\end{loglogaxis}
	\end{tikzpicture}
\hspace{0.4cm}
\begin{tikzpicture}
	\begin{loglogaxis}[width=7cm, height=8cm, grid=major,xlabel=\Large{$-E$},xmin=1e-2,xmax=1.2e0,ymin=1e-18,ymax=6e2, legend pos=south east]
	\addplot [color = blue,mark=no,thick,densely dashed] table[x index=0, y index=1]{E_Diff_vFasymp_N540.txt};
	\addlegendentry{\large{$\mid1-\varv_{F\text{o}}/\varv_{F\text{o}}(-1)\mid$}} 
	\addplot [color = orange,mark=no,thick,solid] table[x index=0, y index=1]{Ec_Diff_vFasymp_N540_trunc.txt};
	\addlegendentry{\large{$\mid1-\varv_{F}/\varv_{F}(-0.9)\mid$}} 
 \node[above,black] at (8e-1,8e-1) {\Large{$(\textbf{b})$}};
	\end{loglogaxis}
	\end{tikzpicture}
	\caption{ $(\textbf{a})$ Comparison of asymptotic behaviors between $\varv_{R}^\text{(m)}$ and $\varv_{R\text{o}}$. On the graph, the relative errors between them and the corresponding end values $\varv_{R}^\text{(m)}(x=-0.9)$ and $\varv_{R\text{o}}(x=-1)$ are shown. $(\textbf{b})$ Comparison of asymptotic behaviors between $\varv_{F\text{o}}$ and $\varv_{F}$. On the graph, the relative errors between them and the corresponding end values $\varv_{F}(x=-0.9)$ and $\varv_{F\text{o}}(x=-1)$ are shown. }
	\label{fig:asymp_vRm_vF}
\end{figure}
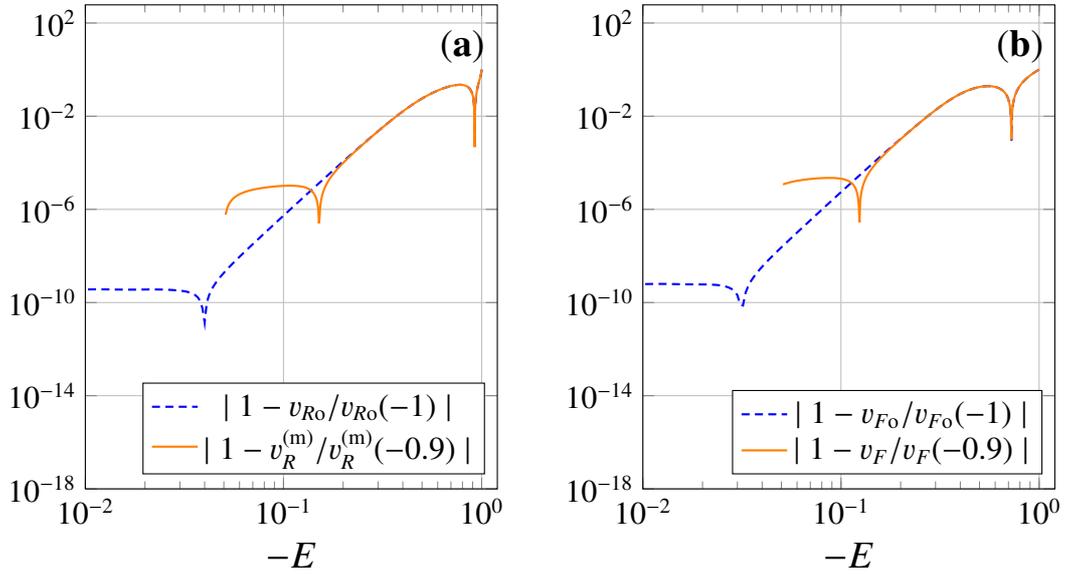

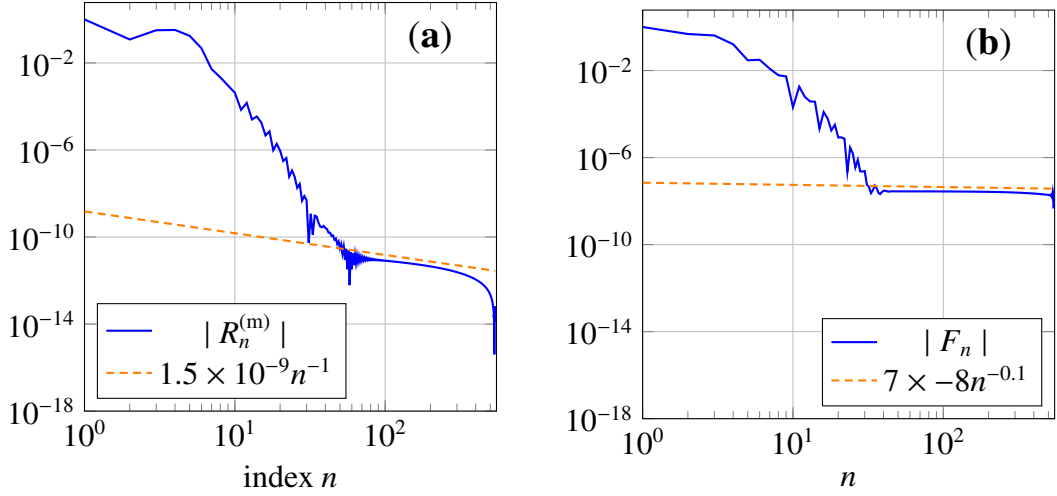
\begin{figure}
	\centering
	\tikzstyle{every node}=[font=\large]
	\begin{tikzpicture}
	\begin{loglogaxis}[width=7cm, height=7cm, grid=major,xlabel=\large{index $n$},xmin=1e-0,xmax=5.5e2,ymin=1e-18,ymax=6e-0, legend pos=south west]
	\addplot [color = blue,mark=no,thick,solid] table[x index=0, y index=1]{n_cR540_guide.txt};
	\addlegendentry{\large{$\mid R^\text{(m)}_{n}\mid$}} 
	\addplot [color = orange,mark=no,thick,densely dashed] table[x index=0, y index=2]{n_cR540_guide.txt};
	\addlegendentry{\large{$ 1.5\times10^{-9}n^{-1}$}} 
   \node[above,black] at (2e2,1e-2) {\Large{$(\textbf{a})$}};
	\end{loglogaxis}
	\end{tikzpicture}
\hspace{0.6cm}
\begin{tikzpicture}
	\begin{loglogaxis}[width=7cm, height=7cm, grid=major,xlabel=\large{$n$},xmin=1e-0,xmax=5.5e2,ymin=1e-18,ymax=6e-0, legend pos=south east]
	\addplot [color = blue,mark=no,thick,solid] table[x index=0, y index=1]{n_cF540_guide.txt};
	\addlegendentry{\large{$\mid F_{n}\mid$}} 
	\addplot [color = orange,mark=no,thick,densely dashed] table[x index=0, y index=2]{n_cF540_guide.txt};
	\addlegendentry{\large{$7\times{-8}n^{-0.1}$}} 
   \node[above,black] at (2e2,1e-2) {\Large{$(\textbf{b})$}};
	\end{loglogaxis}
	\end{tikzpicture}
	\caption{$(\textbf{a})$ Absolute value of Chebyshev coefficients for $\varv_{R}^\text{(m)}$. $(\textbf{b})$ Absolute value of the Chebyshev coefficients for $\varv_{F}$.  Dashed guidelines are also depicted for measure of slow decay.}
	\label{fig:coeff_vRm_vF}
\end{figure}

We found spectral solutions with high degrees (e.g. $\mathcal{N}=540$) for $E_\text{max}=-0.05$. The computed functions $\varv_{F}(x)$ and $\varv_{R}^\text{(m)}(x)$ well explain the feature of discontinuity in the ss-OAFP system. Figures \ref{fig:Rel_err_vRm_vF} shows the maximum relative errors of $\varv_{R}^\text{(m)}(x)$ and $\varv_{F}$ are order of $2\times10^{-5}$ and $5\times10^{-5}$ from their reference solutions $\varv_{R\text{o}}(x)$ and $\varv_{F\text{o}}(x)$. The order of errors well reflects the relative error of the solutions from their asymptotic approximations (Figure \ref{fig:asymp_vRm_vF}). Figure \ref{fig:coeff_vRm_vF} depicts the Chebyshev coefficients of $\varv_{F}(x)$ and $\varv_{R}^\text{(m)}(x)$. A slow decay appears in both the coefficients for $\varv_{F}(x)$ and $\varv_{R}^\text{(m)}(x)$.  The former apparently flattens (more exactly, decays like $7\times10^{-8}\,n^{-0.1}$) and the latter decays like $1.5\times10^{-9}\,n^{-1}$. It is not easy for one to find the cause of the flattening and slow decay due to the mathematically complex structure of the ss-OAFP system. Yet, the asymptotic behavior $a_{n}\sim1/n$ ($n\to\infty$) has approximately the same decay rate as Chebyshev coefficients for discontinuous functions \citep{Boyd_2001,Xiang_2013}. Hence, Appendices \ref{sec:Fixed_cR} and \ref{sec:Fixed_vD} show the numerical results that we obtained by integrating the $Q$ integral for a fixed discontinuous $\varv_{R}$ and also by solving the Poisson equation for a fixed discontinuous $\varv_{D}$. The former provides a slow decay of Chebyshev coefficients like $1/n$ or much slower  (Figure \ref{fig:n_CoeffDiscon})  and the latter a flattening of Chebyshev coefficients for large $n$ (Figure \ref{fig:n_cR_cRm_Poiss_dc}). These unique behaviors occur only when the point of discontinuity is very close to either of endpoints on the domain (See Appendices \ref{sec:Fixed_cR} and \ref{sec:Fixed_vD} for detail).

\begin{figure}[H]
	\centering
	\tikzstyle{every node}=[font=\large]
	\begin{tikzpicture}
	\begin{loglogaxis}[width=9cm, height=7cm, grid=major,xlabel=\Large{$-E$},ylabel=\large{$\mid 1- \varv_{R\text{o}}/\varv^\text{(m)}_{R}\mid$},xmin=3e-2,xmax=1.2e0,ymin=1e-8,ymax=6e-5, legend pos=south west]
	\addplot [color = red,mark=o,thick,solid] table[x index=0, y index=1]{Ec_Diff_vR_N540_N70whole.txt}; 
	\addlegendentry{\large{$\mid 1- \varv_{R\text{o}}/\varv^\text{(m)}_{R}\mid$}};
	\addplot [color = blue,mark=square,thick,solid] table[x index=0, y index=1]{Ec_Diff_vF_N540_N70whole.txt}; 
	\addlegendentry{\large{$\mid 1- \varv_{F\text{o}}/\varv_{F}\mid$}};
	\end{loglogaxis}
	\end{tikzpicture}
	\caption{ Relative error of  $\varv_{R}^\text{(m)}$ and $\varv_{F}$ from their reference solutions $\varv_{R\text{o}}$ and $\varv_{F\text{o}}$. ($E_\text{max}=-0.05$, $\mathcal{N}=540$, $L=1$ and $F_\text{BC}=1$.)}
	\label{fig:Rel_err_vRm_vF}
\end{figure}
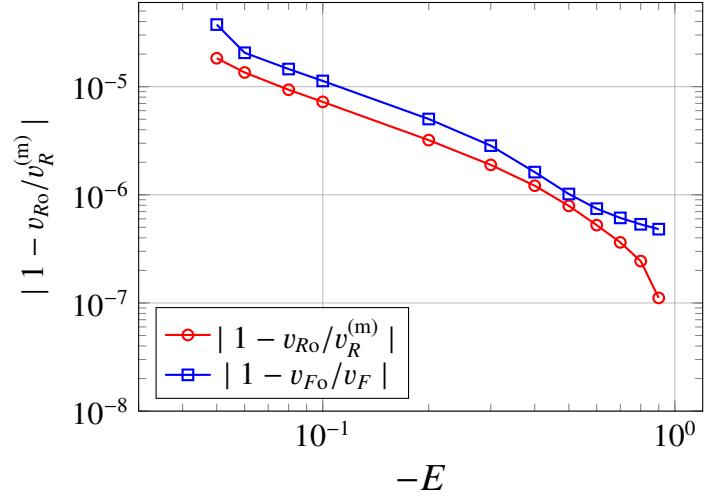

\subsection{Reproducing the HS's solution and eigenvalues with limited degrees}\label{sec:repro_HS_soln}

The present section reproduces the HS's solution with low degrees $(\mathcal{N}<20)$ of polynomials by modifying the regularization of $\varv_{R}$. According to \citep{Heggie_1988}, the numerical values of their solutions are ``thought to be accurate about three significant figures''. On one hand, they described the value of $\chi_\text{esc}$ as ``$\sim13.85$'' and reported three significant figures for the eigenvalues $c_1$, $c_2$, $c_3$ and $c_4$. Due to these ambiguous expressions and lack of detail description for their error analysis in \citep{Heggie_1988}, the present section aims to reproduce \emph{at least two significant figures} of the HS's solutions and eigenvalues. We show the results obtained by reformulating the ss-OAFP system based on $\varv_{R}^{(m)}$ (explained in Section \ref{sec:mod_vR}) and by using the numerical procedure of Section \ref{subsec:numeric_treat}. However, the results reproduced only either of the HS's solution and eigenvalues for a certain $\mathcal{N}$, not both of them. To understand the reproduced solutions, the present section examines two kinds of solutions. In section \ref{sec:rep_HSsoln}, the first kind of solution reproduces the HS's solution but the eigenvalues are the same as only two significant figures of the HS's eigenvalues. In section \ref{sec:rep_HSeigen} the second kind reproduces the HS's eigenvalues but the solution is the same as only two significant figures of the HS's solution. For comparison, the HS's solution is labeled hereafter by subscript `HS', such as $F_\text{HS}$ for stellar DF.

\subsubsection{Reproducing the same solution as HS's work}\label{sec:rep_HSsoln}

We found spectral solutions with low degrees ($\mathcal{N}=13\sim19$) that can provide the same numerical values of $\ln[F(E)]$ as \citep{Heggie_1988}'s work, however the obtained eigenvalues are different from the HS's eigenvalues (Table \ref{table:Repro_HS}). Only two significant figures of the eigenvalues are stable against $E_\text{max}$; $\beta=8.2$, $c_{1}=9.1$ and $c_{4}=3.5$ and three significant figures of the physical parameters; $\chi_\text{esc}=13.8$ and $\alpha=2.23$.  The measures of accuracy, $\text{min}(\{F_{n}\})$ and $\varv_{I}(x=1)$, hold approximately the same order for different $E_\text{max}$ and $\mathcal{N}$, that is, $\text{min}(\{F_{n}\})\approx 10^{-4}$ and $\mid \varv_{I}(x=1)\mid\approx10^{-4}\sim10^{-5}$.

\begin{figure}
	\centering
	\tikzstyle{every node}=[font=\Large]
	\begin{tikzpicture}
	\begin{semilogyaxis}[legend columns=2,width=10cm, height=7cm, grid=major,xlabel=\Large{$-E$},ylabel=\large{$\mid1-\ln[F]/\ln[F_{HS}]\mid$},xmin=0.3,xmax=1,ymin=1e-5,ymax=1e-0, legend pos=north west]
	\addplot [color = black,mark=no,thick,solid] table[x index=0, y index=1]{Ef_Dellogf_Del_N11_N13_N15_N17.txt};
	\addlegendentry{\large{$\mid 0.005/\ln[F_\text{HS}]\mid$}} 
	\addplot [color = green,mark=o,thick,solid] table[x index=0, y index=2]{Ef_Dellogf_Del_N11_N13_N15_N17.txt};
	\addlegendentry{\large{$\mathcal{N}=11$}} 
	\addplot [color = blue,mark=square,thick,solid] table[x index=0, y index=3]{Ef_Dellogf_Del_N11_N13_N15_N17.txt};
	\addlegendentry{\large{$\mathcal{N}=13$}} 
		\addplot [color = orange,mark=triangle,thick,solid] table[x index=0, y index=4]{Ef_Dellogf_Del_N11_N13_N15_N17.txt};
	\addlegendentry{\large{$\mathcal{N}=15$}} 
	\addplot [color = red,mark=asterisk,thick,solid] table[x index=0, y index=5]{Ef_Dellogf_Del_N11_N13_N15_N17.txt};
	\addlegendentry{\large{$\mathcal{N}=17$}} 
	\end{semilogyaxis}
	\end{tikzpicture}
	\caption{Degree-$\mathcal{N}$-dependence of relative error of $\ln[F]$ obtained from the spectral and HS's solutions for $E_\text{max}=-0.275$.}
	\label{fig:Ef_Dellogf_Del_N11_N13_N15_N17}
\end{figure}
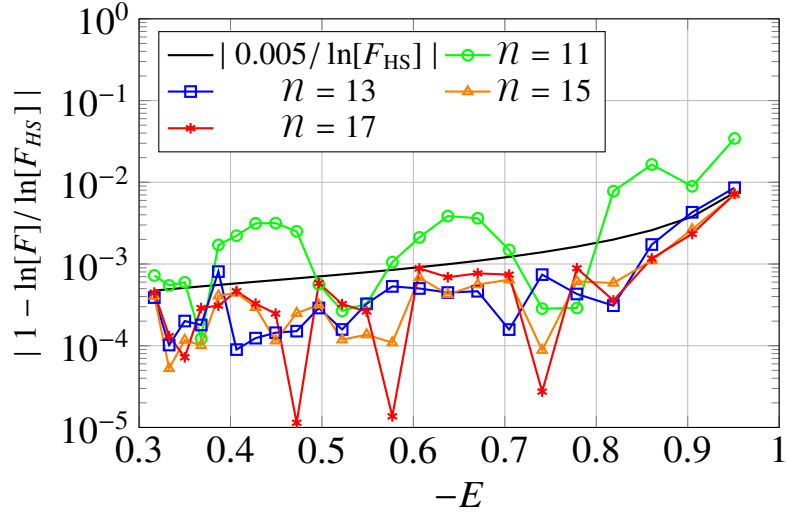

Available degrees $\mathcal{N}$ that can reproduce the HS's solution are limited. Figure \ref{fig:Ef_Dellogf_Del_N11_N13_N15_N17} shows the $\mathcal{N}$-dependence of relative error between the calculated DF and HS's DF for $E_\text{max}=-0.275$. Since the HS's work reported their solution rounded to the second decimal places, we also show the values of $0.005/\ln[F_{HS}]$ in the figure as reference. The spectral solution reproduced the HS's solution for $\mathcal{N}=15$ and $\mathcal{N}=17$; in Figure \ref{fig:Ef_Dellogf_Del_N11_N13_N15_N17} all the relative errors are below $0.005/\ln[F_{HS}]$.  However, beyond $\mathcal{N}=17$, our DF deviates from the HS's DF.

\begin{table*}\centering
	\ra{1.3}
	\scalebox{0.8}{
	\begin{tabular}{@{}c|ccccccc@{}}\toprule
		$E_\text{max}$&$\mathcal{N}$ & $\beta$ & $c_{1}(\times10^{-4})$ & $c_{4}(\times10^{-2})$ & $\chi_\text{esc}$ & $\text{min}(\{F_{n}\})$ & $\mid\varv_{I}(x=1)\mid$ \\ 
		\midrule
		$-0.300$ & $19$ & $8.17370$ & $9.101$ & $3.449$ & $13.838$ & $3.5\times10^{-4}$  & $4.0\times10^{-4}$ \\
		$-0.290$ & $17$ & $8.17050$ & $9.110$ & $3.451$ & $13.837$ & $3.2\times10^{-4}$  & $3.3\times10^{-4}$ \\
		$-0.275$ & $17$ & $8.17316$ & $9.103$ & $3.495$ & $13.837$ & $3.4\times10^{-5}$  & $3.2\times10^{-5}$ \\
		$-0.260$ & $15$ & $8.16900$ & $9.112$ & $3.497$ & $13.835$ & $1.3\times10^{-4}$  & $2.6\times10^{-5}$ \\
		$-0.250$ & $15$ & $8.17188$ & $9.105$ & $3.526$ & $13.836$ & $3.5\times10^{-4}$  & $1.7\times10^{-4}$  \\
		$-0.240$ & $13$ & $8.16110$ & $9.137$ & $3.485$ & $13.832$ & $1.1\times10^{-4}$  & $7.2\times10^{-6}$ \\
		$-0.230$ & $13$ & $8.16060$ & $9.130$ & $3.497$ & $13.832$ & $2.8\times10^{-4}$  & $9.0\times10^{-5}$ \\
		$-0.220$ & $13$ & $8.16020$ & $9.123$ & $3.514$ & $13.832$ & $4.7\times10^{-4}$  & $1.8\times10^{-4}$ \\
		$-0.210$ & $13$ & $8.15800$ & $9.124$ & $3.489$ & $13.835$ & $3.8\times10^{-4}$  & $1.3\times10^{-4}$ \\
		$-0.200$ & $13$ & $8.15850$ & $9.122$ & $3.529$ & $13.834$ & $7.0\times10^{-4}$  & $3.0\times10^{-4}$ \\
		\bottomrule
	\end{tabular}
}
	\caption{ Eigenvalues obtained when the spectral solution reproduced the HS's solution. \cite{Heggie_1988} reported the numerical values of their solution on $-1\lessapprox E \leq -0.317$. They mentioned that their Newton iteration method worked up to $E_\text{max}\approx-0.223$ and it could work beyond -0.223.}
	\label{table:Repro_HS}
\end{table*}

\subsubsection{Finding solution whose eigenvalues are the same as the HS's eigenvalue}\label{sec:rep_HSeigen}

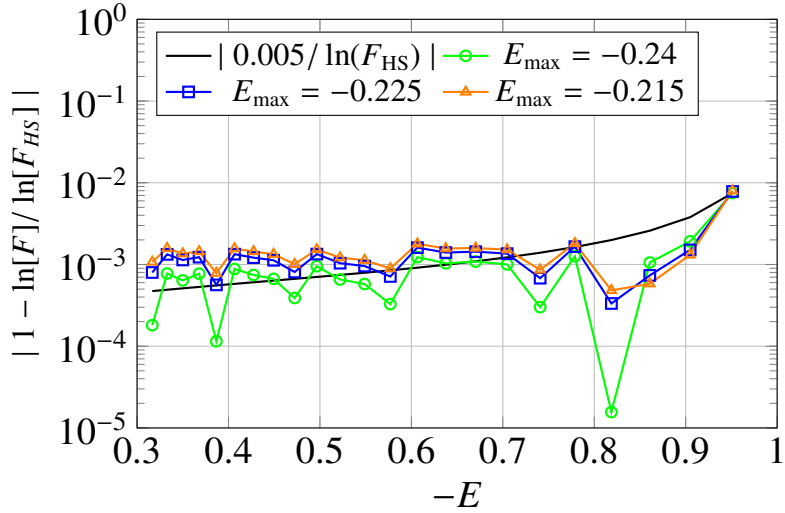
\begin{figure}
	\centering
	\tikzstyle{every node}=[font=\Large]
	\begin{tikzpicture}
	\begin{semilogyaxis}[legend columns=2,width=10cm, height=7cm, grid=major,xlabel=\Large{$-E$},ylabel=\large{$\mid1-\ln[F]/\ln[F_{HS}]\mid$},xmin=0.3,xmax=1,ymin=1e-5,ymax=1e-0, legend pos=north west]
	\addplot [color = black,mark=no,thick,solid] table[x index=0, y index=1]{Ef_Dellogf_Del_xmin_052_055_057_060.txt};
	\addlegendentry{\large{$\mid 0.005/\ln(F_\text{HS})\mid$}} 
	\addplot [color = green,mark=o,thick,solid] table[x index=0, y index=2]{Ef_Dellogf_Del_xmin_052_055_057_060.txt};
	\addlegendentry{\large{$E_\text{max}=-0.24$}} 
	\addplot [color = blue,mark=square,thick,solid] table[x index=0, y index=3]{Ef_Dellogf_Del_xmin_052_055_057_060.txt};
	\addlegendentry{\large{$E_\text{max}=-0.225$}} 
	\addplot [color = orange,mark=triangle,thick,solid] table[x index=0, y index=4]{Ef_Dellogf_Del_xmin_052_055_057_060.txt};
	\addlegendentry{\large{$E_\text{max}=-0.215$}} 
	\end{semilogyaxis}
	\end{tikzpicture}
	\caption{ Relative error between DFs obtained from the spectral solution with $\mathcal{N}=15$ and HS's solution for different truncated domains.}
	\label{fig:Ef_Dellogf_Del_xmin_052_055_057_060}
\end{figure}

We found spectral solutions whose eigenvalues are the same as the HS's eigenvalues ($c_{1}=9.10$ and $c_{4}=3.52$) with $\mathcal{N}=15$ near $E_\text{max}=-0.225$. For the solutions, Table \ref{table:Repro_HS_good_eigen} shows $\beta$, $\chi_\text{esc}$ and measures of accuracy ($\text{min}(\{F_{n}\})$ and $\mid\varv_{I}(x=1)\mid$). The measures of accuracy are approximately the same order as the reproduced HS's solution (shown in Table \ref{table:Repro_HS}); $\text{min}(\{F_{n}\})\sim\mid \varv_{I}(x=1)\mid \sim10^{-4}$ . Interestingly, for $E_\text{max}=-0.225$, $\chi_\text{esc}$ reaches the HS' value ($=13.85$). The numerical values of $\ln[F]$ reproduced $2\sim4$ significant figures of $\ln[F_\text{HS}]$.  The relative error between $\ln[F]$  and $\ln[F_\text{HS}]$ is at most order of $1\times10^{-3}$ for $E\geq-0.9$ (Figure \ref{fig:Ef_Dellogf_Del_xmin_052_055_057_060}). This result would infer that the spectral solution reproduced ``about three significant figures'' of the HS's solution with the same eigenvalues.  

\begin{table*}\centering
	\ra{1.3}
	\scalebox{0.8}{
	\begin{tabular}{@{}c|cccc@{}}\toprule
		$E_\text{max}$ & $\beta$ & $\chi_\text{esc}$ & $\text{min}(\{F_{n}\})$ & $\varv_{I}(x=1)$ \\ 
		\midrule
		$-0.240$ & $8.17310$ & $13.840$ & $4.1\times10^{-4}$  & $2.1\times10^{-4}$ \\
		$-0.225$ & $8.17460$ & $13.845$ & $4.5\times10^{-4}$  & $2.3\times10^{-4}$ \\
		$-0.215$ & $8.17536$ & $13.847$ & $4.7\times10^{-4}$  & $2.4\times10^{-4}$ \\
		$-0.200$ & $8.17560$ & $13.850$ & $4.6\times10^{-4}$  & $2.3\times10^{-4}$ \\
		\bottomrule
	\end{tabular}
    }
	\caption{Eigenvalues of the reproduced HS's solution with eigenvalues $c_{1}=9.10\times10^{-4}$ and $c_{4}=3.52\times10^{-2}$ for $\mathcal{N}=15$. \citep{Heggie_1988} reported the numerical values of their solution on $-1\lessapprox E \leq -0.317$. They mentioned that their Newton iteration method worked up to $E_\text{max}\approx-0.223$ and it could work beyond -0.223.}
	\label{table:Repro_HS_good_eigen}
\end{table*}

\subsubsection{Successfully reproducing HS's solution and accuracy of the reference solution}\label{sec:succes_HS}

We briefly explain the condition to obtain the both reference- and HS's solution on truncated domains based on only a single mathematical formulation of the ss-OAFP model. For brevity the detail discussion is made in Appendix \ref{Appendix_Ref_HS} and we explain only the results. The most important result in Appendix \ref{Appendix_Ref_HS} is that one can find the HS's solution if the absolute value of the coefficients $\{I_{n}\}$ for $\varv_{I}(x)$ reach approximately $10^{-4}\sim10^{-5}$ for $E_\text{max}\approx-0.25$ and also the reference solution if the coefficients reach order of  $10^{-6}\sim10^{-7}$ for $E_\text{max}\approx-0.05$ (Figure \ref{fig:n_cI_HS_Polin_dc3}). We believe the reason why we could not find out the condition in the present section is that the decay rate of the Chebyshev coefficients is too rapid and provided only limited degrees to obtain the HS's solution for  $E_\text{max}\approx-0.25$. Hence, for numerical calculation in the Appendix \ref{Appendix_Ref_HS}, we intentionally included the effect of the non-analytic and non-regular properties into dependent variables by modifying the regularization of $\varv_R$ (with a discontinuity) and $\varv_F$ (with a logarithmic dependence).   

We believe our numerical accuracy of the reference solution is at least four significant figures based on the detail analyses that we carried out for the various formulations in the present section, Sections \ref{sec:ss_soln} and \ref{sec:domain_trunc} and Appendixes \ref{sec:stability} and \ref{Appendix_Ref_HS}. What we made the most efforts in the majority of the present work is to find a truncated-domain solution which is close to HS's solution for small $E_\text{max}$ but still close to the reference solution for large $E_\text{max}$  based on only a single formulation. Among the variant formulations,  the $\varv_{R}^{(m)}$-formulation of the present section not only reproduced both the HS's and reference- solutions but also provided the smallest relative error $(\sim4\times10^{-5})$ from the reference solution (See Figure \ref{fig:Rel_err_vRm_vF}). This error corresponds with the relative error of $c_4$ from the reference eigenvalue. Hence, Table \ref{table:Eigenvalues} lists four significant figures for $c_4$ and five for the rest of eigenvalues (since $c_1$, $c_2$ and $c_3$ were more stable against numerical parameters than $c_4$ for any formulations in our work.)

\section{Conclusion}\label{sec:conclusion}

The self-similar OAFP equation to model core-collapsing star clusters is important in the sense that it provides a conceptual understanding of the late stage of the relaxation evolution of isotropic-spherical dense star clusters and useful physical parameters. The equation, however, has never been solved with an agreeable accuracy and existing solutions were domain-truncated, whose domain is $-0.2<E<1$. Accordingly, the detail physical feature and application of the model have never been discussed; those are the topics we discuss in our follow-up papers. This work is the first paper of our works on the ss-OAFP equation focusing on finding an accurate solution of the equation using a Gauss-Chebyshev pseudo-spectral method.
  
We first applied the pseudo-spectral method to the ss-OAFP equation on the whole domain ($-1<E<0$). Section \ref{sec:ss_soln} provided the whole-domain solution whose degree of Chebyshev polynomials is $70$. The minimum of the normalized Chebyshev coefficients reaches order of $10^{-12}$ for all the regularized independent variables in the equation. We obtained the corresponding eigenvalues more consistently compared to existing works as follows $c_{1}=9.0925\times10^{-4}$, $c_{2}=1.1118\times10^{-4}$, $c_{3}=7.1975\times10^{-2}$ and $c_{4}=3.303\times10^{-2}$. The eigenvalues result in the following physical parameters; the power-law index $\alpha$ is $2.230$, the collapse rate $\xi=3.64\times10^{-3}$ and the scaled escape energy $\chi_\text{esc}=13.89$. Also, we provided a semi-analytical form of the whole-domain solution whose degree of polynomials is at most 18. 

Since the whole-domain solution depends on degree $\mathcal{N}$ of polynomials in an undesirable way, in Section \ref{sec:domain_trunc} we aimed at finding truncated-domain solutions whose accuracy improves with increasing degree $\mathcal{N}$. We obtained truncated-domain solutions whose numbers of significant figures are up to 8 for $-0.08\leq E_\text{max}<-0.04$ and the degrees of the polynomials are only $N\approx25 \sim55$. To find an optimal truncated-domain solution that is close to the whole-domain solution, we obtained the truncated-domain solutions with $\beta=\beta_\text{o}$ for $-0.05\leq E_\text{max}<-0.02$ and those solutions are stable against up to specific degrees of polynomials. At point $E_\text{max}=-0.03$, the truncated-domain solution has the same order of accuracy in $c_{4}^{*}$ as the whole-domain solution. Hence, we compared the reference solution and the truncated-domain solution with $N=65$ at $E_\text{min}=-0.03$; the relative error between the solutions are approximately $10^{-9}$ at certain energy-domain points. 

Also, in Section \ref{sec:mod_change} by modifying the regularization of independent variable $\varv_{J}$, we improved the divergent asymptotic behavior as $E\to0$ in differentiation of the whole-domain and truncated-domain solutions. Also, the new regularization of $\varv_{R}$ and $\varv_{F}$ helped us to reproduce the \citep{Heggie_1988}'s solution around at $E_\text{max}=-0.225$ while it still can provide the whole domain solution around at $E_\text{max}=-0.05$ with accuracy of order of $10^{-5}$. We consider that one can find the HS's solution as a result of low accuracy with small $E_\text{max}$ and that the actual number of significant figures of the HS's solution is one. 

We will discuss the physical properties and application of the ss-OAFP model in the follow-up papers; the second paper \citep{Ito_2020_2} is for thermodynamic property of the model focusing the negative heat capacity in the core and the third \citep{Ito_2020_3} for application of the model to globular clusters in Milky Way. We are also planning to extend our numerical code to post-core-collapse solutions in future work. The present model can be meaningful only to the clusters that (i) have already reached in complete-core-collapsed state (if possible) and (ii) are  undergoing core collapse as an approximation of  more exact models (time-dependent OAFP model and $N$-body simulations). Our numerical code can extend to post-core-collapse models such as the ss-OAFP model \citep{Heggie_1988} and a FP model that follows the approach of self-similar conductive gaseous model \citep{Goodman_1984}.

\section*{Acknowledgements}
The present work is partial fulfillment of the degree of Philosophy at CUNY graduate center. Spectral method and part of regularization for independent variables were encouraged to use by my advisor Carlo Lancellotti. 
\vspace{1cm}
\appendix
\begin{appendices}
\renewcommand{\theequation}{\Alph{section}.\arabic{equation}}

\section{The asymptotic approximations of function in the 4ODEs}

We detail the asymptotic approximations of the regularized functions $\varv_{I}(x)$ and $\varv_{J}(x)$ (Appendix \ref{sec:asymp_vI_vJ}) and $\varv_{J}(x)+1$ (Appendix \ref{sec:eqn_vJp1}).

\subsection{The asymptotic approximation of the functions $\varv_{I}(x)$ and $\varv_{J}(x)$}\label{sec:asymp_vI_vJ}

The function $\varv_{I}(x)$ is important to determine the eigenvalue $\beta$ and the asymptotic approximation of $\varv_I(x)$ is related to the boundary condition of the ss-OAFP system. Equation \eqref{Eq.ss-4ODE-vI} for $\varv_{I}(x)$ does not include $c_1$ at first-order differential equation level and even the asymptotic approximations in first-order differentiation do not include $c_{1}$ around endpoints
\begin{align}
&\varv_{I}(x\to-1)=\frac{4}{2\beta-7}\left(\frac{x+1}{2}\right)^{L}+\cdots,\label{Eq.vI_asymp_1mx}\\
&\varv_{I}(x\to1)=\frac{L}{4}\left[1-\left(\frac{1+x}{2}\right)^{L}\right]+\cdots.\label{Eq.vI_asymp_1px}
\end{align}
On one hand, the eigenvalue $c_1$ is associated with $\varv_{J}(x)$ since equation \eqref{Eq.ss-4ODE-vJ} for $\varv_{J}(x)$ includes $c_1$ in its asymptotic approximation
\begin{align}
&\varv_{J}(x\to-1)=-\left(\frac{x+1}{2}\right)^{L}+\cdots,\label{Eq.vJ_asymp_1mx}\\
&\varv_{J}(x\to1)=\frac{\varv_{F}(x\to 1)}{L}-\frac{\beta}{2}=\frac{1}{2F_\text{BC}}\frac{2\beta-3}{4\beta}\left(\frac{F_\text{BC}-c_1}{c_3}\right)\left[1-\left(\frac{1+x}{2}\right)^{L}\right]+\cdots.\label{Eq.vJ_asymp_1px}
\end{align}
The relation between the eigenvalues and boundary conditions can be confirmed  by fixing the value of $\beta$ during iteration process and by seeing how the value of $\varv_{I}(x)$ reaches the expected boundary numerical value, i.e. 0, for different values of $\beta$ (See Appendix \ref{sec:stability_eigen}).

\subsection{The asymptotic approximation of the factor $[\varv_{J}(x)+1]$}\label{sec:eqn_vJp1}

Careful readers would realize that 4ODEs \eqref{Eq.ss-4ODE-vF} - \eqref{Eq.ss-4ODE-vJ} do not \emph{apparently} include an equation to describe the asymptotic approximation of $\varv_{F}(x)$ in the limit of $x\to-1$ while they include the corresponding approximations of $\varv_I$, $\varv_J$ and $\varv_G$. To see this, take the limit of $x\to-1$ in equation \eqref{Eq.ss-4ODE-vF}; one can see that the factor $\left[1+\varv_{J}(x)\right]$ is proportional to $(1/2+x/2)^{\beta}$. Hence, one may introduce a new dependent variable
\begin{align}
\overline{\varv_{J}}(x)\equiv\frac{1+\varv_{J}(x)}{\left(\frac{1+x}{2}\right)^{\beta L}}.
\end{align}
By the new variable, equations \eqref{Eq.ss-4ODE-vF} and \eqref{Eq.ss-4ODE-vJ} can be rewritten as 
\begin{subequations}
	\begin{align}
	&\left[\frac{1+x}{L}\frac{\,\text{d}\overline{\varv_{F}}}{\,\text{d}x}\left(\frac{1+x}{2}\right)^{\beta L}+\beta\right]\left[\varv_{I}(x)+\varv_{G}(x)\right]+\left(\frac{1+x}{2}\right)^{L}\frac{4\beta}{2\beta-3}\left[\overline{\varv_{J}}(x)\left(\frac{1+x}{2}\right)^{\beta L}-1\right]+\beta c_{2}e^{-\varv_{F}(x)}\overline{\varv_{J}}(x)\left(\frac{1+x}{2}\right)^{L}=0,\label{Eq.Asymp_ss-4ODE-vF}\\
	&\frac{1+x}{L}\varv_{Q}(x)\frac{\,\text{d}\overline{\varv_{J}}}{\,\text{d}x}+\overline{\varv_{J}}(x)\left(\frac{1+x}{2}\right)^{\beta L}\left\{ 3\frac{2\beta+1}{4} \varv_{Q}(x)+\frac{1+x}{2L}\left[\varv_{Q}(x)\frac{\,\text{d}\varv_{F}}{\,\text{d}x}+3\frac{\,\text{d}\varv_{Q}}{\,\text{d}x}\right]\right\}-\frac{1+x}{2L}\varv_{Q}(x)\frac{\,\text{d}\varv_{F}}{\,\text{d}x}\frac{3(2\beta-1)}{2}-\frac{6}{L}\frac{\,\text{d}\varv_{Q}}{\,\text{d}x}=0.\label{Eq.Asymp_ss-4ODE-vJ}
	\end{align}
\end{subequations}
Taking the limit of $x\to-1$ in equation \eqref{Eq.Asymp_ss-4ODE-vF} provides the asymptotic approximation;
\begin{align}
\overline{\varv_{J}}(x\to-1)=-\frac{c^{*}_{4}}{c_{2}}\frac{(2\beta+7)(6\beta-3)}{\beta(2\beta-7)(2\beta-3)(\beta+1)},\label{Eq.Asymp_vJ}
\end{align}
Hence, we can find from equations \eqref{Eq.Asymp_ss-4ODE-vJ} and \eqref{Eq.Asymp_vJ} the equation for $\varv_F$ as $x\to-1$; $-\frac{1+x}{2L}\varv_{Q}(x)\frac{\,\text{d}\varv_{F}}{\,\text{d}x}\frac{3(2\beta-1)}{2}-\frac{6}{L}\frac{\,\text{d}\varv_{Q}}{\,\text{d}x}=0.$ However, this expression is false since our numerical result showed that derivatives $\frac{\,\text{d}\varv_{F}}{\,\text{d}x}$ $\frac{\,\text{d}\varv_{J}}{\,\text{d}x}$ and $\frac{\,\text{d}\varv_{Q}}{\,\text{d}x}$ behave like power-law $(0.5+0.5x)^{\beta-1}$ as $x\to-1$ (such power-law behaviors are shown graphically in Figure \ref{fig:Del_asymFKGL}). This means,  equation \eqref{Eq.Asymp_ss-4ODE-vJ} is still the equation to determine the behavior of $\varv_J$ as $x\to-1$. Accordingly, the expression for the asymptotic approximation of $\varv_J$ (equation \eqref{Eq.Asymp_vJ}) is correct only when the first term in \eqref{Eq.Asymp_ss-4ODE-vF} is greater than order of double precision (as explained in Appendix \ref{sec:Newton_method}); strictly speaking the first term should be always included in numerical calculation to consistently solve the 4ODEs.

\section{Stability analyses of the whole-domain solution}\label{sec:stability}

The present appendix shows the numerical stability of the whole-domain solution. We detail the dependence of the solution on eigenvalue $\beta$ (Appendix \ref{sec:stability_eigen}), the nodes of Fej$\acute{\mathrm{e}}$r's quadrature (Appendix \ref{sec:stability_quadratures}), the boundary condition for $\varv_\text{F}(x)$ (Appendix \ref{sec:stability_BC}) and the numerical parameter $L$ (Appendix \ref{sec:Stability_L}).

\subsection{Stability of the whole-domain solution against the eigenvalue $\beta$}\label{sec:stability_eigen}

Throughout the present work the boundary value $\varv_{I}(x=1)$ is important since it determines the eigenvalue $\beta $; the present appendix shows its stability. We solved the ss-OAFP system for different $\beta$ between $\beta_\text{o}-10^{-8}$ and $\beta_\text{o}+10^{-8}$. Figure \ref{fig:Del_a_vKc1c4} shows the $\beta$-dependence of $\varv_{I}(x=1)$, $\mid 1-c_{1}/c_{1\text{o}}\mid$ and $\mid1 - c^{*}_{4}/c^{*}_{4\text{o}}\mid$. All the values are almost symmetric about $\beta_\text{o}$ and minimized around at $\beta=\beta_\text{o}$.  Also, the eigenvalues consistently converge to their reference values, $c_{1}\approx c_{1\text{o}}$ and $c_{4}^{*}\approx c^{*}_{4\text{o}}$ when $\beta$ reaches $\beta_\text{o}$.  One can find the following approximate relationship in the order of values
\begin{align}
&\mid 1-c_{1}/c_{1\text{o}}\mid\quad\sim\quad\mid1-\beta/\beta_\text{o}\mid\quad\sim\quad \frac{\mid\varv_{I}(x=1)\mid}{10^{2}}\quad\sim\quad\frac{\mid 1-c^{*}_{4}/c^{*}_{4\text{o}}\mid}{10^{5}}.\label{Eq.accuracy_relation}
\end{align}
This relationship implies that one needs $5\sim6$ significant figures of $\beta$ and $c_{1}$ to determine one significant figure of $c^{*}_{4}$. 

The Newton iteration did not work when the value of $\beta$  deviated from the reference value $\beta_\text{o}$ by $1.3\times 10^{-6}\%$ in the lower limit while we gave up at the relative error of $2.5\times10^{-7}\%$ in the upper limit due to an expensive CPU cost\footnote{Over one million iterations were needed when the eigenvalue $\beta$ deviated more than $1\times10^{-7}\%$ above the reference value $\beta_\text{o}$}.  Hence, the condition that Newton iteration method works for the whole-domain formulation is that one must correctly specify the eight or nine significant figures of $\beta$ $(8.17837105\leq \beta \lesssim 8.17837119)$.

\begin{figure}[H]
	\centering
	\tikzstyle{every node}=[font=\large]
	\begin{tikzpicture}[scale=0.9]
	\begin{loglogaxis}[legend columns=2, width=16cm, height=8cm, grid=major,xlabel=\large{$\mid\Delta \beta/\beta_\text{o}\mid$},xmin=1e-15,xmax=1e-7,ymin=1e-14,ymax=1e2, legend pos=north west]
	\addplot [color = bred,mark=no, thick, solid] table[x index=0, y index=1]{Del_am_vKc1c4.txt}; 
	\addlegendentry{\large{$\mid \varv_{I}(x=1)\mid$ $(\Delta \beta<0)$}}
	\addplot [color = bred,mark=no, thick, dashed] table[x index=0, y index=2]{Del_am_vKc1c4.txt}; 
	\addlegendentry{\large{$\mid1-c_{1}/c_{1\text{o}}\mid$ $(\Delta \beta<0)$}}
	\addplot [color = bred,mark=no, thick, dotted] table[x index=0, y index=3]{Del_am_vKc1c4.txt}; 
	\addlegendentry{\large{$\mid1-c^{*}_{4}/c^{*}_{4\text{o}}\mid$ $(\Delta \beta<0)$}}
	\addplot [only marks, color = bgreen,mark=square, thick] table[x index=0, y index=1]{Del_ap_vKc1c4.txt}; 
	\addlegendentry{\large{$\mid \varv_{I}(x=1)\mid$ $(\Delta \beta>0)$}}
	\addplot [only marks, color = bgreen,mark=o, thick] table[x index=0, y index=2]{Del_ap_vKc1c4.txt}; 
	\addlegendentry{\large{$\mid1-c_{1}/c_{1\text{o}}\mid$ $(\Delta \beta>0)$}}
	\addplot [only marks, color = bgreen,mark=triangle, thick] table[x index=0, y index=3]{Del_ap_vKc1c4.txt}; 
	\addlegendentry{\large{$\mid1-c^{*}_{4}/c^{*}_{4\text{o}}\mid$ $(\Delta \beta>0)$}}
	\end{loglogaxis}
	\end{tikzpicture}
	\caption{Values of $\mid \varv_{I}(x=1)\mid $, $\mid 1-c_{1}/c_{1\text{o}}\mid$ and $\mid1 - c^{*}_{4}/c^{*}_{4\text{o}}\mid$ against change $\Delta\beta$ around $\beta_\text{o}$ ($\Delta\beta\equiv\beta-\beta_{\text{o}}$). Numerical parameters $\mathcal{N}=70$, $L=1$ and $F_\text{BC}=1$ are employed.}\label{fig:Del_a_vKc1c4}
\end{figure}

\subsection{Stability of the whole-domain solution against the number of nodes in Fej$\acute{\mathrm{e}}$r's first-rule quadrature}\label{sec:stability_quadratures}

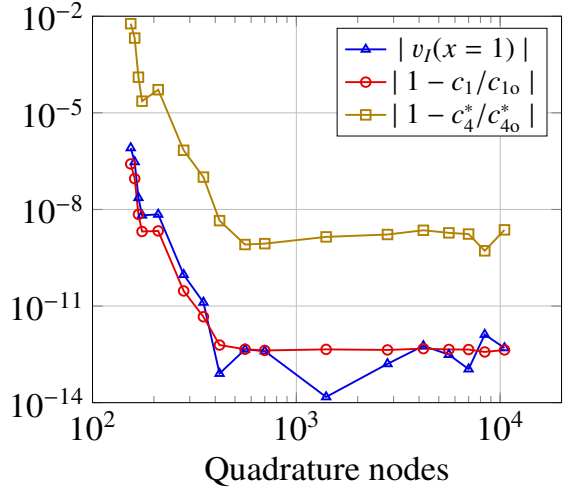
\begin{figure}[H]
	\centering
	\tikzstyle{every node}=[font=\Large]
	\begin{tikzpicture}[scale=0.9]
	\begin{loglogaxis}[ grid=major,xlabel=\Large{Quadrature nodes},xmin=1e2,xmax=2e4,ymin=1e-14,ymax=1e-2, legend pos=north east]
	\addplot [color = bblue ,mark=triangle, thick] table[x index=0, y index=1]{Quad_vK_c1_c2_N70.txt}; 
	\addlegendentry{\large{$\mid\varv_{I}(x=1)\mid$}}
	\addplot [color = bred ,mark=o, thick] table[x index=0, y index=2]{Quad_vK_c1_c2_N70.txt}; 
	\addlegendentry{\large{$\mid 1-c_{1}/c_{1\text{o}}\mid$}}
	\addplot [color = bgold ,mark=square, thick] table[x index=0, y index=3]{Quad_vK_c1_c2_N70.txt}; 
	\addlegendentry{\large{$\mid 1-c^{*}_{4}/c^{*}_{4\text{o}}\mid$}}
	\end{loglogaxis}
	\end{tikzpicture}
	\caption{Dependence of the eigenvalues $c_1$ and $c_4^{*}$ and boundary value $\mid\varv_{I}(x=1)\mid$ on the number of nodes in Fej$\acute{\mathrm{e}}$r's first-rule quadrature. The eigenvalues are compared to their reference eigenvalues $c_{1\text{o}}$ and $c^{*}_{4\text{o}}$ and the following numerical parameters are employed; $\mathcal{N}=70$, $L=1$ and $F_\text{BC}=1$.}
	\label{fig:Quad_vK_c1_c2_N70}
\end{figure}

Figure \ref{fig:Quad_vK_c1_c2_N70} shows the dependence of the eigenvalues $c_1$ and $c_4$ and boundary value $\varv_{I}(x=1)$ on the number of nodes in Fej$\acute{\mathrm{e}}$r's first-rule quadrature. The total number of nodes are chosen between $150$ to $10^{4}$ for fixed $\beta=\beta_\text{o}$ and $\mathcal{N}=70$; the Newton iteration did not work for the number of nodes less than $150$. The eigenvalues get stable for the nodes over $\sim580$ points; $c_1$ and $c_2$ approach the reference eigenvalues $c_{1\text{o}}$ and $c^{*}_{4\text{o}}$. Also, the boundary value $\varv_{I}(x=1)$ is qualitatively similar to $\mid 1-c_{1}/c_{1\text{o}}\mid$ and $\mid 1-c^{*}_{4}/c^{*}_{4\text{o}}\mid$.

\subsection{Stability of the whole-domain solution against $F_\text{BC}$}\label{sec:stability_BC}

\begin{figure}[H]
	\centering
	\tikzstyle{every node}=[font=\Large]
	\begin{tikzpicture}[scale=0.9]
	\begin{loglogaxis}[ grid=major,xlabel=\Large{$F_\text{BC}$},xmin=1e-5,xmax=1e5,ymin=1e-8,ymax=1e18, legend pos=north west]
	\addplot [color = bblue ,mark=triangle, thick] table[x index=0, y index=1]{Fo_cond_c1c4_N70.txt}; 
	\addlegendentry{\large{condition number}}
	\addplot [color = bred ,mark=o, thick] table[x index=0, y index=2]{Fo_cond_c1c4_N70.txt}; 
	\addlegendentry{\large{$c_{1}$}}
	\addplot [color = bgold ,mark=square, thick] table[x index=0, y index=3]{Fo_cond_c1c4_N70.txt}; 
	\addlegendentry{\large{$c^{*}_{4}$}}
	\end{loglogaxis}
	\end{tikzpicture}\hspace{0.6cm}
	\begin{tikzpicture}[scale=0.9]
	\begin{loglogaxis}[ grid=major,xlabel=\Large{$F_\text{BC}$},xmin=1e-5,xmax=1e5,ymin=1e-15,ymax=1e-5, legend pos=north east]
	\addplot [color = bblue ,mark=triangle, thick] table[x index=0, y index=1]{FoSki_Del_c1c4_N70.txt}; 
	\addlegendentry{\large{$\mid 1-c_{1}/F_\text{BC}/c_{1\text{o}}\mid$}}
	\addplot [color = bred ,mark=o, thick] table[x index=0, y index=2]{FoSki_Del_c1c4_N70.txt}; 
	\addlegendentry{\large{$\mid 1-c^{*}_{4}/F_\text{BC}/c^{*}_{4\text{o}}\mid$}}
	\end{loglogaxis}
	\end{tikzpicture}
	\caption{(Left panel) Dependence of the eigenvalues $c_{1}$ and $c^{*}_{4}$ on the boundary condition $F(x=1)$ compared to the condition number of the Jacobian matrix for the 4ODEs. (Right panel) Relative error of the regularized eigenvalues $c_{1}/F_\text{BC}$ and $c^{*}_{4}/F_\text{BC}$ from the reference eigenvalues $c_{1\text{o}}$ and $c^{*}_{4\text{o}}$. ($F_\text{BC}=1$, $L=1$ and $\mathcal{N}=70$)}
	\label{fig:Fo_cond_c1c4_N70}
\end{figure}
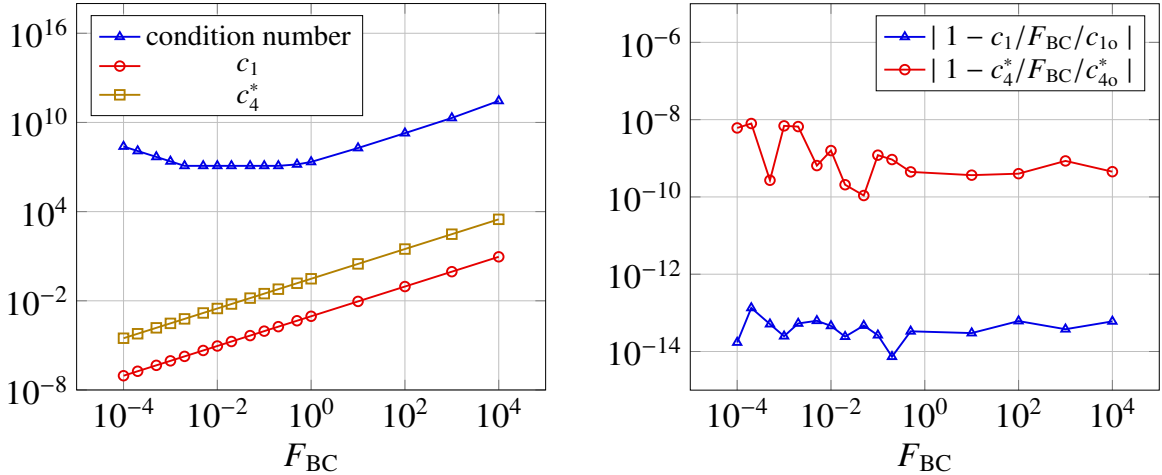

While the boundary condition $F(E=-1)=1$ was employed in \citep{Heggie_1988,Takahashi_1993}, there is no specific reason to choose the value 1 unless one needs to change the central density. Hence, we employed different boundary values of $F(E=-1)$ to see the consistency of the eigenvalues.  The left panel in Figure \ref{fig:Fo_cond_c1c4_N70} shows the values of $c_{1}$ and $c^{*}_{4}$ against the different values of $F_\text{BC}$ between $0.0001$ and $10000$. We found that $c_{1}$ and $c^{*}_{4}$ are proportional to $F_\text{BC}$ while the eight significant figures of $\beta$ and $c_{3}$ are constant. Also, as $F_\text{BC}$ increases, a similar characteristics was found in the condition number of the Jacobian matrix for the 4ODEs and the number reached $\sim10^{12}$ for $F_\text{BC}=10^{4}$.
Due to the linear relation between the eigenvalues and the boundary value, we divided $c_{1}$ and $c^{*}_{4}$ by $F_\text{BC}$ and compared  to the reference values $c_{1\text{o}}$ and $c^{*}_{4\text{o}}$ obtained for $F_\text{BC}=1$. We confirmed the eigenvalues ($c_{1}$ and $c^{*}_{4}$ ) are proportional to $F_\text{BC}$ with a relative accuracy of $\sim10^{-8}$ for $c^{*}_{4}$ and $\sim10^{-13}$ for $c_{1}$ while the high condition number did not interfere the accuracies (Figure \ref{fig:Fo_cond_c1c4_N70}, right panel).

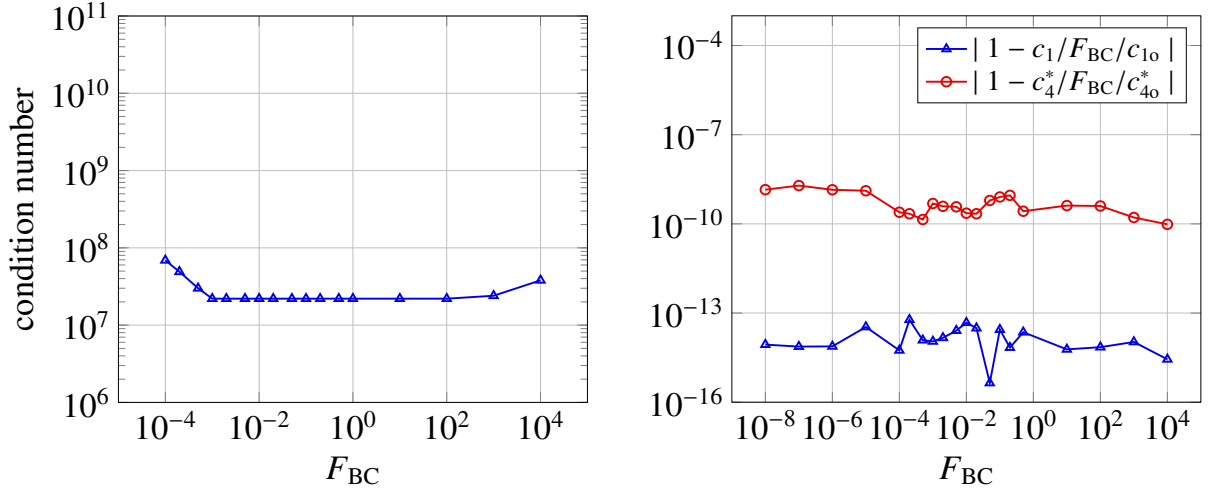
\begin{figure}[H]
	\centering
	\tikzstyle{every node}=[font=\Large]
	\begin{tikzpicture}[scale=0.9]
	\begin{loglogaxis}[ grid=major,xlabel=\Large{$F_\text{BC}$}, ylabel=\Large{condition number}, xmin=1e-5,xmax=1e5,ymin=1e6,ymax=1e11, legend pos=north west]
	\addplot [color = bblue ,mark=triangle, thick] table[x index=0, y index=1]{Do_cond_N70.txt}; 
	\end{loglogaxis}
	\end{tikzpicture}\hspace{0.6cm}
	\begin{tikzpicture}[scale=0.9]
	\begin{loglogaxis}[ grid=major,xlabel=\Large{$F_\text{BC}$},xmin=1e-9,xmax=1e5,ymin=1e-16,ymax=1e-3, legend pos=north east]
	\addplot [color = bblue ,mark=triangle, thick] table[x index=0, y index=1]{DoSki_Del_c1c4_N70.txt}; 
	\addlegendentry{\large{$\mid 1-c_{1}/F_\text{BC}/c_{1\text{o}}\mid$}}
	\addplot [color = bred ,mark=o, thick] table[x index=0, y index=2]{DoSki_Del_c1c4_N70.txt}; 
	\addlegendentry{\large{$\mid 1-c^{*}_{4}/F_\text{BC}/c^{*}_{4\text{o}}\mid$}}
	\end{loglogaxis} 
	\end{tikzpicture}
	\caption{(Left panel) Condition number of the Jacobian matrix for the $Q$-integral and 4ODEs regularized by dividing $F(E)$ by $F_\text{BC}$. (Right panel) Relative error of the regularized eigenvalues $c_{1}/F_\text{BC}$ and $c^{*}_{4}/F_\text{BC}$ from the reference eigenvalues $c_{1\text{o}}$ and $c^{*}_{4\text{o}}$ for the 4ODEs regularized through $F(E)/F_\text{BC}$. ($F_\text{BC}=1$ , $L=1$ and $N=70$.)}
	\label{fig:Do_cond_c1c4_N70}
\end{figure}

To avoid the significant change in the condition number for high values of $F_\text{BC}$, we regularized the ss-OAFP system by dividing the function $F(E)$ by $F_\text{BC}$. This regularization corresponds with that only the density $D(E)$ in the system is proportional to $F_\text{BC}$. We again solved the regularized ss-OAFP system for different $F_\text{BC}$. As expected, the condition number does not change significantly against change in $F_\text{BC}$ (Figure \ref{fig:Do_cond_c1c4_N70}). Also, the eigenvalues are stable against $F_\text{BC}$;  $\mathcal{O}(10^{-14})<\mid 1-c_{1}/F_\text{BC}/c_{1\text{o}}\mid<\mathcal{O}(10^{-13})$  and $\mathcal{O}(10^{-10})<\mid 1-c^{*}_{4}/F_\text{BC}/c^{*}_{4\text{o}}\mid<\mathcal{O}(10^{-8})$ .

In conclusion, the eigenvalues are less sensitive to high condition number and the eigenvalues $\beta$ (or $\alpha$) and $c_{3}$ have a numerically intrinsic property against change in $F_\text{BC}$ while $c_{1}$, $c_{2}$ and $c^{*}_{4}$ are extrinsic;
\begin{align}
c_{1}(F_\text{BC})\propto \left(c_{1\text{o}} +\mathcal{O}\left(10^{-13}\right)\right)F_\text{BC}, \qquad c^{*}_{4}(F_\text{BC})\propto\left(c^{*}_{4\text{o}} +\mathcal{O}\left(10^{-8}\right)\right)F_\text{BC},\qquad   \beta(F_\text{BC})=\beta_\text{o}+\mathcal{O}(10^{-8}),
\end{align}
where $-10^{-4}<F_\text{BC}<10^{4}$.

\subsection{Stability of the whole-domain solution against the numerical parameter $L$}\label{sec:Stability_L}

The parameter-$L$-dependence of the solutions provides an understanding of the ss-OAFP equation. We found spectral solutions of the ss-OAFP system with the mapping parameters $L=1/2$ and $L=3/4$ (Table \ref{table:trunc_contract}) while Newton method with $L>1$ was hard to work\footnote{Choosing high numbers for $L$ (e.g. $L=1.5$ and $L=2$) resulted in much more difficulty in Newton interaction convergence. We had to shorten the Newton steps from $1$ to a fraction less than $0.01$.  On one hand, low numbers of $L$ less than $1/2$ did not work; this is perhaps because contracted-domain formulation provides slow decay of Chebyshev coefficients, accordingly low accuracy of the solutions. As discussed in Section \ref{sec:repro_HS_soln} solutions with low accuracy can not provide the reference solution.}. In this sense, we call a solution with $L<1$ the 'contracted-domain' solution of the ss-OAFP system. The contracted-domain solutions provide some advantages over the reference solution; they are still whole-domain solutions (since they are not truncated) while they need less degrees of polynomials and are compatible to the reference solution. The convergence rate of Chebyshev coefficients for large $n$ is apparently\footnote{The slow convergence does not originates from the branch point. This is since the regularized function $\varv_{F}$ behaves like $c_{4}^{*}(1+b(0.5+0.5x)^\beta)$ as $x\to-1$ where $b$ is a constant. In fact, as we increased the digits of $\beta$ by correctly specifying the value, the coefficients decayed rapidly and reached order of$10^{-13}$ at the maximum degree ($\mathcal{N}=65$).} characterized by $a_{n}\propto n^{-1-2L}$ due to the end-point singularity $(1\pm x)^{L}$ at branch points $x\pm1$. The characteristics of the low convergence rate for the function $\varv_{F}(x)$ clearly appears when the degree $\mathcal{N}$ is greater than $65$ and $75$ for $L=0.75$ and $L=0.5$ respectively. The Newton iteration converged only when the Chebyshev coefficients reach as low as order of $10^{-9}$ for $L=3/4$ and $10^{-6}$ for $L=1/2$. Recalling the Newton iteration worked only when the Chebyshev coefficients of the whole-domain solution with $L=1$ reach order of $10^{-12}$ (Table \ref{fig:cFKGLRQ_N70}), we infer a \emph{rule-of-thumb} for the relationship between the coefficients and iteration method that \emph{Newton iteration method could work when the minimum absolute value of Chebyshev coefficients reaches as low as order of $10^{-12L}$}.

\begin{table*}\large
	\ra{1.3}
	\scalebox{0.7}{
		\begin{tabular}{@{}clccc@{}}
			\large{$L=0.75$} &                   &                                         &                                        &  \\
			\toprule
			\large{$\mathcal{N}$} & Eigenvalue $\beta$ & $\mid1-c_{1}/c_{1\text{o}} \mid$ &  $\mid1-c^{*}_{4}/c^{*}_{4\text{o}} \mid$  & $\varv_{I}(x=1)$  \\ 
			\midrule
			$60$ & $8.178371160$ & $2.1\times10^{-10}$& $6.6\times10^{-7}$  & $9.4\times10^{-9}$\\
			$55$ & $8.178371160$ & $2.1\times10^{-10}$& $1.3\times10^{-6}$  & $1.7\times10^{-9}$\\
			$50$ & $8.178371160$ & $8.0\times10^{-10}$& $9.8\times10^{-6}$  & $1.3\times10^{-9}$\\
			\bottomrule
	\end{tabular}}\hspace{0.3cm}
	\scalebox{0.7}{
		\begin{tabular}{@{}clccc@{}}
			\large{$L=0.5$} &                   &                                         &                                        &  \\
			\toprule
			\large{$\mathcal{N}$} & Eigenvalue $\beta$ & $\mid1-c_{1}/c_{1\text{o}} \mid$ &  $\mid1-c^{*}_{4}/c^{*}_{4\text{o}} \mid$  & $\varv_{I}(x=1)$  \\ 
			\midrule
			$35$ & $8.17837104$ & $4.9\times10^{-8}$& $2.3\times10^{-6}$  & $ 2.1\times10^{-10}$\\
			\bottomrule
	\end{tabular}}
	\caption{Numerical results for the contracted-domain formulation with $L=1/2$ and $L=3/4$ and $F_\text{BC}=1$.}
\end{table*}\label{table:trunc_contract}

\begin{figure}[H]
	\centering
	\tikzstyle{every node}=[font=\Large]
	\begin{tikzpicture}[scale=0.9]
	\begin{loglogaxis}[ grid=major,xlabel=\Large{$F_\text{BC}$},xmin=1e-0,xmax=70,ymin=1e-10,ymax=1e1, legend pos=south west]
	\addplot [color = bblue ,mark=no, thick] table[x index=0, y index=1]{n_cF_N60_Lc075.txt}; 
	\addlegendentry{\large{$\mathcal{N}=60$}}
	\addplot [color = black ,mark=no, dashed, thick] table[x index=0, y index=1]{n_cF_N65_Lc075.txt}; 
	\addlegendentry{\large{$\mathcal{N}=65$}}
	\addplot [color = bred,mark=no, densely dotted, thick] table[x index=0, y index=2]{n_cF_N65_Lc075.txt}; 
	\addlegendentry{guide line \large{$n^{-2L-1}$}}
	\node[above,red] at (32,1e-1) {$L=3/4$};
	\end{loglogaxis}
	\end{tikzpicture}\hspace{0.5cm}
	\begin{tikzpicture}[scale=0.9]
	\begin{loglogaxis}[ grid=major,xlabel=\Large{$F_\text{BC}$},xmin=1e-0,xmax=80,ymin=1e-10,ymax=1e1, legend pos=south west]
	\addplot [color = bblue ,mark=no, thick] table[x index=0, y index=1]{n_cF_N35_Lc05.txt}; 
	\addlegendentry{\large{$\mathcal{N}=35$}}
	%\addplot [color = black ,mark=no, dashed, thick] table[x index=0, y index=1]{n_cF_N50_Lc05.txt}; 
	%\addlegendentry{\large{$\mathcal{N}=50$}}
	\addplot [color = black ,mark=no, dashed, thick] table[x index=0, y index=1]{n_cF_N75_Lc05.txt}; 
	\addlegendentry{\large{$\mathcal{N}=75$}}
	\addplot [color = bred,mark=no, densely dotted, thick] table[x index=0, y index=2]{n_cF_N75_Lc05.txt}; 
	\addlegendentry{guide line \large{$n^{-2L-1}$}}
	\node[above,red] at (32,1e-1) {$L=1/2$};
	\end{loglogaxis}
	\end{tikzpicture}
	\caption{(Left Panel) Chebyshev coefficients of $\varv_{F}(x)$ for $L=0.75$ in the following cases (a) $\mathcal{N}=60$ and $\beta=8.178371160$ and (ii) $\mathcal{N}=65$ and $\beta=8.178371275$. The iteration method for the latter did not work satisfactorily since $\mid\{a\}^\text{new}-\{a\}^\text{old}\mid\approx 7\times10^{-10}$  (resulting in $\mid1-c^{*}_{4}/ c^{*}_{4\text{o}}\mid\approx6.0\times10^{-3}\mid$  and $\varv_{I}=4.4\times10^{-6}$). Yet, it is shown here for comparison.  (Right panel) Chebyshev coefficients of $\varv_{F}(x)$ for $L=0.5$ in the following cases (a) $\mathcal{N}=35$ and $\beta=8.178371160$ and (ii) $\mathcal{N}=75$ and $\beta=8.1783712$.}
	\label{fig:n_cF_Lc05_075}
\end{figure}

\section{Stability of the truncated-domain solution against change in extrapolated DF}\label{sec:Stability_extrapol}

We found that the truncated-domain solutions is little sensitive to the the expression of the extrapolated DF (equation \eqref{Eq.vF_extrap}). We compared the effects of change in the extrapolated DF on the eigenvalues (Table \ref{table:trunc_estrapol}). The set of parameters $(c,d)=(1,10)$ provided the best accuracy in the sense that $\mid\varv_{I}(x)\mid$ reaches the minimum value ($7.3\times10^{-12}$) among the chosen parameters $(c,d)$, hence we compared the eigenvalues obtained for $(c,d)$ to the eigenvalues for $(c,d)=(1,10)$. For combinations of different sets of parameters among $0.01<c<10$ and $0.01<d<10$, the relative error of eigenvalues are order of $10^{-13}$ in $c_{1}$ compared to its reference value and $10^{-9}$ in $c_{4}^{*}$ at most, holding small values of $\mid \varv_{I}(x=1)\mid\approx1\times10^{-11}$.  Even the effect of discontinuity in derivative of the extrapolated DF at $E=E_\text{min}$ ($c\to\infty$) is not significant compared to the effect of large value of $c (=5,10)$. 

\begin{table*}\large\centering
	\ra{1.3}
	\scalebox{0.7}{
		\begin{tabular}{@{}clccc@{}}
			\toprule
			\large{$c$} &  $d$ & $\mid1-c_{1}/c_{1\text{ex}} \mid$ &  $\mid1-c^{*}_{4}/c^{*}_{4\text{ex}} \mid$  & $\varv_{I}(x=1)$  \\ 
			\midrule
			$\infty$ & $N/A$ & $1.2\times10^{-13}$& $5.3\times10^{-10}$  & $1.0\times10^{-11}$\\
			$10$ & $1$ & $1.2\times10^{-13}$& $5.3\times10^{-10}$  & $1.0\times10^{-11}$\\
			$5$ & $1$ & $7.7\times10^{-14}$& $5.3\times10^{-10}$  & $1.0\times10^{-11}$\\
			$1$ & $1$ & $1.0\times10^{-13}$& $4.5\times10^{-10}$  & $9.8\times10^{-12}$\\
			$0.1$ & $1$ & $1.6\times10^{-13}$& $9.6\times10^{-10}$  & $1.2\times10^{-11}$\\
			$0.01$ & $1$ & $1.4\times10^{-13}$& $9.1\times10^{-10}$  & $1.1\times10^{-11}$\\
			\bottomrule
	\end{tabular}}\hspace{0.5cm}
	\scalebox{0.7}{
		\begin{tabular}{@{}clccc@{}}
			\toprule
			\large{$c$} &  $d$ & $\mid1-c_{1}/c_{1\text{ex}} \mid$ &  $\mid1-c^{*}_{4}/c^{*}_{4\text{ex}} \mid$  & $\varv_{I}(x=1)$  \\ 
			\midrule
			$10$ & $10$ & $9.4\times10^{-14}$& $5.3\times10^{-10}$  & $1.0\times10^{-11}$\\
			$2$ & $10$ & $2.1\times10^{-14}$& $3.1\times10^{-11}$  & $7.5\times10^{-12}$\\
			$1$ & $5$ & $4.0\times10^{-14}$& $2.6\times10^{-10}$  & $8.8\times10^{-12}$\\
			$1$ & $0.1$ & $1.1\times10^{-13}$& $8.7\times10^{-10}$  & $1.1\times10^{-11}$\\
			$1$ & $0.01$ & $1.2\times10^{-13}$& $9.6\times10^{-10}$  & $1.1\times10^{-11}$\\
			$0.1$ & $0.1$ & $1.2\times10^{-13}$& $8.4\times10^{-10}$  & $1.0\times10^{-11}$\\
			\bottomrule
	\end{tabular}}
	\caption{ Numerical results for different extrapolated DF ($L=1$ and $F_\text{BC}=1$).The eigenvalues are compared to $c_{1\text{ex}}\equiv c_{1\text{o}}$ and $c^{*}_{4\text{ex}}\equiv3.03155223\times10^{-1}(=c^{*}_{4\text{o}}+1\times10^{-1})$ obtained for $(c,d)=(1,10)$ and the value of $\mid\varv_{I}(x=1)\mid$ is $7.3\times10^{-12}$. The combination $(c,d)=(\infty,N/A)$ means the extrapolated function is constant.}
	\label{table:trunc_estrapol}
\end{table*}

\section{Why is Newton iteration method hard to work for the ss-OAFP system?}\label{sec:Newton_method}

The difficulty in numerical integration of the ss-OAFP system may originate from the complicated mathematical structure of the 4ODEs \eqref{Eq.ss-4ODE-vF} - \eqref{Eq.ss-4ODE-vJ}. To understand the structure, one must refer to the values of the infinity norms of the difference between `new' and `old' Chebyshev coefficients associated with 4ODEs in the process of Newton iteration method. We found the following values were universally output for all the truncated-domain-, whole-domain- and contracted-domain- formulations
\begin{subequations}
\begin{align}
&\left\|\{F_n\}^\text{new}-\{F_n\}^\text{old}\right\|_{\infty}\approx\mathcal{O}\left(10^{-13}\right), \\
&\left\|\{G_n\}^\text{new}-\{G_n\}^\text{old}\right\|_{\infty}\sim\left\|\{I_n\}^\text{new}-\{I_n\}^\text{old}\right\|_{\infty}\sim\left\|\{J_n\}^\text{new}-\{J_n\}^\text{old}\right\|_{\infty}\approx\mathcal{O}\left(10^{-16}\right)\approx eps.\\
&\left\|c_1^\text{new}-c_1^\text{old}\right\|_{\infty}\sim\mathcal{O}\left(10^{-16}\right),\hspace{1cm}\left\|c_4^\text{*new}-c_4^\text{*old}\right\|_{\infty}\approx\mathcal{O}\left(10^{-13}\right), 
\end{align}\label{Eq.Inf_norm}
\end{subequations}
where $eps$ means the machine precision of MATLAB ($\approx2.2\times10^{-16}$). Only the norms for $\{F_{n}\}$ and $c_{4}^{*}$ are approximately $10^{3}$ higher than the others, implying that equation \eqref{Eq.ss-4ODE-vF} associated with $\varv_{F}$ may have a mathematically internal conflict. Equation \eqref{Eq.ss-4ODE-vF} has the following mathematical structure
\begin{align}
4L\left(\frac{1+x}{2}\right)^{\beta+1}\frac{\,\text{d}\varv_{F}(x)}{\,\text{d}x}\mathcal{\eta}(x;\varv_{G},\varv_{I})+\left(\frac{1+x}{2}\right)^{\beta}\mathcal{\mu}(x;\varv_{G},\varv_{I})+c_{1}e^{-\varv_{F}(x)}(\varv_{J}+1)=0,\label{Eq.1ODE}
\end{align}
where $\mathcal{\eta}(x;\varv_{G},\varv_{I})$ and $\mathcal{\mu}(x;\varv_{G},\varv_{I})$ are functionals of $\varv_{G}(x)$ and $\varv_{I}(x)$ and their absolute values are order of unity on the whole domain. We explain possible relationships of the Newton's method with the mathematical structures focusing on problems in equation \eqref{Eq.1ODE} in the limit of $x\to1$ (Section \ref{sec:eqn_vf_xp1}) and $x\to-1$ (Section \ref{sec:eqn_vf_xm1}) for $\varv_F$, and in the derivative of $\varv_F$ (Section \ref{sec:eqn_vf_deriv}). Also, we show equation \eqref{Eq.1ODE} is important in integration of the 4ODE at equation level (Section \ref{sec:4ODE_machine}) and explain some other numerical difficulties in integrating the ss-OAFP system (Section \ref{sec:prpb_ss_OAFP}). 

\subsection{A problem in solving equation for $\varv_F$ in the limit of $x\to1$}\label{sec:eqn_vf_xp1}

A problem in solving equation \eqref{Eq.1ODE} is that the factor $c_{1}$ forms a numerical gap between terms at equation level. First, take the limit of $x\to1$ in equation \eqref{Eq.1ODE}
\begin{align}
4c_{3}\frac{\,\text{d}\varv_{F}(x\to1)}{\,\text{d}x}+\beta c_{3}-1+\frac{c_{1}}{F_\text{BC}}=0. \label{Eq.1ODE_xp1}
\end{align}
where $L=1$ is chosen for simplicity.  In equation \eqref{Eq.1ODE_xp1}, since $\beta c_{3}$ is approximately unity ($\approx 0.59$), the largest gap is order of $10^{-3}$ between the third and fourth terms regardless of the value of $F_\text{BC}$ (since $c_{1}$ is proportional to $F_\text{BC}$ as explained in Appendix \ref{sec:stability_BC}).  Hence, the equation can turn into an overdetermined problem at equation level greater than order of $10^{-3}$, which would be one of the reasons why the Newton method is hard to work. Also, the smallness of the gap could explain the large value of the norms for $c_{4}^{*}$ and $\{F_n\}$  (equation \eqref{Eq.Inf_norm}); the boundary value $F_\text{BC}(=\ln(\varv_{DF}(x=1)))$ is effective only up to 13 digits in the sense that it consistently determine the differentiation $\frac{\,\text{d}\varv_{F}(x)}{\,\text{d}x}$; digits more than 13 would be counted as rounding error due to the gap in $1-c_{1}/F_\text{BC}$. Due to this mathematical structure, we call order of $10^{-13}$ 'practical' machine precision at equation level as comparison to Matlab machine precision $\approx2.2\times10^{-16}$. 

\subsection{A problem in solving equation for $\varv_F$ in the limit of $x\to-1$}\label{sec:eqn_vf_xm1}

Another problem in solving equation \eqref{Eq.1ODE} is that the factors $\left(\frac{1+x}{2}\right)^{\beta}$, $\frac{\,\text{d}\varv_{F}(x)}{\,\text{d}x}$ and $(\varv_{J}+1)$ form power law profiles as $x\to-1$, which limits the effective domain on which we can consistently solve the 4ODE. Take the limit of $x\to-1$ in equation \eqref{Eq.1ODE}
\begin{align}
&\frac{6(2\beta-1)}{(2\beta-7)(\beta+1)}\left(\frac{1+x}{2}\right)^{\beta+1}\frac{\,\text{d}\varv_{F}(x)}{\,\text{d}x}(x\to-1)+\left(\frac{1+x}{2}\right)^{\beta}\frac{4\beta^{2}-4\beta+37}{(2\beta-7)(\beta+1)(2\beta-3)}+\frac{c_{1}}{c_{4}^{*}}\left[\varv_{J}(x\to-1)+1\right]=0. \label{Eq.1ODE_xm1}\\
&\hspace{3cm}\sim\left(\frac{1+x}{2}\right)^{2\beta}\hspace{3cm} \sim\left(\frac{1+x}{2}\right)^{\beta}\hspace{3cm}\sim\left(\frac{1+x}{2}\right)^{\beta}\nonumber
\end{align}
where the second line represents the power-law dependence of each term; the differentiation $\frac{\,\text{d}\varv_{F}(x)}{\,\text{d}x}(x\to-1)$ behaves like $\left(\frac{1+x}{2}\right)^{\beta-1}$ according to the result of Section \ref{sec:ss_soln_asymp} and $\left[\varv_{J}(x\to-1)+1\right]$ is explicitly proportional to $\left(\frac{1+x}{2}\right)^{\beta}$ as explained in Section \ref{sec:eqn_vJp1}. The first term in equation \eqref{Eq.1ODE_xm1} describes the 'time-evolution' equation with respect to $x$ in the sense that the equation is first order in differentiation or an initial value problem. Hence, one may consider the first term in equation \eqref{Eq.1ODE_xm1} is important to determine the interval on which one can solve the equation satisfactorily beginning from $x=1$. The factor $\left(\frac{1+x}{2}\right)^{2\beta}$, of course, does not contribute to the numerical integration of equation \eqref{Eq.1ODE_xm1} if it reaches order of machine precision $\sim10^{-16}$. Hence, by equating the first term to machine precision of Matlab $\frac{6(2\beta-1)}{(2\beta-7)(\beta+1)}\left(\frac{1+x}{2}\right)^{\beta+1}\frac{1-c_1}{c_3}=2.2\times10^{-16}$, where $\frac{1-c_1}{c_3}=\frac{\,\text{d}\varv_{F}(x)}{\,\text{d}x}(x=-1)$, we can estimate the lower limit of the interval is $x_\text{const}\approx-0.82$ (or the upper limit $E_\text{const}\approx -0.09$). This discussion implies that one can not effectively determine the value of $c_{4}^{*}$ at equation level with a numerical accuracy better than $10^{-9}$ $\left(=(0.5+0.5x_\text{const})^{\beta}\right)$ (Since $c_4^*$ is related to the third term in equation \ref{Eq.1ODE_xm1}). This order of values well reflects the result in Figure \ref{fig:Del_a_vKc1c4} in which $\mid 1-c_4^{*}/c_{4\text{o}}^{*}\mid$ is stable at order of $10^{-9}$ at best against change in $\beta$. Also, it may explain the reason that the relative error of the optimal truncated solution to the reference solution is at best $\sim10^{-9}$ as shown in Figure \ref{fig:Stab_Deg_F_xmin_094_N65}.   

\subsection{Absolute values of terms in equation for $\varv_F$ and classification of truncated-domain solutions}\label{sec:eqn_vf_deriv}

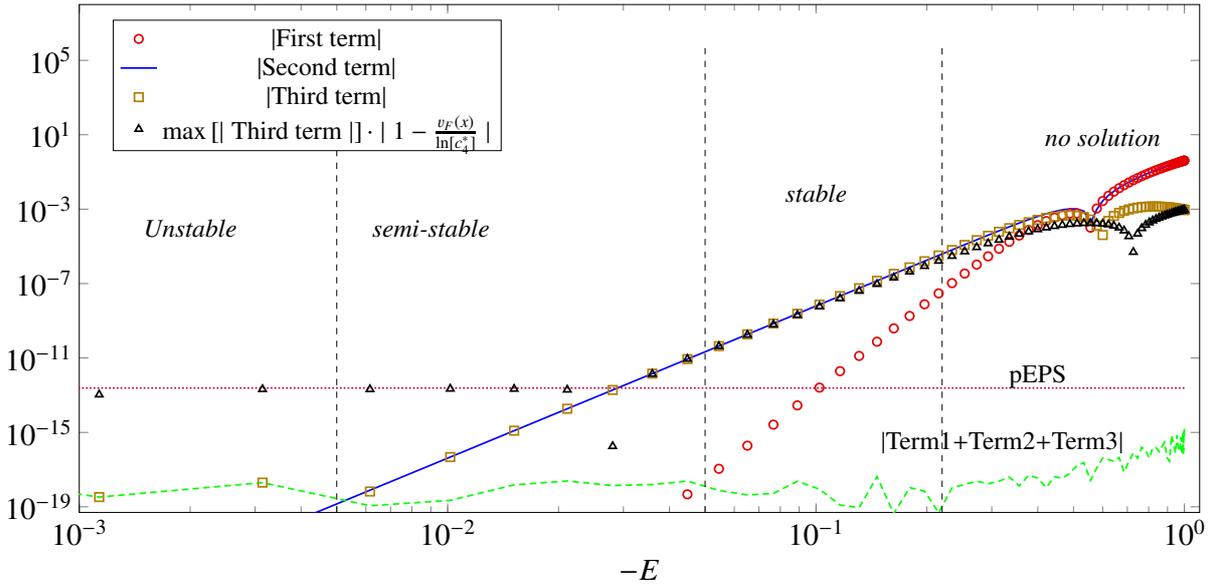
\begin{figure}[H]
	\centering
	\tikzstyle{every node}=[font=\Large]
	\begin{tikzpicture}[scale=0.8]
	\begin{loglogaxis}[width=20cm,height=10cm,xlabel=\Large{$-E$},xmin=1e-3,xmax=1.1,ymin=5e-20,ymax=1e8, legend pos=north west]
	\addplot [only marks, color = bred ,mark=o, thick] table[x index=0, y index=1]{Terms_Eq1_vF.txt}; 
	\addlegendentry{\large{$\mid$First term$\mid$}}
	\addplot [ color = blue ,mark=no,solid, thick] table[x index=0, y index=2]{Terms_Eq1_vF.txt}; 
	\addlegendentry{\large{$\mid$Second term$\mid$}}
	\addplot [only marks, color = bgold,mark=square, solid, thick] table[x index=0, y index=3]{Terms_Eq1_vF.txt}; 
	\addlegendentry{ \large{$\mid$Third term$\mid$}}
	\addplot [only marks,color = black,mark=triangle, thick] table[x index=0, y index=5]{Terms_Eq1_vF.txt};
	\addlegendentry{ \large{$\max\left[\mid \text{Third term}\mid\right]\cdot\mid 1- \frac{\varv_{F}(x)}{\ln[c_{4}^{*}]}\mid$}} 
	\addplot [color = purple,mark=no, densely dotted, thick] table[x index=0, y index=4]{Terms_Eq1_vF.txt}; 
	\addplot [color = green,mark=no, thick, densely dashed] table[x index=0, y index=6]{Terms_Eq1_vF.txt}; 
	\draw[dashed] (2.2e-1,5e-20) -- (2.2e-1,1e6);
	\draw[dashed] (5e-2,5e-20) -- (5e-2,1e6);
	\draw[dashed] (5e-3,5e-20) -- (5e-3,1e6);
	% \filldraw[draw=black,fill=lightgray] (1e-2,5e-14) rectangle (4e-4,1.2e0);
	\node[above,black] at (3.2e-1,3e-17) {\large{$\mid$Term1+Term2+Term3$\mid$}};
	\node[above,black] at (4e-1,1e-13) {\large{pEPS}};
	%\node[above,black] at (4e-3,3e-10) {\large{$\mid 1- \frac{\varv_{F}(x)}{\ln[c_{4}^{*}]}\mid$}};
	\node[black] at (2e-3,1e-4) {\large{\emph{Unstable}}};
	\node[black] at (9e-3,1e-4) {\large{\emph{semi-stable}}};
	\node[above,black] at (1e-1,1e-3) {\large{\emph{stable}}};
	\node[above,black] at (6e-1,1e-0) {\large{\emph{no solution}}};
	\end{loglogaxis}
	\end{tikzpicture}
	\caption{Absolute values of terms appearing in equation \eqref{Eq.ss-4ODE-vF}. The horizontal line represents the limit of precision and the deviation $\mid 1- \varv_{F}(x)/\ln[c_{4}^{*}]\mid$.}
	\label{fig:Terms_Eq1_vF}
\end{figure}

The present appendix compares the orders of absolute values of terms in equation \eqref{Eq.ss-4ODE-vF} to detail the mathematical structures and explains the classification of the truncated-domain solutions employed in Section \ref{sec:domain_trunc}. Figure \ref{fig:Terms_Eq1_vF} depicts the absolute values of the first through third terms in equation  \eqref{Eq.ss-4ODE-vF} together with relative error $\mid 1- \varv_{F}(x)/\ln[c_{4}^{*}]\mid$ and practical machine precision ($\sim10^{-13}$). Also, the sum of the three terms is depicted. The absolute value of the first term reaches the total of the three terms approximately at $E=-0.05$ while the second and third terms reach it at $E=-0.005$. Since we expect that we can satisfactorily solve equation \eqref{Eq.ss-4ODE-vF} at $E<-0.05$, we name the solutions that we can obtain on interval $E<-0.05$ as the 'stable solution' (The incorrect solution is discussed later in the present appendix). This well reflects the result for the reference solution in Figure \ref{fig:vJ_slope_index} in which the asymptotic behavior in differentiation of $\varv_J(x)$ loses accuracy at $E>-0.05$. Also, the truncated-domain solution holds accuracy beyond $E=-0.05$ as shown in Figure \ref{fig:Best_xmin_094_N65}. On one hand, we call solutions obtained for $-0.05<E_\text{max}<-0.005$ as 'semi-stable' solutions. This is since as $x\to-1$ the second and third terms, in place of the derivative of $\varv_F$, can determine the value of $\varv_F$, which results in that the accuracy of $c_4$ does not change with increasing $E_\text{max}$. The practical machine precision well describes the constancy of the accuracy of $c_4$. Lastly, beyond $E_\text{max}>-0.005$ there does not exist a meaningful term below machine precision, hence we can not solve the equation consistently. Since we could reasonably solve 4ODEs with fixed $Q$-integral (Appendix \ref{sec:Fixed_cQ}), we believe the essential cause of the numerical instability originates from the relationship between $\varv_F$ and the integrals $\varv_Q$ and $\varv_D$; one can not integrate the integrals for $E_\text{max}>-0.005$ holding a high accuracy since one needs more than double-precision to integrate them on their domains $(-1,E)$ and $(-E,0)$\footnote{One may consider the similarity in mathematical structure between the $\varv_Q$ and $\varv_D$ integrals and Dawson's integral; the latter exponentially loses accuracy \citep[e.g.][]{Cody_1970,Boyd_2008_2} and the former algebraically with increasing argument of them. As a result one needs more than double precision to find numerical values on broader truncated-domain (corresponding $E\to0$).}. We call solutions that we could obtain for $E_\text{max}>-0.005$ as the 'unstable' solution.

We can show Figure \ref{fig:Terms_Eq1_vF} has two more important characteristics of equation \eqref{Eq.ss-4ODE-vF} focusing on the second and third terms. First, we can obtain solutions that are close to the reference solution and HS's solution only for $E_\text{max}<-0.25$ (Section \ref{sec:stable_trunc_solns}). This nature appears when the second and third terms reach the same order of value and cancel out each other. Around at $E=-0.25$, the absolute values of the terms are order of $10^{-5}$. This reflects the order of value under which $\mid 1-c_4/c^{*}_4 \mid$ is stable against change in $\beta$ (Figure \ref{sec:stability_eigen}). Also, this infers that, in order to make Newton's method work, one must prepare an accurate 'initial guess' for solution whose accuracy is order of $10^{-5}$ to effectively determine the \emph{first} digit of the eigenvalue $c_4$, which would make the Newton method hard to work. Another important characteristics is that the value of $\mid 1- \varv_{F}(x)/\ln[c_{4}^{*}]\mid$ multiplied by the maximum value of the third term is greater than that of the first term. This means the reason why $\varv_F$ can behave like a constant function as $E\to0$ is \emph{not} because the first term reaches machine precision and lose its significance. This property is important to secure the consistency of our solution.

\subsection{A problem in solving the 4ODE at machine precision level}\label{sec:4ODE_machine}

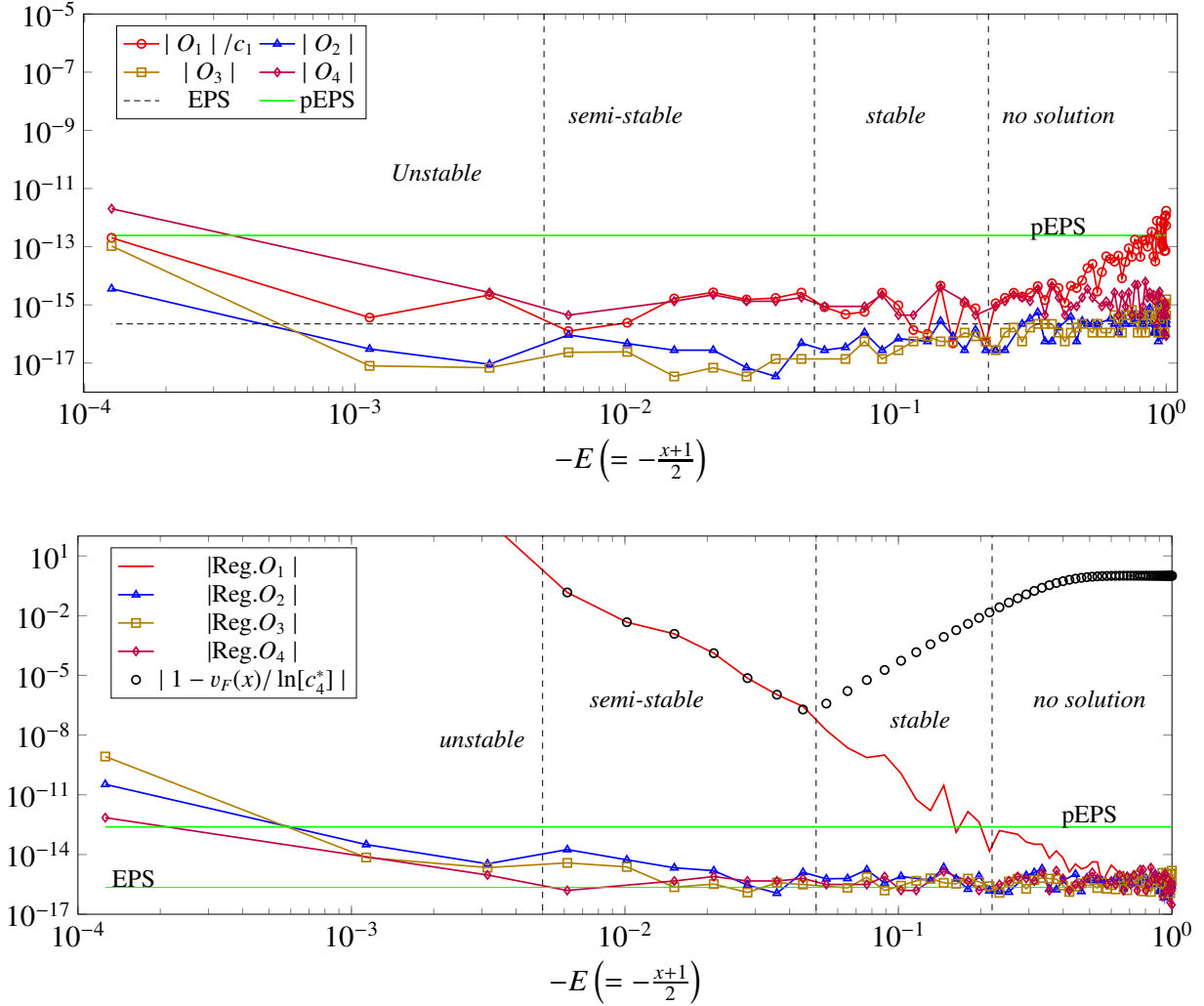
\begin{figure}[H]
	\centering
	\tikzstyle{every node}=[font=\Large]
	\begin{tikzpicture}[scale=0.8]
	\begin{loglogaxis}[width=20cm,height=8cm,xlabel=\Large{$-E\left (=-\frac{x+1}{2}\right)$},xmin=9.9e-5,xmax=1.1,ymin=1e-18,ymax=1e-5, legend pos=north west, legend columns=2]
	\addplot [color = bred ,mark=o, thick] table[x index=0, y index=1]{Four_ODEs_EPS.txt}; 
	\addlegendentry{\large{$\mid O_1\mid/c_{1}$}}
	\addplot [ color = blue ,mark=triangle,solid, thick] table[x index=0, y index=2]{Four_ODEs_EPS.txt}; 
	\addlegendentry{\large{$\mid O_2\mid$}}
	\addplot [color = bgold,mark=square, solid, thick] table[x index=0, y index=3]{Four_ODEs_EPS.txt}; 
	\addlegendentry{ \large{$\mid O_3\mid$}}
	\addplot [color = purple,mark=diamond, solid, thick] table[x index=0, y index=4]{Four_ODEs_EPS.txt}; 
	\addlegendentry{ \large{$\mid O_4\mid$}}
	\addplot [color = black,mark=no,densely dashed] table[x index=0, y index=5]{Four_ODEs_EPS.txt};
	\addlegendentry{ \large{EPS} }
	\addplot [color = green,mark=no, thick,  thick] table[x index=0, y index=6]{Four_ODEs_EPS.txt}; 
	\addlegendentry{ \large{pEPS} }
	\draw[dashed] (2.2e-1,5e-20) -- (2.2e-1,1e6);
	\draw[dashed] (5e-2,5e-20) -- (5e-2,1e6);
	\draw[dashed] (5e-3,5e-20) -- (5e-3,1e6);
	% \filldraw[draw=black,fill=lightgray] (1e-2,5e-14) rectangle (4e-4,1.2e0);
	\node[above,black] at (4e-1,1e-13) {\large{pEPS}};
	%\node[above,black] at (4e-3,3e-10) {\large{$\mid 1- \frac{\varv_{F}(x)}{\ln[c_{4}^{*}]}\mid$}};
	\node[above,black] at (2e-3,1e-11) {\large{\emph{Unstable}}};
	\node[above,black] at (1e-2,1e-9) {\large{\emph{semi-stable}}};
	\node[above,black] at (1e-1,1e-9) {\large{\emph{stable}}};
	\node[above,black] at (4e-1,1e-9) {\large{\emph{no solution}}};
	\end{loglogaxis}
	\end{tikzpicture}
	
	\vspace{0.6cm}
	
	\begin{tikzpicture}[scale=0.8]
	\begin{loglogaxis}[width=20cm,height=8cm,xlabel=\Large{$-E\left(=-\frac{x+1}{2}\right)$},xmin=1e-4,xmax=1,ymin=1e-17,ymax=1e2, legend pos=north west, x tick label style={
		/pgf/number format/fixed,/pgf/number format/precision=5}, scaled ticks=false] 
	\addplot [color = bred ,mark=no, thick,solid] table[x index=0, y index=1]{Reg_Four_ODEs_EPS.txt}; 
	\addlegendentry{\large{$\mid$Reg.$O_1\mid$}}
	\addplot [ color = blue ,mark=triangle,solid, thick] table[x index=0, y index=2]{Reg_Four_ODEs_EPS.txt}; 
	\addlegendentry{\large{$\mid$Reg.$O_2\mid$}}
	\addplot [color = bgold,mark=square, solid, thick] table[x index=0, y index=3]{Reg_Four_ODEs_EPS.txt}; 
	\addlegendentry{ \large{$\mid$Reg.$O_3\mid$}}
	\addplot [color = purple,mark=diamond, solid, thick] table[x index=0, y index=4]{Reg_Four_ODEs_EPS.txt}; 
	\addlegendentry{ \large{$\mid$Reg.$O_4\mid$}}
	\addplot [only marks, color = black,mark=o,thick,solid] table[x index=0, y index=2]{vJ_slope_index.txt}; 
	\addlegendentry{ \large{$\mid 1- \varv_{F}(x)/\ln[c_{4}^{*}]\mid$} }
	\addplot [color = green,mark=no,solid] table[x index=0, y index=5]{Reg_Four_ODEs_EPS.txt};
	%\addlegendentry{ \large{EPS} }
	\addplot [color = green,mark=no, thick,  solid] table[x index=0, y index=6]{Reg_Four_ODEs_EPS.txt}; 
	%\addlegendentry{ \large{pEPS} }
	\draw[dashed] (2.2e-1,5e-20) -- (2.2e-1,1e6);
	\draw[dashed] (5e-2,5e-20) -- (5e-2,1e6);
	\draw[dashed] (5e-3,5e-20) -- (5e-3,1e6);
	% \filldraw[draw=black,fill=lightgray] (1e-2,5e-14) rectangle (4e-4,1.2e0);
	\node[above,black] at (1.6e-4,1e-16) {\large{EPS}};
	\node[above,black] at (5e-1,1e-13) {\large{pEPS}};
	\node[above,black] at (3e-3,1e-9) {\large{\emph{unstable}}};
	\node[above,black] at (1.2e-2,1e-7) {\large{\emph{semi-stable}}};
	\node[above,black] at (1.2e-1,1e-8) {\large{\emph{stable}}};
	\node[above,black] at (5e-1,1e-7) {\large{\emph{no solution}}};
	%\node[above,black] at (4e-3,3e-10) {\large{$\mid 1- \frac{\varv_{F}(x)}{\ln[c_{4}^{*}]}\mid$}};
	\end{loglogaxis}
	\end{tikzpicture}
	\caption{Values of the regularized functions $O_1$ through $O_4$ in 4ODEs \eqref{Eq.ss-4ODE-vF}- \eqref{Eq.ss-4ODE-vJ} at Gauss-Chebyshev nodes. (Top) only $O_1$ is divided by $c_{1}$ (Bottom panel) All the functions are normalized so that the largest value of terms in each equation approaches unity as $E\to-1$.  The horizontal lines represent limits of precision. On the bottom panel, $\mid 1- \varv_{F}(x)/\ln[c_{4}^{*}]\mid$ is further regularized by the first term of equation \eqref{Eq.ss-4ODE-vF}.}
	\label{fig:Mag_4ODEs}
\end{figure}

Appendixes \ref{sec:eqn_vf_xp1}, \ref{sec:eqn_vf_xm1} and \ref{sec:eqn_vf_deriv} only focuses on equation \eqref{Eq.ss-4ODE-vF} among the 4ODEs; to emphasize the importance of equation \eqref{Eq.ss-4ODE-vF} we compare the equation to the rest of the equations.  To analyze the mathematical structures of the 4ODEs, we rewrite the 4ODE with new functions for convenience
\begin{align}
O_1(E)\equiv0, \hspace{1cm} O_2(E)\equiv0, \hspace{1cm} O_3(E)\equiv0, \hspace{1cm} O_4(E)\equiv0,
\end{align}
where $O_1(x)$ through $O_4(x)$ are functions that read the left hand sides of equations \eqref{Eq.ss-4ODE-vF}-\eqref{Eq.ss-4ODE-vJ}. Figure \ref{fig:Mag_4ODEs} (top panel) depicts the absolute values of $O_1$ through $O_4$ at Gauss-Chebyshev nodes on the whole domain. In the figure only $O_1$ is regularized by dividing $O_1$ by $c_{1}$. All the functions $O_1(x)$ through $O_4(x)$ lose accuracy on the unstable region increasing their absolute values as $x\to-1$. One can see the absolute values of $O_1(x)$ and $O_4(x)$ are very alike around the semi-stable region, which well describes the fact that $O_1$ and $O_4$ 'switch' their roles; they determine $\varv_J$ and $\varv_F$ respectively as $x\to-1$ while $\varv_F$ and $\varv_J$ as $x\to+1$, as explained in Appendix \ref{sec:eqn_vJp1}. Since the absolute values of the functions in Figure \ref{fig:Mag_4ODEs} are not regularized consistently to compare their absolute values, Figure \ref{fig:Mag_4ODEs} (Bottom panel) shows the regularized functions $O_1(x)$ - $O_4(x)$; we regularized the absolute values of $O_1 (x)$ - $O_4 (x)$ by dividing each function by the term whose value is the largest in the corresponding equation in the limit of $E\to-1$. As expected, $O_2(x)$, $O_3(x)$ and $O_4(x)$ stalls near the machine precision except for the unstable region. On one hand, $O_1 (x)$ significantly loses accuracy as $x$ approaches $-1$ and it well reflects the relation of $O_1 (x)$  with $\mid 1- \varv_{F}(x)/\ln[c_{4}^{*}]\mid$ in a similar way to Figure \ref{fig:vJ_slope_index}. In Figure \ref{fig:Mag_4ODEs} (Bottom Panel) $\mid 1- \varv_{F}(x)/\ln[c_{4}^{*}]\mid$ is also regularized by the same term for equation \eqref{Eq.ss-4ODE-vF}. This result highlights the dominant effect of equation \eqref{Eq.ss-4ODE-vF} to determine the accuracy of the 4ODE as $x\to-1$. 

\subsection{Some problems in numerical integration of ss-OAFP system}\label{sec:prpb_ss_OAFP}

Lastly, we summarize the three more difficulties that we faced in numerical integration of the ss-OAFP system. (i) The effect of discontinuity in solutions was an issue for truncated-domain formulation (see some discussion in Appendix \ref{sec:fixed_func}), which would have made harder guessing a `good' initial solutions in Newton iteration process. (ii) We also employed the Radau-Chebyshev spectral method and boundary condition $\varv_{I}(x=1)=0$ so that we can determine a spectral solution when the value $\varv_{I}(x=-1)$ is minimized by changing the value of $\beta$, but such solution included \emph{very} strong discontinuous property in both whole- and truncated-domain solutions. This could be due to the gap $\mid1-c_{1}/c_{1\text{ex}} \mid/\varv_{I}(x=1)\approx10^{-2}$ that prevents us from imposing the boundary condition $\varv_{I}(x=1)=0$. If one would like to determine 15 significant digits for $\varv_{I}(x=1)$, one must find 17 significant digits of eigenvalue $c_\text{1}$, which is beyond the limit of double-precision. (iii) The Newton iteration was hard to work for truncated-domain solutions for $-0.1<E_\text{max}<-0.4$. This would simply reflect the fact that an extrapolation of DF by the power-law profile on the domain is not a proper treatment.

\section{Solving part of the ss-OAFP system with a fixed independent variable}\label{sec:fixed_func}

The present appendix shows the results of numerical integration of part of the ss-OAFP system that we solved including some fixed independent variables (without self-consistently solving the entire system).  Appendices \ref{sec:Fixed_cR} and \ref{sec:Fixed_vD} show the effect of discontinuities in independent variable on the convergence rate of Chebyshev coefficients for integration of the Poisson equation and $Q$-integral respectively. The results possibly explain the slow convergence rate of the truncated-domain solutions (Section \ref{sec:domain_trunc}). Also, Appendix \ref{sec:Fixed_cQ} shows that the numerical instability (reported in Section \ref{sec:inst_whole}) does not occur for integration of 4ODE with a fixed  $\{Q_{n}\}$. This infers that the instability may originate from the relation between the 4ODE and the integrals $Q$ and $D$ rather than 4ODE itself.  

\subsection{Solving Q-integral with fixed discontinuous $\varv_{R}$}\label{sec:Fixed_cR}

\begin{figure}[H]
	\centering
	\tikzstyle{every node}=[font=\Large]
	\begin{tikzpicture}
	\begin{loglogaxis}[width=7cm, height=7cm, grid=major,xlabel=\Large{index $n$},ylabel=\large{$\mid Q_{n}\mid$},xmin=1e0,xmax=2.1e2,ymin=1e-9,ymax=1e0, legend pos=south west]
	\addplot [color = orange,mark=o,thick,solid] table[x index=2, y index=3]{E_vr_n_CoeffDiscon_01_guide.txt}; 
	\addlegendentry{\large{$ x_\text{trans}=0.1$}} 
	\addplot [color = black,mark=no,thick,dashed] table[x index=2, y index=4]{E_vr_n_CoeffDiscon_01_guide.txt}; 
	\addlegendentry{\large{$ 1/n^{2}$}} 
	\addplot [color = blue,mark=triangle,thick,solid] table[x index=2, y index=3]{E_vr_n_CoeffDiscon_0001.txt}; 
	\addlegendentry{\large{$x_\text{trans}=0.001$}} 
	\end{loglogaxis}
	\end{tikzpicture}
	\hspace{0.5cm}
	\begin{tikzpicture}
	\begin{axis}[width=7cm, height=7cm, grid=major,xlabel=\Large{index $-E$},ylabel=\large{$\mid \varv_{R}(E)\mid$},xmin=9.2e-1,xmax=1.e0,ymin=9e-1,ymax=1.15e0, legend pos=south east]
	\addplot [color = orange,mark=o,thick,solid] table[x index=0, y index=1]{E_vr_n_CoeffDiscon_01_guide.txt}; 
	\addlegendentry{\large{$ x_\text{trans}=0.1$}} 
	\addplot [color = blue,mark=triangle,thick,solid] table[x index=0, y index=1]{E_vr_n_CoeffDiscon_0001.txt}; 
	\addlegendentry{\large{$x_\text{trans}=0.001$}} 
	\end{axis}
	\end{tikzpicture}
	\caption{ Chebyshev coefficients of $Q$-integral for a discontinuous test function $\varv_{R}^{\text{(tes)}}=0.1\Theta(0.50+0.5[x-x_\text{trans}])+1$. Recall $E=-(0.5+0.5x)^{L}$, here $L=1$.}
	\label{fig:n_CoeffDiscon}
\end{figure}
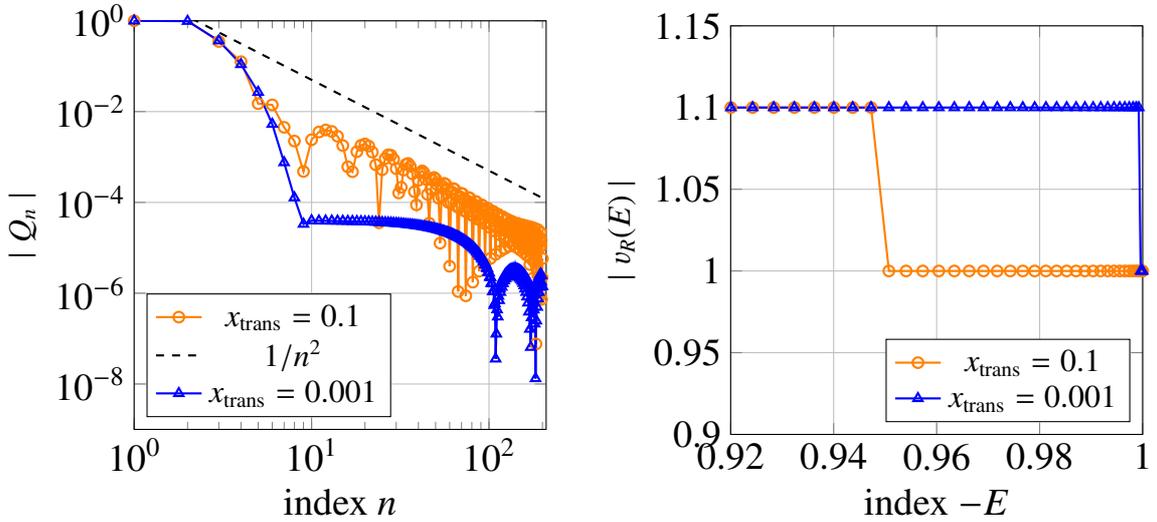

In the present work, all the spectral solutions that we obtained with \emph{truncated-domain} formulations include a certain flattening in their Chebyshev coefficients as index $n$ becomes large. To find a possible cause of the flattening, we calculated the Chebyshev coefficients of the $Q$-integral for the following discontinuous test function $\varv_{R}$
\begin{align}
\varv_{R}^\text{(tes)}=0.1\,\Theta\left(\frac{1+x-x_\text{trans}}{2}\right)+1,
\end{align}
where $x_\text{trans}$ is a small positive number and $\Theta(\cdot)$ the Heaviside function. When the point of discontinuity is relatively close to order of unity, say $x_\text{trans}=0.1$, the Chebyshev coefficients for $Q$-integral slowly decay like $\sim 1/n^{2}$ for large $n$ (Left panel, Figure \ref{fig:n_CoeffDiscon}) in a similar way to Chebyshev coefficients for discontinuous functions and for the integral of them \citep[e.g.][]{Boyd_2001,Xiang_2013}. However, once the discontinuity point more closely approaches the end point of the domain such as  $x_\text{trans}=0.001$ (Right panel, Figure \ref{fig:n_CoeffDiscon}), the coefficients show a flattening with large $n$. Since for $x_\text{trans}=0.001$ the majority of domain is covered by a constant function, one can find a rapid decay for small $n$. One can also observe for very large $n$ that the coefficients reach the same order of value regardless of the value of $x_\text{trans}$.

\subsection{Solving Poisson equation with fixed discontinuous $\varv_{D}$}\label{sec:Fixed_vD}

\begin{figure}[H]
	\centering
	\tikzstyle{every node}=[font=\Large]
	\begin{tikzpicture}
	\begin{loglogaxis}[legend columns=1,width=7cm, height=11cm, grid=major,xlabel=\Large{$n$},ylabel=\large{$\mid R_{n}\mid$},xmin=1,xmax=200,ymin=1e-7,ymax=1e-0, legend pos=south west]
	\addplot [color = orange,mark=no,thick,solid] table[x index=0, y index=1]{n_cR_cRm_guide_Poiss_more_dc.txt};
	\addlegendentry{\large{$A_\text{trans}=0.1$, $\varv_{R}$}} 
	\addplot [color = blue,mark=o,thick,solid] table[x index=0, y index=2]{n_cR_cRm_guide_Poiss_more_dc.txt};
	\addlegendentry{\large{$A_\text{trans}=0.1$, $\varv_{R}^{(m)}$}} 
	\addplot [color = black,mark=no,thick,densely dashed] table[x index=0, y index=3]{n_cR_cRm_guide_Poiss_more_dc.txt};
	\addlegendentry{\large{$0.4 n^{-1}$}} 
	\end{loglogaxis}
	\end{tikzpicture}
	\hspace{0.3cm}
	\begin{tikzpicture}
	\begin{loglogaxis}[legend columns=1,width=7cm, height=11cm, grid=major,xlabel=\Large{$n$},ylabel=\large{$\mid R_{n}\mid$},xmin=1,xmax=200,ymin=1e-7,ymax=1e-0, legend pos=south west]
	\addplot [color = orange,mark=no,thick,solid] table[x index=0, y index=1]{n_cR_cRm_guide_Poiss_less_dc.txt};
	\addlegendentry{\large{$A_\text{trans}=10^{-4}$, $\varv_{R}$}} 
	\addplot [only marks, color = blue,mark=o,thick,solid] table[x index=0, y index=2]{n_cR_cRm_guide_Poiss_less_dc.txt};
	\addlegendentry{\large{$A_\text{trans}=10^{-4}$, $\varv_{R}^{(m)}$}} 
	\addplot [color = black,mark=no,thick,densely dashed] table[x index=0, y index=3]{n_cR_cRm_guide_Poiss_less_dc.txt};
	\addlegendentry{\large{$5 n^{-2}$}} 
	\end{loglogaxis}
	\end{tikzpicture}
	\caption{ Chebyshev coefficients of $\sqrt{\varv_{R}^{(m)}}$ and $\exp(\varv_{R})\sqrt{0.5+0.5x}$ for discontinuous test function $\varv_{D}^\text{(tes)}$. The dashed guidelines are added only for measure of slow decays.}
	\label{fig:n_cR_cRm_Poiss_dc}
\end{figure}
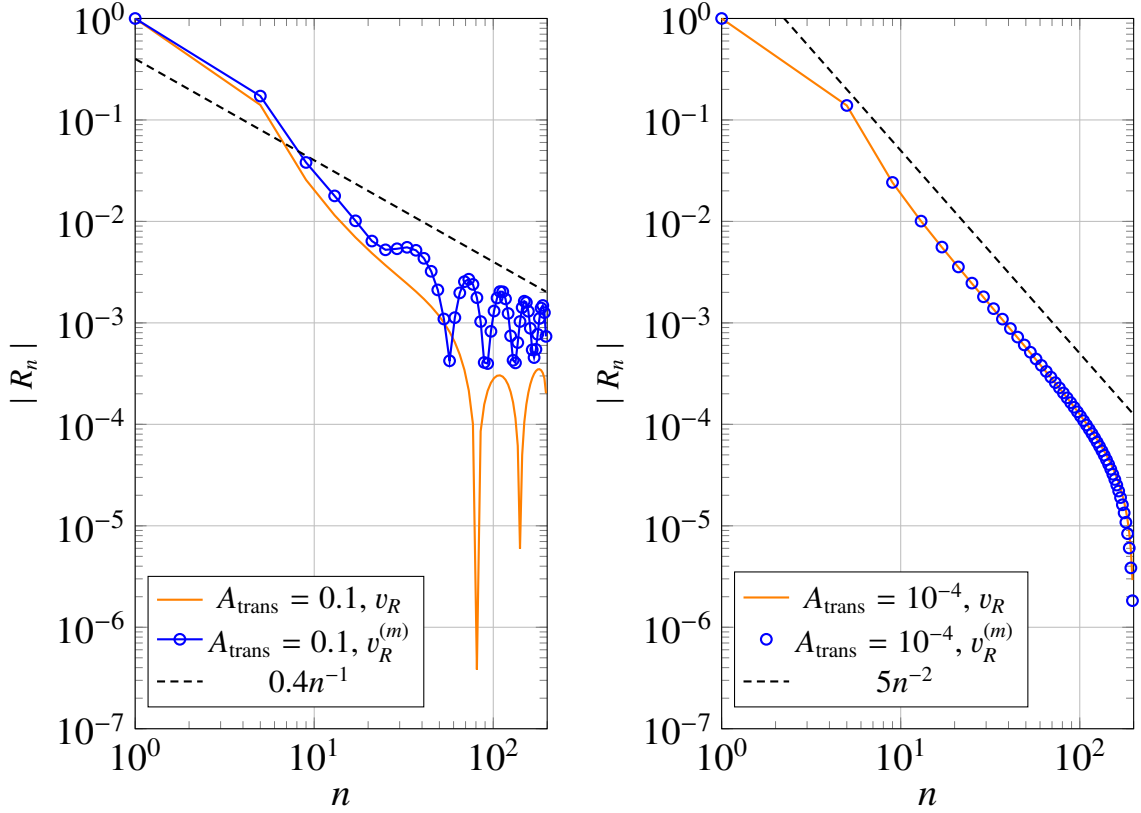

In Section \ref{sec:mod_vR} the modification of function from $\varv_{R}$ to $\varv_{R}^{(m)}$ changes the numerical result significantly; especially, a slow decay of the Chebyshev coefficients is observed. This also may be associated with the effect of discontinuous behavior of independent variable $\varv_{D}$ on $\varv_{R}^{(m)}$ in Poisson equation. We tested the following test function
\begin{align}
\varv_{D}^\text{(tes)}=\varv_{D\text{o}}(x)\left(A_\text{trans}\,\Theta\left[\frac{1+x-0.001}{2}\right]+1\right),
\end{align}
where $A_\text{trans}$ is a small positive number and $\varv_{D\text{o}}(x)$ is the regularized density of the reference solution. We solved the Poisson equation with the fixed $\varv_{D}^\text{(tes)}$ and different $A_\text{trans}$. When the value of $A_\text{trans}$ is very small such as 0.00001, Figure \ref{fig:n_cR_cRm_Poiss_dc}(Right panel) compares the solutions $\sqrt{\varv_{R}^{(m)}}$ and $\exp(\varv_{R})\sqrt{0.5+0.5x}$ (that are supposed to be the same if the Poisson equation is successfully integrated) and shows the difference appears only at order of $10^{-4}$. On one hand, when $A_\text{trans}$ is close to unity such as 0.1, not only the difference appears in the value of coefficients at order of 0.1 but also $\varv_{R}^{(m)}$ shows a slower decay compared to $\varv_{R}$ (Left Panel, Figure \ref{fig:n_cR_cRm_Poiss_dc}).

\subsection{Solving the ss-OAFP equation with Fixed $\{Q_{n}\}$}\label{sec:Fixed_cQ}

\begin{figure}[H]
	\centering
	\tikzstyle{every node}=[font=\Large]
	\begin{tikzpicture}
	\begin{loglogaxis}[legend columns=1,width=13cm, height=10cm, grid=major,xlabel=\Large{$\mathcal{N}$},xmin=10,xmax=2000,ymin=1e-17,ymax=1e-1, legend pos=north east]
	\addplot [color = red,mark=triangle,thick,solid] table[x index=0, y index=1]{N_Delc1_Delc4_vKmin_Fixed_cQ_betao.txt};
	\addlegendentry{\large{$\mid 1-c_{1}(\mathcal{N})/c_{1}(\mathcal{N}=1000)\mid$}} 
	\addplot [color = orange,mark=o,thick,solid] table[x index=0, y index=2]{N_Delc1_Delc4_vKmin_Fixed_cQ_betao.txt};
	\addlegendentry{\large{$\mid 1-c_{4}(\mathcal{N})/c_{4}(\mathcal{N}=1000)\mid$}} 
	\addplot [color = blue,mark=square,thick,solid] table[x index=0, y index=3]{N_Delc1_Delc4_vKmin_Fixed_cQ_betao.txt};
	\addlegendentry{\large{$\mid \varv_{I}(x=1)\mid$}} 
	\end{loglogaxis}
	\end{tikzpicture}
	\caption{Relative error in $c_{1}$ and $c_{4}$ and the value of $\mid\varv_{I}(x=-1)\mid$ obtained by numerical integration of the 4ODE and $Q$-integral for the fixed $\{Q_{n}\}$. $c_{1}$ and $c_{4}$ are compared to their values at $\mathcal{N}=1000$.}
	\label{fig:N_Delc1_Delc4_vKmin_Fixed_cQ_betao}
\end{figure}
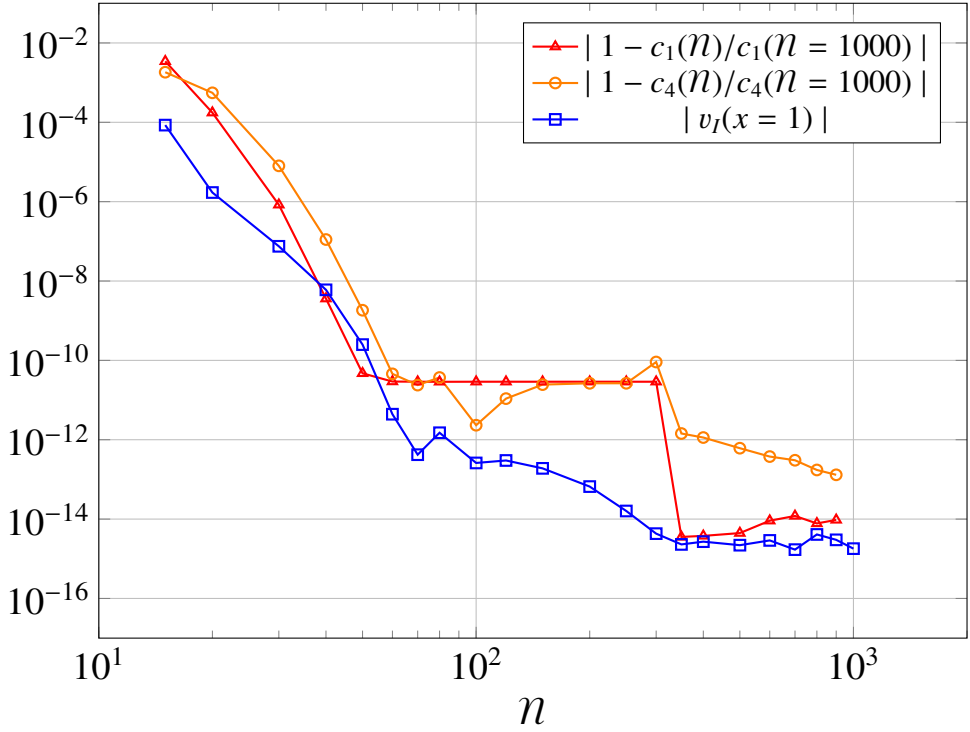

To test whether the origin of numerical instability in integration of the ss-OAFP system is \emph{only} from the large change in independent variables due to the factor $(-E)^{\beta}$ in 4ODE \eqref{Eq.ss_OAFP_F} - \eqref{Eq.ss_OAFP_J}, the present appendix shows a result of solving the 4ODE and $Q$-integral for fixed coefficients $\{Q_{n}\}$. As test coefficients, we used the Chebyshev coefficients for the reference solutions (depicted in Figure \ref{fig:cFKGLRQ_N70}). We found that, for the fixed $\{Q_{n}\}$, the Chebyshev coefficients of spectral solutions show very stable behaviors (Figure \ref{fig:n_cFGKL_N1000_fixed_cQ}); coefficients $\{F_{n}\}$, $\{G_{n}\}$, $\{I_{n}\}$ and $\{J_{n}\}$ reach order of $10^{-15}$ around at $n=90$ and show flattenings even at large index ($n\approx1000$) that are due to the rounding error. Also, Figure \ref{fig:N_Delc1_Delc4_vKmin_Fixed_cQ_betao} compares the values of $c_{1}$ and $c_{4}$ at different degrees $\mathcal{N}=10, 20, \dots, 900$ to the corresponding values at $\mathcal{N}=1000$. The relative error in $c_{1}$ and the value of $\mid\varv_{I}(x=-1)\mid$ reach machine precision around at $\mathcal{N}=400$ while the error in $c_4$ reaches order of $10^{-13}$ at $\mathcal{N}=1000$. The relative error in $c_{1}$ flattens at order of $10^{-13}$ that appears on degrees $70 \leq \mathcal{N} \leq 300$.  This result would reflect the fact that the minimum absolute value of test coefficients $\{Q_{n}\}$ is order of $10^{-13}$ and to gain more accurate solution one needs more Gauss-Chebyshev nodes near the endpoints. 

The result of the present appendix is important to consider the cause of the numerical instability. The difference between the 4ODE with fixed $\{Q_n\}$ and those with unfixed may appear in equation \eqref{Eq.Asymp_ss-4ODE-vJ}. In the equation as $x\to-1$ the differentiations of $\varv_F$, $\varv_Q$ and $\varv_J$ becomes significant compered to the rest of factors and terms. For fixed $\varv_Q$, one can determine $\varv_F$ in the equation while $\varv_J$ is determined from equation \eqref{Eq.ss-4ODE-vF}. In case of non-fixed $\varv_Q$, as one can see the form of the $Q$-integral, the value of the integral is undetermined beyond $E\approx-0.06$ at which $(-E)^\sigma$ reaches machine precision. This infers $\varv_Q$ must be also further determined as $E\to0$ with an extra equation. Hence, for non-fixed $\varv_Q$ equation \eqref{Eq.Asymp_ss-4ODE-vJ} becomes an underdetermined problem at $E<-0.06$; a possible remedy would be to enhance machine precision.

\begin{figure}[H]
	\centering
	\tikzstyle{every node}=[font=\Large]
	\begin{tikzpicture}
	\begin{loglogaxis}[legend columns=1,width=6cm, height=8cm, grid=major,xlabel=\Large{$n$},title=\large{$\mid F_{n}/F_{1} \mid$},xmin=1,xmax=1000,ymin=1e-17,ymax=1e-0, legend pos=north east]
	\addplot [color =orange,mark=no,thick,solid] table[x index=0, y index=1]{n_cFGKL_N1000_fixed_cQ.txt};
	\end{loglogaxis}
	\end{tikzpicture}
	\begin{tikzpicture}
	\begin{loglogaxis}[legend columns=1,width=6cm, height=8cm, grid=major,xlabel=\Large{$n$},title=\large{$\mid G_{n}/G_{1} \mid$},xmin=1,xmax=1000,ymin=1e-17,ymax=1e-0, legend pos=north east]
	\addplot [color =red,mark=no,thick,solid] table[x index=0, y index=2]{n_cFGKL_N1000_fixed_cQ.txt};
	\end{loglogaxis}
	\end{tikzpicture}
	
	\vspace{0.4cm}
	
	\begin{tikzpicture}
	\begin{loglogaxis}[legend columns=1,width=6cm, height=8cm, grid=major,xlabel=\Large{$n$},title=\large{$\mid I_{n}/I_{1} \mid$},xmin=1,xmax=1000,ymin=1e-17,ymax=1e-0, legend pos=north east]
	\addplot [color =blue,mark=no,thick,solid] table[x index=0, y index=3]{n_cFGKL_N1000_fixed_cQ.txt};
	\end{loglogaxis}
	\end{tikzpicture}
	\begin{tikzpicture}
	\begin{loglogaxis}[legend columns=1,width=6cm, height=8cm, grid=major,xlabel=\Large{$n$},title=\large{$\mid J_{n}/J_{1} \mid$},xmin=1,xmax=1000,ymin=1e-17,ymax=1e-0, legend pos=north east]
	\addplot [color =green,mark=no,thick,solid] table[x index=0, y index=4]{n_cFGKL_N1000_fixed_cQ.txt};
	\end{loglogaxis}
	\end{tikzpicture}
	\caption{Absolute values of Chebyshev coefficients for regularized functions for which ss-OAFP system (only 4ODE and $Q$-integral) was solved with fixed $\{Q_{n}\}$ on the whole domain ($\mathcal{N}=1000$, $L=1$, $F_\text{BC}=1$).}
	\label{fig:n_cFGKL_N1000_fixed_cQ}
\end{figure}
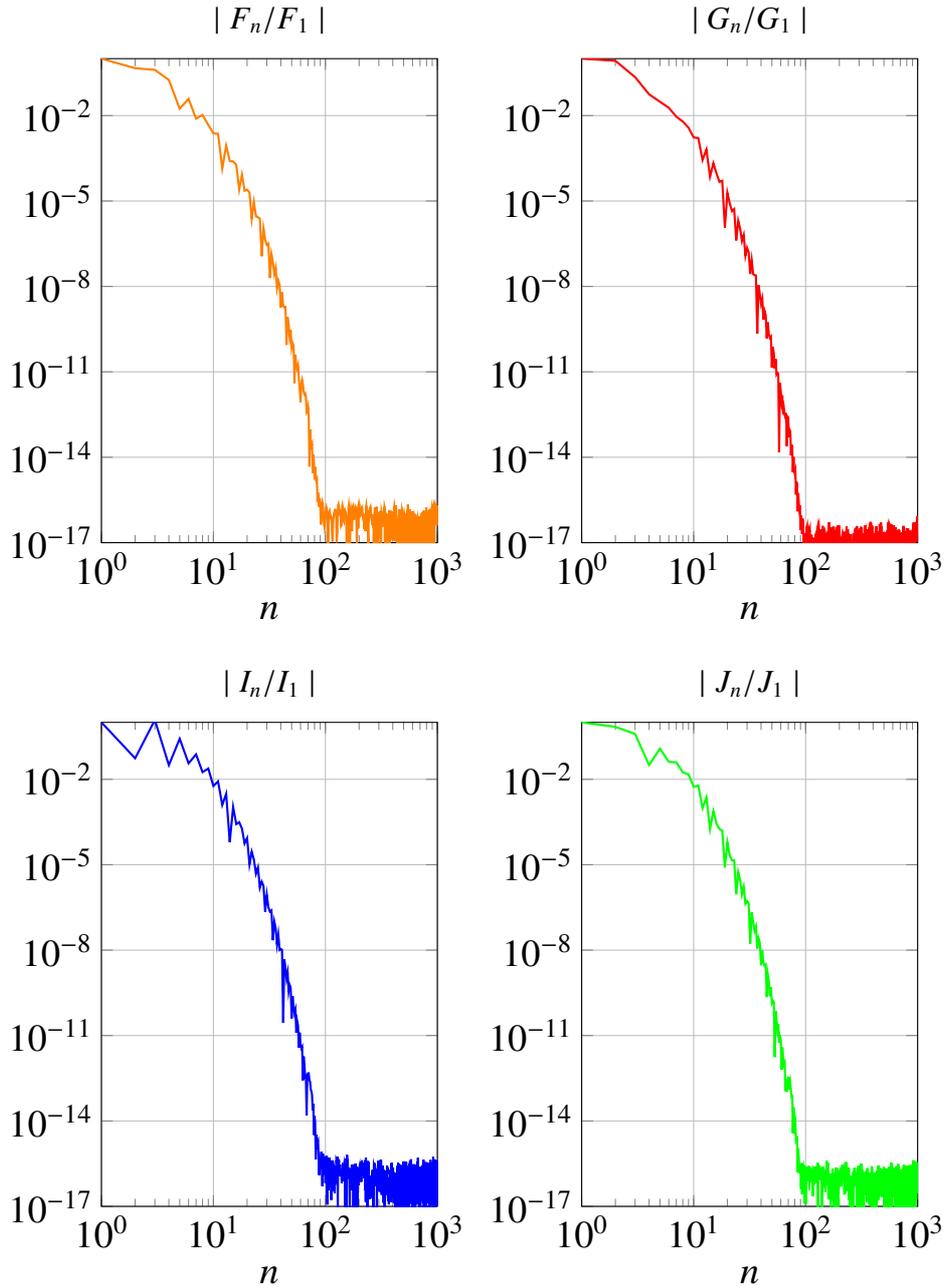

\section{Relation between the reference- and Heggie-Stevenson's solutions}\label{Appendix_Ref_HS}

The reproduced HS's solution in Section \ref{sec:repro_HS_soln} is not satisfactory due to the limited available degrees $\mathcal{N}$, hence the present appendix tests variants of modified independent variables aiming to detail a distinct condition to systematically find the reference solution and the HS's solution even for high degree of polynomials.  For this appendix, we employed formulations similar to the formulation of \citep{Heggie_1988}. Appendix \ref{sec:mod_R_I_G_J} shows the results that we obtained by solving the ss-OAFP system after modifying the regularization of variables $\varv_R$, $\varv_I$, $\varv_G$, and $\varv_J$ and Appendix \ref{sec:mod_F} after modifying that of $\varv_F$. The latter modification provided the reference- and HS's solutions with reasonable accuracy by controlling $E_\text{max}$. However, the available degrees of polynomials are limited in the same way as in Section \ref{sec:repro_HS_soln}. This motivated us to apply to the ss-OAFP system combinations of modified variables employed in Appendix  \ref{sec:mod_F} and Section \ref{sec:repro_HS_soln} (Appendix \ref{sec:mod_double_R_F}). This combination reproduced the HS's and reference- solutions even for high ($\sim200$) degree of polynomials.

\subsection{Modifying the regularization of variables $\varv_R$, $\varv_I$, $\varv_G$, and $\varv_J$ in the ss-OAFP system}\label{sec:mod_R_I_G_J}

We first examined formulations similar to \citep{Heggie_1988}'s formulation but they were not useful, rather they  increased the condition number for the 4ODEs and $Q$-integral. First, we introduced the following modified independent variables 
\begin{subequations}
	\begin{align}
	&\varv_{R}^\text{(m2)}(x)\equiv(0.5+0.5x)^{\nu}\,\varv_{R}(x),\\
	&\varv_{I}^\text{(m)}(x)\equiv(0.5+0.5x)^{\beta}\,\varv_{I}(x),\\
	&\varv_{G}^\text{(m)}(x)\equiv(0.5+0.5x)^{\beta}\,\varv_{G}(x),\\
	&\varv_{J}^\text{(m2)}(x)\equiv(0.5+0.5x)^{b}\varv_{J}(x).\hspace{1cm}(b\geq 0)
	\end{align}
\end{subequations}
If one applies all the modified functions to the ss-OAFP model, the new system is very similar to the HS's formulation. We found the whole and truncated-domain- solutions for the ss-OAFP system with the modified functions using the procedure of Section \ref{subsec:numeric_treat} after we
tested many different combinations of the modified variables. The first three modified independent variables ($\varv_{R}^\text{(m2)}$, $\varv_{I}^\text{(m)}$ and $\varv_{G}^\text{(m)}$) did not change the results almost at all compared to the reference solution. On one hand, the fourth modification ($\varv_{J}^\text{(m2)}$) provided very high condition numbers. For $b=1$, we found a spectral solution on whole domain and it is almost identical to the reference solution, while the condition number was high $\sim 10^{11}$. For $b>1$, the Newton method was hard to work due to higher condition numbers on whole domain. On one hand, we found solutions on truncated domain with $b=\beta $ near $E_\text{max}=-0.225$. These numerical parameters are close to those used in HS's work. However, the condition number is still high ($\sim10^{13}$). Solutions with high condition numbers (close to a reciprocal of machine precision) are generally less trustful \citep[e.g. Section 3.3 of][]{Walter_2014}. Also, as done in \citep{Heggie_1988}, we had to shorten the size of Newton step to less than 0.1 to find those solutions using Newton iteration method, which costed an unfeasible CPU time.

\subsection{Modifying the regularization of variable $\varv_F$ in the ss-OAFP system }\label{sec:mod_F}

We employed the following modification that provides a sensible condition to find the both HS's and reference- solutions only by controlling $E_\text{max}$
\begin{align}
\varv_{F}^\text{(m)}=\ln\left(\exp[\varv_{F}]\left[\frac{1+x}{2}\right]^{\beta}\right).
\end{align}
We solved the ss-OAFP system for $\varv_{F}^\text{(m)}$ and unmodified variables $\varv_{S}, \varv_{Q},\varv_{G},\varv_{I},\varv_{J}$ and $\varv_{R}$ using the procedure of Section \ref{subsec:numeric_treat}. In a similar way to the $\varv_{R}^\text{(m)}$-formulation (Sections \ref{sec:mod_vR} and \ref{sec:repro_HS_soln}),  the spectral solution based on $\varv_{F}^\text{(m)}$-formulation is close to the HS's solution for small $E_\text{max}$ while it also can be close to the reference solution for small  $E_\text{max}$ (See Table \ref{table:Repro_HS_variants} in which $(\varv_{F}^{\text{(m)}},\varv_{R},)$ is the corresponding result).  Due to the logarithmic endpoint singularity of $\varv_{F}^\text{(m)}$, the Chebyshev coefficients $\{F_{n}^\text{(m)}\}$ for $\varv_{F}^\text{(m)}$  show slow decays for both large- and small- $E_\text{max}$ (Figure \ref{fig:n_cF_HS_Polin_dc3}).  A more distinct slow decay appears in Chebyshev coefficients $\{I_{n}\}$ for $\varv_{I}$ especially when $E_\text{max}$ is large (Figure \ref{fig:n_cI_HS_Polin_dc3}). Interestingly, the value of $\varv_{I}(x=1)$ is still order of $10^{-4}$ for large  $E_\text{max}$ that is the same order as the value given by the modified function $\varv_{R}^\text{(m)}$ in Sections \ref{sec:mod_vR} and \ref{sec:repro_HS_soln}. This infers that the HS's solution may be obtained when a numerical scheme has a low accuracy and $E_\text{max}$ is small ($\approx-0.225$). This condition occurred to our spectral solutions when we intentionally included the non-analytic and non-regular properties in the solutions and so Chebyshev coefficients decayed slowly. The modified function $\varv_{F}^\text{(m)}$ provides the HS's solution only for small $\mathcal{N}$, hence one may further be able to find the HS's solution with larger $\mathcal{N}$ by controlling the singularities in independent variables.

\subsection{Combination of modified variables $(\varv_{R}^\text{(m)},\varv_{F}^\text{(m)})$ to find the HS's solution with high degree of Chebyshev polynomials }\label{sec:mod_double_R_F}

Double modification $(\varv_{R}^\text{(m)},\varv_{F}^\text{(m)})$ provides a proper feature of ss-OAFP solutions in the sense that one can obtain the reference- and HS's solutions for high degrees ($\mathcal{N}=80\sim 200$).  The results of Appendix \ref{sec:mod_F} shows that slowing the rapid decay in Chebyshev coefficients is also a key to find the both HS's and reference solutions based on a single formulation. Hence, we combined the two formulations of Appendix  \ref{sec:mod_F} and Section \ref{sec:repro_HS_soln}. As expected, we found the HS's and reference- solutions only by controlling the value of $E_\text{max}$ based on double modification $\left(\varv_{R}^\text{(m)},\varv_{F}^\text{(m)}\right)$. This double modification provided spectral solutions that can reach high degree, such as $\mathcal{N}=200$ for $E_\text{max}=-0.225$, while it also provided a spectral solution close to the reference solution for $E_\text{max}=-0.04$ and $\mathcal{N}=80$ (Table \ref{table:Repro_HS_variants}). One may conclude that the HS's solution can be found around for small $E_\text{max}(\approx-0.225)$ with low accuracy ($\varv_{I}(x=1)=\mathcal{O}(10^{-4})$) while the reference solution can be found for large $E_\text{max}(\approx-0.05)$ with high accuracy (at least $\varv_{J}(x=1)=\mathcal{O}(10^{-6})$).  Also, the Chebyshev coefficients $\{F_{n}\}$ and $\{I_{n}\}$ show a distinctive difference between the two solutions. The coefficients $\{F_{n}\}$ decay in different fashions depending on the combination of modifications for $\varv_{F}$ and $\varv_{R}$ (Figure \ref{fig:n_cF_HS_Polin_dc3}) while the absolute values of $\{I_{n}\}$ stall approximately at $10^{-4}\sim10^{-5}$ for $E_\text{max}\approx-0.25$ and at  $10^{-6}\sim10^{-7}$ for $E_\text{max}\approx-0.05$ (Figure \ref{fig:n_cI_HS_Polin_dc3}). The latter would well reflect the fact that $\{I_\text{n}\}$ is directly associated with $\varv_{I}$ and $\beta$ that are more stable against numerical parameters compared to $c_4$, accordingly $\varv_F$.

\begin{table*}\centering
	\ra{1.3}
	\begin{tabular}{@{}c|clllll@{}}\toprule
		function &$\mathcal{N}$ &	$E_\text{max}$ & $\beta$ & $c_{1}$ & $c_{4}$ &$\mid \varv_{I}(x=1)\mid $ \\   
		\midrule
		$\left(\varv_{F}^{\text{(m)}},\varv_{R}\,\,\right)$& 15  &	$-0.24$ & $8.181$ & $9.1014\times10^{-4}$ &  $3.516\times10^{-1}$   & $3.3\times10^{-4}$ \\
		$\left(\varv_{F}\,\,,\varv_{R}^{\text{(m)}}\right)$& 15  &	$-0.24$ & $8.1731$ & $9.1023\times10^{-4}$ &  $3.524\times10^{-1}$   & $2.1\times10^{-4}$ \\
		$\left(\varv_{F}^{\text{(m)}},\varv_{R}^{\text{(m)}}\right)$& 200 &	$-0.225$ & $8.175860$ & $9.1018\times10^{-4}$& $3.523\times10^{-1}$  & $2.9\times10^{-4}$ \\
		\midrule
		$\left(\varv_{F}^{\text{(m)}},\varv_{R}\,\right)$ & 70  &	$-0.00525$& $8.1783712$ & $9.0925\times10^{-4}$ &  $3.304\times10^{-1}$   
		& $2.2\times10^{-6}$ \\
		$\left(\varv_{F}\,,\varv_{R}^{\text{(m)}}\right)$ & 200  &	$-0.04$ & $8.178371129$ & $9.0925\times10^{-4}$ &  $3.301\times10^{-1}$   
		& $4.9\times10^{-7}$\\
		$\left(\varv_{F}^{\text{(m)}},\varv_{R}^{\text{(m)}}\right)$ & 80 &	$-0.04$ & $8.1783683$ & $9.0926\times10^{-4}$ & $3.301\times10^{-1}$  & $8.9\times10^{-7}$ \\
		\bottomrule
	\end{tabular}\\
	\caption{ Eigenvalues and $\mid \varv_{I}(x=1)\mid$ for combinations of $\varv_{R}^\text{(m)}$ and $\varv_{F}^\text{(m)}$. The upper three rows are the data that reproduced the HS's eigenvalues while the lower three rows are the data that reproduced three significant figures of the reference eigenvalues. In the modified ss-OAFP systems, the variables $\varv_{S}, \varv_{Q},\varv_{G},\varv_{I}$ and $\varv_{J}$ are not modified}
	\label{table:Repro_HS_variants}
\end{table*}

\begin{figure}[H]
	\centering
	\tikzstyle{every node}=[font=\large]
	\begin{tikzpicture}
	\begin{loglogaxis}[legend columns=1,width=7cm, height=8cm, grid=major,xlabel=\Large{$n$},title=\large{$\mid F_{n}^\text{(m)}\mid$},xmin=1,xmax=150,ymin=1e-11,ymax=1e1, legend pos=south west]
	\addplot [color = blue,mark=o,thick,solid] table[x index=0, y index=1]{n_cF_cK_HS_Polin_dc3_N200_xmin_055_tes.txt};
	\addlegendentry{\large{$\left( \varv_{F}^\text{(m)}, \varv_{R}^\text{(m)}\right)$}} 
	\addplot [color = red,mark=square,thick,solid] table[x index=0, y index=1]{n_cF_cK_HS_dc3_N15_xmin_052_tes.txt};
	\addlegendentry{\large{$\left( \varv_{F}^\text{(m)}, \varv_{R}\,\,\right)$}}
	\node[above,black] at (30,1e-3) {\small{$E_\text{max}=-0.225$}};
	\node[above,black] at (3e0,1e-7) {\small{$E_\text{max}=-0.24$}};
	\draw[thin, dashed] (40,2e-3) -- (70,1e-6);
	\draw[thin, dashed] (4,1e-6) -- (6,4e-4);
	\end{loglogaxis}
	\end{tikzpicture}
	\hspace{0.5cm}
	\begin{tikzpicture}
	\begin{loglogaxis}[legend columns=1,width=7cm, height=8cm, grid=major,xlabel=\Large{$n$},title=\large{$\mid F_{n}^\text{(m)}\mid$},xmin=1,xmax=150,ymin=1e-11,ymax=1e1, legend pos=south west]
	\addplot [color = black,mark=no,thick,solid] table[x index=0, y index=1]{n_cF_cK_HS_Polin_dc3_N80_xmin_092_tes.txt};
	\addlegendentry{\large{$\left( \varv_{F}^\text{(m)}, \varv_{R}^\text{(m)}\right)$}} 
	\addplot [color = orange,mark=triangle,thick,solid] table[x index=0, y index=1]{n_cF_cK_HS_dc3_N70_xmin_09895_tes.txt};
	\addlegendentry{\large{$\left( \varv_{F}^\text{(m)}, \varv_{R}\,\,\right)$}}
	\node[above,black] at (5e0,1e-6)  {\small{$E_\text{max}=-0.04$}};
	\node[above,black] at (4e1,4e-2)  {\small{$E_\text{max}=-0.00525$}};
	\draw[thin, dashed] (50,1e-1) -- (40,1e-4);
	\draw[thin, dashed] (5,1e-5) -- (10,1e-3);
	\end{loglogaxis}
	\end{tikzpicture}
	\caption{ Absolute values of Chebyshev coefficients $\{F_{n}^\text{(m)}\}$ for $\varv_{F}^\text{(m)}$.  In the modified ss-OAFP system, $\varv_{S}, \varv_{Q},\varv_{G},\varv_{I}$ and $\varv_{J}$ are not modified. The maximum values $E_\text{max}$ of the truncated domains are also depicted in the figure.}
	\label{fig:n_cF_HS_Polin_dc3}
\end{figure}

\begin{figure}[H]
	\centering
	\tikzstyle{every node}=[font=\large]
	\begin{tikzpicture}
	\begin{loglogaxis}[legend columns=1,width=7cm, height=8cm, grid=major,xlabel=\Large{$n$},title=\large{$\mid I_{n}\mid$},xmin=1,xmax=150,ymin=1e-12,ymax=1e-0, legend pos=south west]
	\addplot [color = blue,mark=o,thick,solid] table[x index=0, y index=2]{n_cF_cK_HS_Polin_dc3_N200_xmin_055_tes.txt};
	\addlegendentry{\large{$\left( \varv_{F}^\text{(m)}, \varv_{R}^\text{(m)}\right)$}} 
	\addplot [color = red,mark=square,thick,solid] table[x index=0, y index=2]{n_cF_cK_HS_dc3_N15_xmin_052_tes.txt};
	\addlegendentry{\large{$\left( \varv_{F}^\text{(m)}, \varv_{R}\,\,\right)$}}
	\node[above,black] at (40,1e-8) {\small{$E_\text{max}=-0.225$}};
	\node[above,black] at (4e0,9e-8) {\small{$E_\text{max}=-0.24$}}; 
	\draw[thin, dashed] (50,5e-8) -- (60,5e-5);
	\draw[thin, dashed] (4,1e-6) -- (9,3e-4);
	\end{loglogaxis}
	\end{tikzpicture}
	\hspace{0.5cm}
	\begin{tikzpicture}
	\begin{loglogaxis}[legend columns=1,width=7cm, height=8cm, grid=major,xlabel=\Large{$n$},title=\large{$\mid I_{n}\mid$},xmin=1,xmax=150,ymin=1e-12,ymax=1e-0, legend pos=south west]
	\addplot [color = black,mark=no,thick,solid] table[x index=0, y index=2]{n_cF_cK_HS_Polin_dc3_N80_xmin_092_tes.txt};
	\addlegendentry{\large{$\left( \varv_{F}^\text{(m)}, \varv_{R}^\text{(m)}\right)$}} 
	\addplot [color = orange,mark=triangle,thick,solid] table[x index=0, y index=2]{n_cF_cK_HS_dc3_N70_xmin_09895_tes.txt};
	\addlegendentry{\large{$\left( \varv_{F}^\text{(m)}, \varv_{R}\,\,\right)$}}
	\node[above,black] at (4e1,7e-2)  {\small{$E_\text{max}=-0.04$}};
	\node[above,black] at (9e0,5e-9)  {\small{$E_\text{max}=-0.00525$}};
	\draw[thin, dashed] (80,8e-6) -- (40,5e-2);
	\draw[thin, dashed] (7,4e-8) -- (50,1e-6);
	\end{loglogaxis}
	\end{tikzpicture}
	\caption{Absolute values of Chebyshev coefficients $\{I_{n}\}$ for $\varv_{I}$.   In the modified ss-OAFP system, $\varv_{S}, \varv_{Q},\varv_{G},\varv_{I}$ and $\varv_{J}$ are not modified. The maximum values $E_\text{max}$ of the truncated domains are also depicted in the figure.}
	\label{fig:n_cI_HS_Polin_dc3}
\end{figure}

\end{appendices}


\bibliographystyle{elsarticle-harv} 
\bibliography{science}


\end{document}